\documentclass{aa}  
\usepackage[varg]{txfonts}
\usepackage{graphicx}
\usepackage{xcolor}

\usepackage{afterpage}
\usepackage{soul}
\usepackage{natbib}
\bibpunct{(}{)}{;}{a}{}{,}
\usepackage[bookmarksopen=false,bookmarksnumbered=true,breaklinks=true,colorlinks=true,linkcolor=blue,citecolor=blue]{hyperref}
\usepackage{booktabs}
\usepackage{threeparttable}
\usepackage{comment}

\usepackage{longtable}
\usepackage{supertabular}
\usepackage{array}                                      
\usepackage{booktabs}                           
\usepackage{multicol,multirow}

\newcolumntype{C}{>{\begin{math}}c<{\end{math}}}
\newcolumntype{L}{>{\begin{math}}l<{\end{math}}}

\newcommand{\pc}{\ensuremath{\mathrm{pc}}}

\newcommand{\kelvin}{\ensuremath{\mathrm{K}}}

\newcommand{\myr}{\ensuremath{\mathrm{Myr}}}

\newcommand{\au}{\ensuremath{\mathrm{au}}}

\newcommand{\kms}{\ensuremath{\mathrm{km/s}}}

\newcommand{\mm}{\ensuremath{\mathrm{mm}}}
\newcommand{\mic}{\ensuremath{\mathrm{\mu m}}}

\newcommand{\pixs}{\ensuremath{\mathrm{pixels}}}

\newcommand{\jy}{\ensuremath{\mathrm{Jy}}}
\newcommand{\mjy}{\ensuremath{\mathrm{mJy}}}
\newcommand{\microjy}{\ensuremath{\mu\mathrm{Jy}}}

\newcommand{\dego}{\ensuremath{^\circ}}

\newcommand{\pa}[1]{\left(#1\right)}    
\newcommand{\pac}[1]{\left[#1\right]}

\newcommand{\mj}{\ensuremath{\mathrm{M_{Jup}}}}

\newcommand{\mearth}{\ensuremath{\mathrm{M}_\oplus}}

\newcommand{\msun}{\ensuremath{\textnormal{M}_\odot}}
\newcommand{\lsun}{\ensuremath{\textnormal{L}_\odot}}

\definecolor{lagoon}{rgb}{0.3, 0.5, 0.7}
\definecolor{greenjuju}{rgb}{0.1, 0.5, 0.5}

\makeatletter
\let\do@mlinenumbers\relax

\let\linenumbers\relax
\makeatother

\begin{document}

   \title{Dust populations from 30 to 1000 au in the debris disk of HD\,120326}
   \subtitle{Panchromatic view with VLT/SPHERE, ALMA, and HST/STIS}
   \authorrunning{Desgrange et al.}
   \author{C.~Desgrange\inst{1} \fnmsep\inst{2}\fnmsep\inst{3}, J.~Milli\inst{2}, G.~Chauvin\inst{3}, M.~Bonnefoy\inst{2}, Th.~Henning\inst{3}, J.~Miley\inst{4} \fnmsep\inst{5}\fnmsep\inst{6}, G.~Kennedy\inst{7} \fnmsep\inst{8}, S.~Juillard\inst{9}, J.~Olofsson\inst{10}, J.-C.~Augereau\inst{2}, V.~Faramaz\inst{11}, V.~Christiaens\inst{9} \fnmsep\inst{12}, A.~A.~Sefilian\inst{11}, 
   J.~Mazoyer\inst{13}, T.~D.~Pearce\inst{7}, H.~Beust\inst{2}, F.~Ménard\inst{2}, M.~Booth\inst{14}
        }

   \institute{
    $^{1}$ European Southern Observatory, Alonso de Córdova 3107, Vitacura, Santiago, Chile \\
    $^{2}$ Univ. Grenoble Alpes, CNRS, IPAG, F-38000 Grenoble, France;  e-mail: celia.desgrange@univ-grenoble-alpes.fr \\
    $^{3}$ Max Planck Institute for Astronomy, K\"onigstuhl 17, D-69117 Heidelberg, Germany\\ 
    $^{4}$ Departamento de Física, Universidad de Santiago de Chile, Av. Victor Jara 3659, Santiago, Chile \\
    $^{5}$ Millennium Nucleus on Young Exoplanets and their Moons (YEMS)  \\
    $^{6}$ Center for Interdisciplinary Research in Astrophysics Space Exploration (CIRAS), Universidad de Santiago de Chile, Chile \\
    $^{7}$ Department of Physics, University of Warwick, Gibbet Hill Road, Coventry, CV4 7AL, UK\\
    $^{8}$ Centre for Exoplanets and Habitability, University of Warwick, Gibbet Hill Road, Coventry CV4 7AL, UK\\
    $^{9}$ Space sciences, Technologies and Astrophysics Research (STAR) Institute, Universit\'e de Li\`ege, 19 All\'ee du Six Août, 4000 Li\`ege, Belgium\\
    $^{10}$ European Southern Observatory, Karl-Schwarzschild-Strasse 2, 85748 Garching bei München, Germany \\
    $^{11}$ Department of Astronomy/Steward Observatory, The University of Arizona, 933 North Cherry Avenue, Tucson, AZ 85721, USA \\
    $^{12}$ KU Leuven, Institute for Astronomy, Celestijnenlaan 200D, Leuven, Belgium \\
    $^{13}$\;LESIA, Observatoire de Paris, Universit\'e PSL, CNRS, Sorbonne Université, Université de Paris, 5 place Jules Janssen, 92195 Meudon, France \\ 
    $^{14}$ UK Astronomy Technology Centre, Royal Observatory Edinburgh, Blackford Hill, Edinburgh EH9 3HJ, UK \\
    }

   \date{}

  \abstract
   {To date, more than a hundred debris disks have been spatially resolved. Among them, the young system HD\,120326 stands out, displaying 
   different disk substructures on both intermediate (30--150~au) and large (150--1\,000~au) scales.}
   {We present new VLT/SPHERE (1.0--1.8~$\mu$m) and ALMA (1.3~mm) data of the debris disk around HD\,120326. By combining them with archival HST/STIS ($0.2$--$1.0~\mic$) and archival SPHERE data, we have been able to examine the morphology and photometry of the debris disk, along with its dust properties. }
   {We present the open-access code \texttt{MoDiSc} (Modeling Disks
in Scattered light) to model the inner belt jointly using   the SPHERE polarized and total intensity observations. 
   Separately, we modeled the ALMA data and the spectral energy distribution (SED).
   We combined the results of both these analyses with the STIS data to determine the global architecture of HD\,120326.
   }
   {For the inner belt, identified as a planetesimal belt, we derived a semi-major axis of 43$~\au$,  fractional luminosity of 1.8$\,\times\,10^{-3}$, and maximum degree of polarization of $51\,\%\pm6\,\%$ at $1.6~\mic$. 
   The spectral slope of its reflectance spectrum is red between 1.0 and 1.3~$\mic$ and  gray between 1.3 and 1.8~$\mic$. 
   Additionally, the SPHERE data show that there could be a halo of small particles or a second belt at distances $\leq$150~au. 
   Using ALMA, we derived in the continuum ($1.3$~mm) an integrated flux of $561\pm 20~\mu$Jy. We did not detect any $^{12}$CO emission. 
   At larger separations ($>$150~au), we highlight a spiral-like feature spanning hundreds of astronomical units in the STIS data. 
   }
   {Further data are needed to confirm and better constrain the dust properties and global morphology of HD\,120326.
   }
   {}

    \keywords{Instrumentation: adaptive optics, high angular resolution -- Methods: observational -- Debris disk}
    \titlerunning{Dust populations from 30 to 1\,000 au in HD\,120326's debris disk}
    \maketitle

\section{Introduction}

The direct imaging technique is able to constrain planetary system architectures by probing regions at large ($>$~5 au) separations from the star. This approach can detect the near- and mid-infrared (NIR-MIR) emission of young self-luminous giant planets and resolve circumstellar disks, such as debris disks, made up of leftover material from planet formation processes.  In volume, they consist mainly of small, dust particles resulting from the collisional cascade of larger (i.e., $\sim$\,kilometer-sized)  planetesimals \citep[e.g., reviews from][]{wyatt_evolution_2008,krivov_debris_2010,hughes_debris_2018,pearce_debris_2024}. 

To date, space telescopes and ground-based facilities equipped with extreme adaptive optics instruments have spatially resolved the starlight scattered off dust grains in debris disks down to separations
of $0.1$\arcsec and up to about $10$\arcsec. Such facilities include HST \citep{schneider_nicmos_1999,schneider_probing_2014,augereau_hst_1999,choquet_archival_2014,choquet_hd_2018,ren_postprocessing_2017,ren_debris_2023}, JWST/NIRCam \citep{rebollido_jwsttst_2024,lawson_jwst_2024}, VLT/SPHERE \citep{wahhaj_shardds_2016,boccaletti_observations_2018,xie_referencestar_2022,olofsson_vertical_2022}, and Gemini/GPI \citep{esposito_debris_2020,crotts_uniform_2024}.
Such images require specific strategies of observation (e.g., adaptive optics and coronagraphy) and processing to block the stellar light and reveal the faint disk signal. Scattered-light images enable a better understanding of very small ($\lesssim$ a few $\mic$) dust particles by determining the degree of forward scattering and linear polarization of the dust grains and their color. This provides constraints on the properties of the dust grains, such as their size, shape, and composition.

On the other hand, images of debris disks from mid-infrared to millimeter provide a complementary view because they are sensitive mainly to the cold thermal emission of larger (sub-mm/mm) dust particles; for instance, with JWST/MIRI \citep{gaspar_spatially_2023,rebollido_jwsttst_2024}, \textit{Spitzer} \citep{su_debris_2009}, \textit{Herschel} \citep{lohne_modelling_2012,wyatt_herschel_2012,kennedy_discovery_2013,kennedy_kuiper_2015,kennedy_kuiper_2018}, and ALMA 
\citep{macgregor_millimeter_2013,macgregor_complete_2017,su_alma_2017,marino_alma_2017,marino_gap_2019,faramaz_scatteredlight_2019,faramaz_detailed_2021}. For (sub)mm observations, the location of the large grains seen is thought to track more reliably the parent belt of large, colliding planetesimals, as small grains are subject to stellar radiation effects that make them drift from the parent belt where they are produced, thereby smearing any asymmetry gravitationally induced by massive perturbers \citep{thebault_collisional_2007}.
The properties of debris disks, including their morphology and the  properties (size, shape, and composition) of the constituent dust grains can therefore be studied more comprehensively by combining multi-wavelength observations.

In total, more than one hundred debris disks have been spatially resolved\footnote{see the catalogs: \url{https://www.astro.uni-jena.de/index.php/theory/catalog-of-resolved-debris-disks.html} and \url{https://circumstellardisks.org/}} so far. Among them, there exists a wide diversity of morphologies: narrow or broad rings, circular, eccentric or misaligned rings, arcs, warps, clumps, swept-back wings, needles, forks, or spiral arms  \citep{hughes_debris_2018}. 
A natural question is what causes this diversity \citep{wyatt_how_1999,lee_primer_2016}, whether this is related mostly to the influence of planetary companions, 
or to other physical mechanisms such as disk self-gravity \citep[which may or may not require the presence of planets; e.g.,][]{ward_dynamics_1998,hahn_secular_2003,jalali_density_2012,sefilian_formation_2021,sefilian_formation_2023}, giant impacts \citep{jones_giant_2023}, collisional dust avalanches \citep{grigorieva_collisional_2007}, stellar flybys \citep[e.g.,][]{kenyon_collisional_2002,reche_investigating_2009,lagrange_narrow_2016},  interstellar medium wind \citep[][]{maness_hubble_2009}, or processes occurring during planet formation in protoplanetary disks \citep[e.g.,][]{najita_pebbles_2022}.

In this context, the system HD\,120326 (HIP~67497), located at $113.3\pm0.4~\pc$ \citep{gaiacollaboration_gaia_2021} belongs to the Upper Centaurus Lupus sub-group in the Sco-Cen association, and hosts a debris disk around the F0V-type young \citep[$16~\myr$;][]{mamajek_post_2002} star of mass of $1.6~\msun$ \citep{chen_spitzer_2014}. Using VLT/SPHERE near-infrared imaging in scattered light, \citet{bonnefoy_belts_2017} reported the existence at intermediate scales ($\lesssim~150~\au$) of a dust belt and a putative second dust structure, which could be another belt or a halo of dust grains. In addition, optical data seems to show scattered light of an extended halo, up to a projected separation of $700~\au$ \citep[][using HST/STIS]{padgett_warm_2016,bonnefoy_belts_2017}. 

Previously, the spectral energy distribution (SED) of HD\,120326 has been modeled with either a single-  \citep{jang-condell_spitzer_2015}, or a double-belt \citep{chen_spitzer_2014} architecture, by using in particular measurements of the Infrared Spectrograph (IRS) data of \textit{Spitzer}. From their SED modeling, \citet{chen_spitzer_2014} found a bright ($L_\text{IR}/L_\star=1.1\times10^{-3}$), relatively cold ($127\pm5~\kelvin$) belt at $13.9~\au$, and a dimmer ($L_\text{IR}/L_\star=1.4\times10^{-4}$), colder ($63\pm5~\kelvin$) belt at $116.5~\au$, by assuming two blackbody models. Alternatively, \citet{jang-condell_spitzer_2015} modeled the data with a single belt made of a single grain population of a given size, temperature, and composition (a mixture of olivine and pyroxene). Their best fit was that of a belt with fractional luminosity of $L_\text{IR}/L_\star=1.5\times10^{-3}$ located at $8.8\pm1.0~\au$, dust particles of temperature of $124\pm5~\kelvin$, size of $15.5~\mu$m, and with a composition converging to pure olivine (instead of a mixture of olivine and pyroxene).

However, both locations of the dust belt are underestimated. This is common for representative radii of dust belts inferred via SED modeling \citep[e.g.,][]{pawellek_dust_2015} because small dust particles are inefficient emitters, resulting in higher temperatures than expected based on their separation from their host star. By using SPHERE  total intensity observations, for the first time \citet{bonnefoy_belts_2017} imaged the dust belt and derived a larger reference radius ($r_0$) of $59\pm3~\au$, using a forward-modeling approach,. They also reported the possible existence of a second, outer disk component, which could be another belt ($r_0\sim130\pm8~\au$) or a halo. However, \citet{bonnefoy_belts_2017} were cautious about this second component, as it is faint, and could also be related to self-subtraction effects \citep{milli_impact_2012} caused by their data processing based on the angular differential imaging \citep[ADI;][]{marois_angular_2006} technique. This purported second structure was not detected in polarized intensity light using the same instrument, SPHERE, at the same wavelength \citep[$1.6~\mic$;][]{olofsson_vertical_2022}. This further complicates the interpretation of the dust distribution around HD\,120326.

In this study, we carried out a comprehensive, multi-wavelength study of the morphological and dust properties of the HD\,120326 debris disk. 
We used a combination of optical, near infrared and millimeter data, including new observations using VLT/SPHERE and ALMA and archival data (HST/STIS and SPHERE).

In Sect. \ref{sec:obs_data_reduction}, we present the data used and our reduction process: SPHERE observations (Sect.~\ref{sec:obs_sphere}; with SPHERE/IRDIS and SPHERE/ZIMPOL polarized intensity data in Sect.~\ref{sec:obs_polar_only} and SPHERE/IRDIS and SPHERE/IFS total intensity data in Sect.~\ref{sec:obs_ti_only}), ALMA observations (Sect. \ref{sec:obs_alma}), and STIS observations (Sect. \ref{sec:obs_hst}). 
In Sect.~\ref{sec:modelling_results_sphere}, we present our joint modeling (Sect.~\ref{sec:disk_modelling}) and analysis of the SPHERE total and polarized intensity observations. We derived new constraints on the properties of the inner dust belt, including one at $1.6~\mic$ its scattering phase function (Sect.~\ref{sec:SPF_DP}) and  maximum fraction of polarization (Sect.~\ref{sec:max_dolp}), and between $1.0~\mic$ and $1.8~\mic$ its reflectance spectrum (Sect.~\ref{sec:disk_reflectance}). We also investigate the nature of the second, faint component at larger distances (Sect.~\ref{sec:second_structure_nature}) first mentioned by \citet{bonnefoy_belts_2017}, which we firmly detect in our new SPHERE total intensity observations. In Sect.~\ref{sec:modelling_results_alma}, we report our study of the ALMA data, including the morphological analysis of the disk (Sect.~\ref{sec:alma_morpho}), as well as the modeling of the SED (Sect.~\ref{sec:sed}) and its dust mass (Sect.~\ref{sec:dust_mass}). In Sect.~\ref{sec:discussion}, we discuss the dust properties contextualized with other debris disks and the global morphology of the debris disk around HD\,120326, based on the STIS, SPHERE, and ALMA images. We present our conclusions in Sect.~\ref{sec:ccl}.

\section{Observations and processing of the SPHERE, ALMA, and STIS data \label{sec:obs_data_reduction}}

Here, we  present the data used and our processing:  VLT/SPHERE (Sect.~\ref{sec:obs_sphere}), ALMA  (Sect.~\ref{sec:obs_alma}), and HST/STIS  (Sect.~\ref{sec:obs_hst}).

\subsection{SPHERE NIR observations \label{sec:obs_sphere}}

The system HD\,120326 (HIP\,67497) was observed both in polarimetry and total intensity with the
Spectro-Polarimetric High-contrast Exoplanet REsearch instrument \citep[SPHERE;][]{beuzit_sphere_2019} using its dual imager IRDIS  \citep{dohlen_infrared_2008} and its Integral Field Spectrograph \citep[IFS,][]{claudi_sphere_2008} at the Very Large Telescope (VLT).
The field of view of IRDIS is $11$”$\times11$” and the plate scale of its detector is $12.25$~mas \citep{maire_first_2016}; whereas for the IFS, these are $1.73$”$\times1.73$” and $7.46$~mas, respectively. The astrometric calibration of both IRDIS and IFS is based on past regular observations of the star crowded field 47~Tuc and includes measurements on the true north, distortion, and plate scale \citep{maire_lessons_2021}. 

We summarize the log of the observations in Table~\ref{tab:obslog}. The seeing and atmospheric coherence time ($\tau_0$) were estimated by the Differential Image Motion Monitor \citep[DIMM,][]{sarazin_eso_1990} and  Multi-Aperture Scintillation Sensor \citep[MASS,][]{kornilov_combined_2007}  turbulence monitor at the Paranal Observatory.
The Strehl ratio is computed by the real-time computer named Standard Platform for Adaptive optics Real Time Applications \citep[SPARTA,][]{fedrigo_sparta_2006} of the SPHERE eXtreme Adaptive Optics system \citep[SAXO,][]{petit_sphere_2014}.
In Sect.~\ref{sec:obs_polar_only}, we present the polarized intensity observations and our data processing. In Sect.~\ref{sec:obs_ti_only}, we report the total intensity observations and our data processing. We compare the polarized and total intensity observations in Sect.~\ref{sec:obs_polar_and_ti}.

\subsubsection{Polarimetric observations \label{sec:obs_polar_only}}

As dust grains scatter light, they  polarize it -- and this process will depend on their intrinsic properties. 
Thus, observing circumstellar disks with polarimetry enables the study of dust grain properties. 
Despite the stellar light being very bright, this technique is ideally suited to reveal the much dimmer signal from the disk. 
Indeed, since stellar light is mostly unpolarized, polarimetric observations facilitate the subtraction of the stellar light via polarization differential imaging (PDI). 

Two polarimetric datasets are available on HD\,120326: NIR data using SPHERE/IRDIS and optical data using SPHERE/ZIMPOL. Below, we first describe the IRDIS data, where the disk signal has  clearly been detected, and then the ZIMPOL data, where no disk signal has been detected.

\medskip

By using the NIR instrument SPHERE/IRDIS in its dual-beam polarimetric imaging \citep[DPI;][]{deboer_polarimetric_2020} mode, the host star of HD\,120326 was observed on the night of the June 1, 2018 (PI: Boccaletti) under well-suited observation conditions, with a seeing of $0.45$” and a coherence time of the atmosphere of $4.2$~ms. The observation was carried out in the field-stabilized mode in the broad band filter H ($\lambda=1.625~\mic$, $\Delta\lambda=0.29~\mic$). 

The IRDIS-DPI mode allows us to image simultaneously the linearly polarized light in two orthogonal directions, by using polarizers with orthogonal transmission axes. 
A polarimetric observation with SPHERE/IRDIS is made of polarimetric cycles consisting of images acquired at four half-wave plate (HWP) angles to measure the Stokes parameters $Q^+$, $Q^-$, $U^+$, and $U^-$. These HWP angles are  $0^\circ$, $45^\circ$, $22.5^\circ$, $67.5^\circ$, respectively, with the $Q^+$ vector aligned with a preferred orientation, usually the meridian on-sky \citep[see Fig.~1 from][]{deboer_polarimetric_2020}.
This allows us to retrieve the two linear polarization contributions, $Q$ and $U$ expressed as 
\begin{align}
    Q \;&=\; \frac{1}{2} (Q^+-Q^-)\,, \\
    U \;&=\; \frac{1}{2} (U^+-U^-)\,.
\end{align}
Expressing $Q$ and $U$ as such enables the removal of the instrumental polarization effects caused by the reflections in the instrument downstream of the HWP. From them, one can reconstruct the linearly polarized intensity via the equation $pI=\sqrt{Q^2+U^2}$, and also the degree of linear polarization following $DoLP=pI/I$, where $I$ is the total intensity \citep{deboer_polarimetric_2020}. Tackling the instrumental polarization generated by the reflections upstream of the HWP require additional corrections, which are described in \citet{canovas_datareduction_2011}. To process the polarimetric observations, we used the public pipeline IRDAP\footnote{\url{https://github.com/robvanholstein/IRDAP}} \citep{vanholstein_polarimetric_2020}, which corrects for instrumental polarization both upstream and downstream the HWP.

Finally in practice, we work with the azimuthal Stokes parameters $Q_\phi$ and $U_\phi$ \citep{schmid_limb_2006}, expressed as
\begin{align}
    Q_\phi \;&=\; -Q \cos(2\phi) - U \sin(2\phi) \,,\\
    U_\phi \;&=\; +Q \sin(2\phi) - U \cos(2\phi)
    \label{eq_Stokes}
\end{align}
where $\phi$ is the angle between the north and the point of interest (from the north toward the easterly direction). A positive  $Q_\phi$ signal corresponds to polarization oriented in azimuthal direction (with respect to the position of the star), while a negative  $Q_\phi$ corresponds to radial polarization. The $U_\phi$ signal corresponds to polarization angles oriented at $\pm45\dego$ with respect to the radial (or azimuthal) direction.

\begin{figure*}[h]
    \centering
    \includegraphics[height=3.95cm]{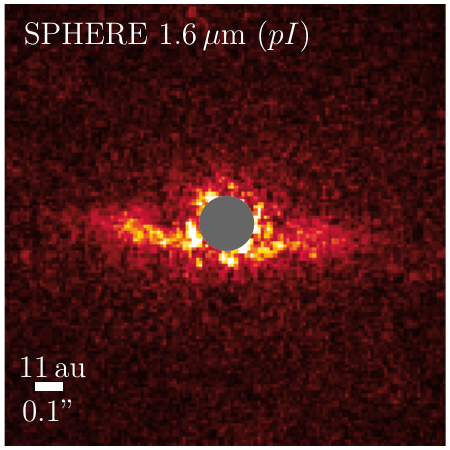} 
    \includegraphics[height=3.95cm]{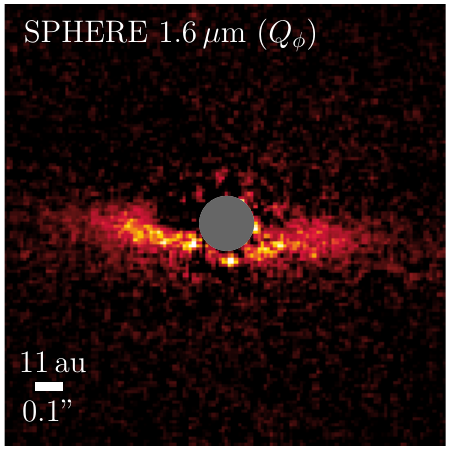}
    \includegraphics[height=3.95cm]{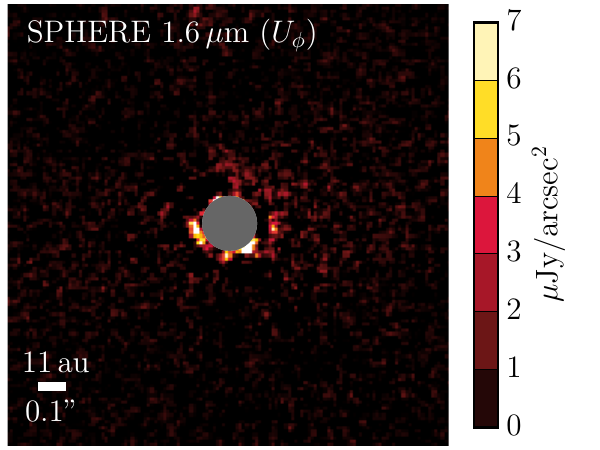} \hfill
    \includegraphics[height=3.95cm]{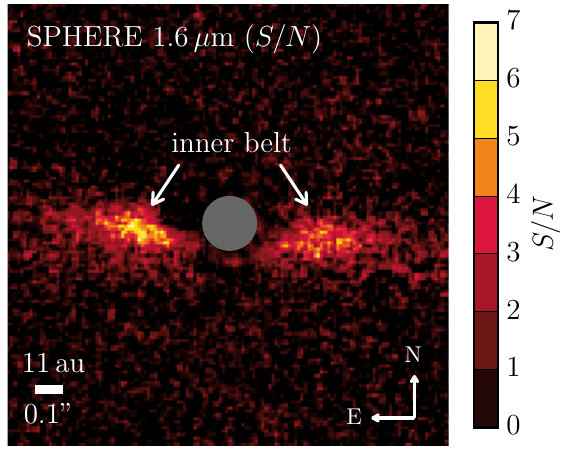} 
    \caption{Polarimetric images of the system HD\,120326 using VLT/SPHERE at $1.6~\mic$ (broad band H; scattered light; epoch 2018-06-01). We processed the data by using the pipeline IRDAP. From left to right, we show the Stokes vector components $pI$ (intensity), $Q_\phi$ (azimuthal or radial polarization), and $U_\phi$ (polarization oriented at $\pm45\dego$ with respect to the azimuth) calibrated in $\mu$Jy/arcsec$^2$, and the signal-to-noise ($S/N$) map, which is derived  on the basis of the $Q_\phi$ and $U_\phi$ images.
    For these images, as all the ones in this paper: north is up and east is left, while the gray circular mask corresponds to the region within the inner working angle of the coronagraph.}
    \label{fig:reduced_images_polar_Jy_snr}
\end{figure*}

In Fig.~\ref{fig:reduced_images_polar_Jy_snr}, we show the final Stokes vector components $pI$, $Q_\phi$ and $U_\phi$ resulting from the polarimetric observation acquired with $10$~HWP cycles, expressed  in $\mu$Jy/arcsec$^2$. 
Stellar light has been subtracted in this image, which was evaluated to $0.06~\pm0.07\,\%$ at $1\sigma$ by the IRDAP pipeline.
All the signal of the disk is seen positively in $Q_\phi$, i.e., this corresponds to azimuthal polarization of the light scattered by the debris disk, 
but it is not seen in $U_\phi$. Thus, we used this polarization to estimate the noise in the polarimetric data. To convert from detector units into flux unit, we normalized the coronagraphic images with the total flux of the star computed in a circular region of radius $25$~pixels summed over the two channels and corrected from respective frame time exposure and transmission of the neutral density of the coronagraphic and non-coronagraphic observations. We then multiplied this by a stellar flux density of $0.977\pm0.044$~Jy \cite[from Johnson H filter;][]{bourges_jmmc_2014} and divided by the plate scale area.
 
To derive the signal-to-noise ($S/N$) map of the disk seen in polarimetry, we divided the $Q_\phi$ component containing the signal of the disk by the azimuthal mean of the $U_\phi$ representing the noise (see Fig.~\ref{fig:reduced_images_polar_Jy_snr}). The disk is easily detected in the IRDIS polarimetric data, with a $S/N$ of an average value of  $4.6$/pixel$^2$ for the eastern part, and $3.5$/pixel$^2$ for the western part of the disk.

Our processing of the SPHERE/IRDIS polarized intensity dataset yields results similar to those of \citet{olofsson_vertical_2022}, if not identical. However, in contrast to their work, we provide additional outputs, such as the calibration of the reduced image in Jy/arcsec$^2$ and the signal-to-noise ($S/N$)  map.
In addition, we also processed the SPHERE/ZIMPOL polarimetric observations acquired on the night of August 1, 2015. The ZIMPOL data were reduced by the High-Contrast Data Centre\footnote{\url{https://hc-dc.cnrs.fr/}} \cite[HC-DC;][]{delorme_sphere_2017}, which implements the pipeline developed at ETH Zürich and described in \citet{hunziker_hd_2021}.  In particular, we measured and corrected the relative beam shift of the two orthogonal polarization states, subtracted the frame transfer smearing, and corrected for the residual telescope polarization and the intrinsic polarization of the star. The Stokes Q and U images were then transformed into the $Q_\phi$ and $U_\phi$ images following Eq. \ref{eq_Stokes}. Unfortunately, no circumstellar structure is visible in the reduced $Q_\phi$ image. Therefore, we do not discuss these data further in this paper.

\subsubsection{Total intensity observations \label{sec:obs_ti_only}}
In terms of the total intensity, the star HD\,120326 was observed in several modes with VLT/SPHERE: IRDIFS H23+YJ (2015-04-09, 2016-04-05, 2016-06-03), IRDIS J23 (2016-06-13), and IRDIFS BBH+YJ (2019-06-26, 2019-07-09), see Table~\ref{tab:obslog}. The IRDIFS mode enables simultaneous observations of the dual-band imager IRDIS in the filter doublet H2H3 ($\lambda_{H2}=1.593\pm0.055~\mathrm{\mu m}$, $\lambda_{H3}=1.667\pm0.056~\mathrm{\mu m}$) and of the low-resolution integral field spectrograph IFS in the band YJ ($0.95$--$1.35\,\mu$m). The filter doublet J2J3 corresponds to $\lambda_{J2}=1.190\pm0.042~\mathrm{\mu m}$, $\lambda_{J3}=1.273\pm0.046~\mathrm{\mu m}$. 
The filters J3 and H2 are supposed to be sensitive to the main emission peaks of cold companions, whereas the filters J2 and H3 should be sensitive to molecular absorptions.
The broad band H (BBH) filter is the same filter as that used in polarimetric mode (Sect.~\ref{sec:obs_polar_only}). All observations were carried out in the pupil-tracking mode to enable the processing of the data with the ADI technique 
and achieve higher contrast at sub-arcsecond separations. In pupil-tracking observations, circumstellar signal (e.g., disks or exoplanets) rotates through the sequence of observation, while the stellar halo remains quasi-static as a function of the parallactic angles. 
The ADI technique therefore enables the removal of most of the stellar halo and the retrieval of faint circumstellar signal. 
For one epoch of observation, a dataset consists of the non-coronagraphic image of the star (to derive the stellar flux and have the point-spread function -PSF- of the instrument), the coronagraphic cube acquired at two (IRDIS) or thirty-nine (IFS) channels ($x$,~$y$,~$t$,~$\lambda$), and the list of the parallactic angles corresponding to each temporal frame.

\begin{figure*}[h!]
    \centering
    \includegraphics[height=4.2cm]
    {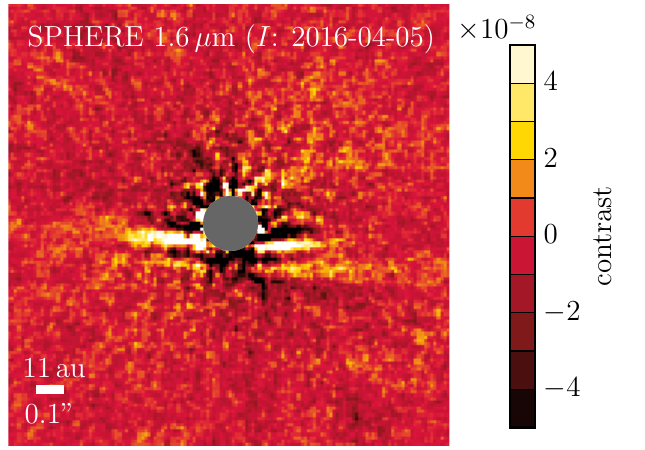} 
    \includegraphics[height=4.2cm]{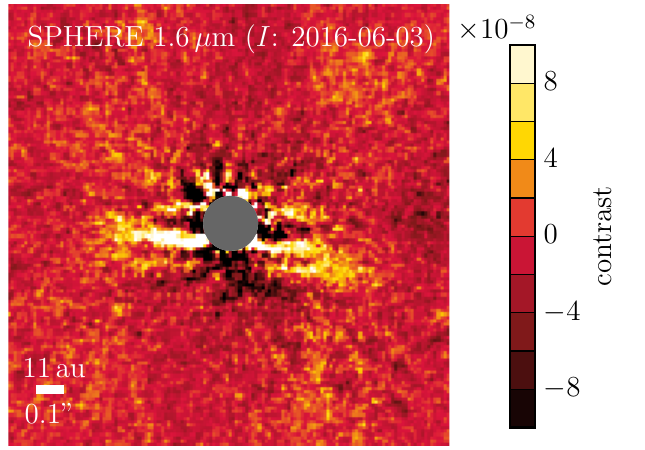} 
    \includegraphics[height=4.2cm]{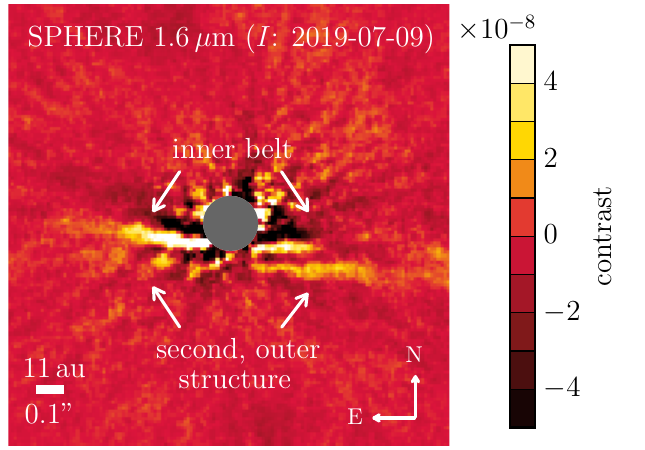} 
    
    \includegraphics[height=4.2cm]{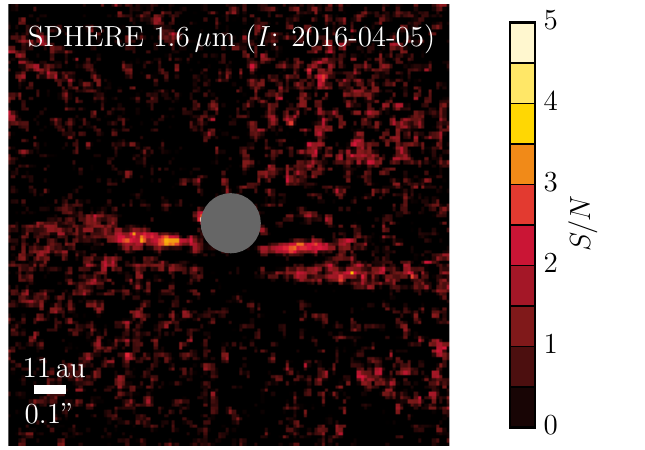} 
    \includegraphics[height=4.2cm]{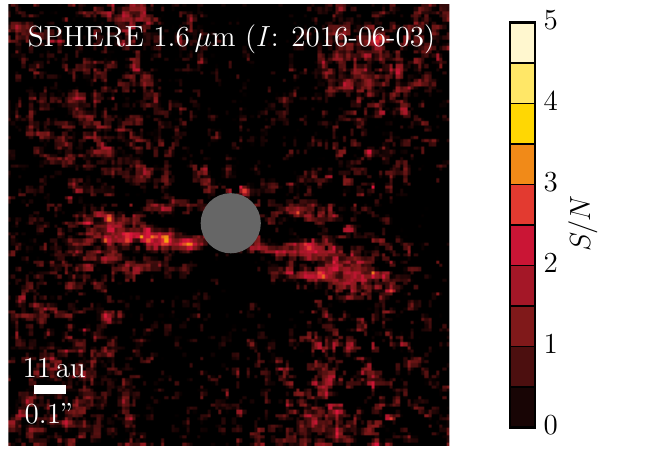} 
    \includegraphics[height=4.2cm]{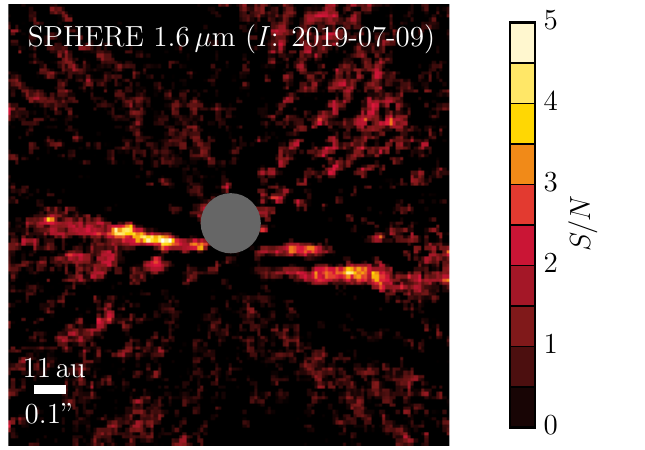} 
    \caption{Total intensity ("$I$") images at about $1.6~\mic$ reduced using the processing PCA~ADI on the system HD\,120326, calibrated in contrast unit (top) or in $S/N$ (bottom). Two structures can be seen: the inner belt and a second, outer structure (see the labels in the top-right image). All images have the same orientation, with north up and east on the left. 
    }
    \label{fig:im_snr_red_ti}
\end{figure*}
Among the total intensity data, we mainly focus on the unpublished high-quality epoch of observation 2019-07-09, which has a total time exposure of $119$~min, with an average seeing of $0.54$'', and a variation of parallactic angles of $58\dego$, as displayed in Table~\ref{tab:obslog}. In Fig.~\ref{fig:im_snr_red_ti} we also present results on the epochs of 2016-04-05 \citep[published in][]{bonnefoy_belts_2017} and 2016-06-03 (unpublished), but in which the circumstellar signal is fainter. Regarding the other datasets (see Table~\ref{tab:obslog}), we do not consider the observation of 2015-04-09 as the focal plane mask was missing during the observation sequence. We processed the observation 2019-06-26, but we do not present the associated results in this paper as the data are of significantly lower quality than those of 2019-07-09, acquired in the same band, and in which the disk is hardly detectable. We processed the epoch 2016-06-03 acquired in the J23 band, but almost no disk signal was recovered.

\medskip

The data are preprocessed by the High-Contrast Data Center \citep[HC-DC,][]{delorme_sphere_2017}, using the SPHERE Data Reduction and Handling pipeline \citep{pavlov_data_2008}. This preprocessing corrects the bad pixels, dark current, flat non-uniformity, and sky background in both IRDIS and IFS data. In addition to the calibration of the IFS, the pre-processing corrects for the wavelength and cross-talk between the spectral channels. Coronagraphic images are centered via four satellite spots used to determine the accurate position of the star hidden behind the coronagraphic mask. 

We homogeneously postprocessed the observations in total intensity with a principal component analysis \citep[PCA;][]{amara_pynpoint_2012,soummer_detection_2012} algorithm as implemented in the open library \texttt{VIP}\footnote{\url{https://github.com/vortex-exoplanet/VIP/}} \citep{gomezgonzalez_vip_2017,christiaens_vip_2023}. For the SPHERE/IRDIS data, we tested several methods to process the data. First, we considered either one of the two channels (combined before or after the PCA processing) and we applied between 1 and 20 components to model the stellar halo before its subtraction with the PCA algorithm. The ultimate goal was to have the cleanest image to retrieve the circumstellar signal. Although the circumstellar signal is drowned in the stellar light, the number of PCA components used to model the stellar halo has to be limited, because the model will capture circumstellar signal too and remove it. This effect is named self-subtraction and it is well known \citep[][]{milli_impact_2012}.
By evaluating the $S/N$ of the disk and examining the residuals, we concluded that applying PCA with ten components to each of the two channels, before averaging them into a final image, is a relatively satisfying compromise; however, we do note that some other options give comparable results. 

We show the reduced SPHERE/IRDIS observations for the epochs 2016-04-05,  2016-06-03 and 2019-07-09 in Fig.~\ref{fig:im_snr_red_ti}, calibrated in contrast or in $S/N$. To convert to contrast, we normalized the coronagraphic observation by the stellar flux estimated in a circular aperture of $25$-pixel radius, as for the polarimetric observations. 
To obtain the $S/N$ map, we built the noise map by processing with PCA the coronagraphic cube 
for the opposite parallactic angles to retain a similar temporal dependence of the stellar residuals and average any circumstellar signals, as done in \citet{pairet_stim_2019}.
We can see that the disk has a higher $S/N$ in the last observation (2019-07-09, H band), reaching (on average) about $3.9$/pixel$^2$ for the eastern part of the disk and   $2.5$/pixel$^2$ for the western part.  Our postprocessed image of the epoch 2016-04-05 is similar to that of \citet{bonnefoy_belts_2017}. However, the flux scales of their images tend to be more saturated, in an attempt to highlight the disk signal (see their Figs.~1 and~3), and they do not show any $S/N$ maps.

From Fig.~\ref{fig:im_snr_red_ti}, we can see two disk structures from the scattered total intensity light observations. 
These two structures will be referred in this work as (i) “the inner belt”, which was already reported by \citet{bonnefoy_belts_2017} based on their epoch 2016-04-05, and (ii) “the second, outer structure” (see Fig.~\ref{fig:im_snr_red_ti}), which is also reported by \citet{bonnefoy_belts_2017}, but which appeared particularly faint in their observation of 2016-04-05.

 For the second, outer structure, the western part is brighter than the eastern part. Using the new observation epoch of 2019-07-09, we retrieved the western part more effectively than \citet{bonnefoy_belts_2017}, who used the epoch 2016-04-05. However, they retrieved the eastern part more effectively compared to our results. For the western part, we derived a median $S/N$ of $3.4\pm0.6$ per pixel using the data of 2019-07-09, while using the data of 2016-04-05 we obtained a median $S/N$ of $1.5\pm1.0$ per pixel.

\begin{figure*}[h!]
    \centering
    \includegraphics[height=4.5cm]{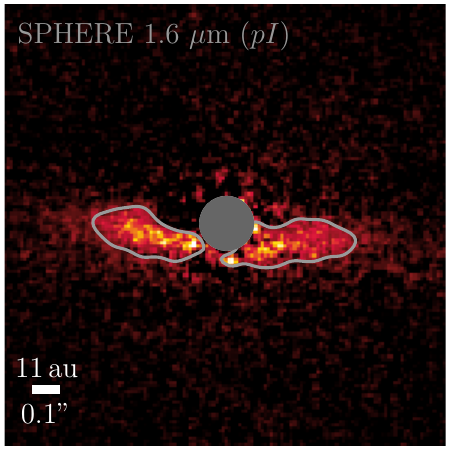} \hspace{2mm}  \includegraphics[height=4.5cm] {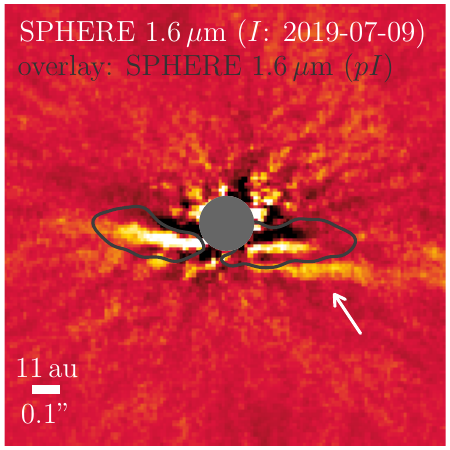} \hspace{2mm} 
    \includegraphics[height=4.5cm]{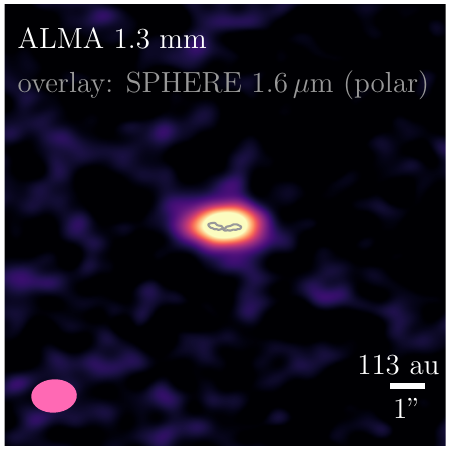}
    \caption{Comparison of the disk structures in HD\,120326 detected in scattered light (SPHERE, $1.6~\mic$) and in thermal emission (ALMA, $1.3~\mm$). On the left, we show the polarized intensity SPHERE image with in overlay the contours of the inner belt determined for a signal level of $2$~mJy/arcsec$^2$ on this image convolved with a gaussian to obtain smooth contours. The contours are added on the total intensity SPHERE image (middle) and on the zoom-out ALMA reconstructed image (right).  In the middle image, we added a white arrow to point out the southwestern part of the second structure seen in total intensity, which does not overlap with the signal seen in polarized intensity.
    }
    \label{fig:im_contours_ti_polar_alma}
\end{figure*}

In addition, we processed the epochs 2019-07-09 and 2016-04-05 with an iterative version of PCA \citep{pairet_mayonnaise_2021,juillard_inverseproblem_2023}, using ADI only (I-PCA~ADI), reference differential imaging \citep[RDI;][]{lafreniere_hst_2009} only (I-PCA RDI) or combined \citep[I-PCA ARDI;][]{juillard_combining_2024}. 
Reference differential imaging has proved to be efficient to remove the stellar halo and retrieve extended circumstellar signal of dozen of ground-based observations \citep{xie_referencestar_2022,ren_protoplanetary_2023}. Using an iterative approach for PCA helps limit the effects of self-subtraction, over-subtraction and deformations of the disk. To apply I-PCA ARDI, we used a library of reference frames obtained from observations of other stars with SPHERE/IRDIS in the same filter. On HD\,120326, contrary to I-PCA RDI, the processing I-PCA~ADI and I-PCA ARDI retrieve the second, outer structure on the epochs 2016-04-05 and 2019-07-09, see Fig.~\ref{fig:im_ipca}. In particular, the eastern and western parts of the second, outer structure are more visible than in the simple PCA~ADI reductions shown in Fig.~\ref{fig:im_snr_red_ti}.

Concerning IFS observations covering Y and J bands, we processed them using PCA~ADI and ten modes. Since the disk is faint, we consider only the best epoch, 2019-07-09, and we binned the spectral channels per six, to increase the signal-to-noise ratio.
Over the 39 spectral channels of the IFS, we kept 30 channels, removing the first and last ones, and a few that were more impacted by artifacts. 
In these images (Fig.~\ref{fig:im_red_ti_IFS}, from top-left to bottom-middle), the first belt is visible, but not the second, outer disk structure. 
We recovered the western part of the second structure only by taking the average of the 30 spectral channels and applying a spatial binning  (Fig.~\ref{fig:im_red_ti_IFS}, bottom-right). The second structure was previously also hardly detected by \citet{bonnefoy_belts_2017}, see their Fig.~1, using the epoch 2016-04-05.

\medskip

In the rest of this work, we will mainly use for the SPHERE/IRDIS and SPHERE/IFS total intensity observations, the epoch 2019-07-09 because its signal-to-noise ratio is higher than the other exploitable observations (2016-04-05, 2016-06-03), and the IRDIS dataset is acquired in the exact same filter than the SPHERE/IRDIS polarized intensity data (see Table~\ref{tab:obslog}). We will use the PCA ADI processing to have a fast-computing processing algorithm (compared to I-PCA ADI/ARDI) when modeling the data to derive the best-fitting disk parameters (see  Sects.~\ref{sec:disk_modelling} and \ref{sec:disk_reflectance}).

\subsubsection{Comparison of polarized versus total intensity data \label{sec:obs_polar_and_ti}}

The striking result from the previous Sects.~\ref{sec:obs_polar_only} and~\ref{sec:obs_ti_only}  is the differences between the disk structures imaged in polarized and total intensity. Only one disk belt is seen in polarized intensity, while two components are detected in total intensity. In Fig.~\ref{fig:im_contours_ti_polar_alma}, we compare the location of both structures, and we show that the first structure in total intensity matches the location of the ring imaged in  polarized intensity. The second structure is located further away.

\subsection{ALMA millimeter observations \label{sec:obs_alma}}

ALMA observations of HD\,120326 were acquired as part of project 2022.1.00968.S (PI J. Miley). In total five separate executions were made on October 19 and 21, and  December 17, 18, and 27, 2022. The observations use a spectral set-up using Band~6 receivers, which include two relatively coarse resolution spectral windows with 128~channels of width 15.6~MHz placed at central frequencies of $218.0$~GHz ($1.375~\mm$) and $233.0$~GHz ($1.287~\mm$). Two further spectral windows were placed to cover any potential bright molecular lines. The first was positioned with central frequency $230.5$~GHz ($1.301~\mm$) containing $3\,840$ channels of width $244$~kHz ($1.4~\mic$). The final spectral window was positioned at central frequency $220.0$~GHz ($1.363~\mm$) with $1\,920$~channels of width $976$~kHz ($6.0~\mic$). We reduced the raw data by using the pipeline script provided by ALMA observatory using CASA version $6.4$ \citep{mcmullin_casa_2007}. 

Although there was no emission from molecular lines  detected, we flagged channels corresponding to $\pm$15~km/s of the rest frequency of the bright molecular line transitions covered by channels within the spectral windows; specifically: $^{12}$CO at $230.538$~GHz, $^{13}$CO at $220.4$~GHz, and $^{18}$CO at $219.6$~GHz. This ensures that any potential contamination of the continuum image by line flux is avoided. The observed data was then averaged in frequency to produce consistent channel widths of $15.625$~MHz before imaging and the five executions were then concatenated into a single dataset for imaging. Images were reconstructed using \texttt{tclean} task in CASA using the Hogbom deconvolver \citep{hogbom_aperture_1974} and natural weighting\footnote{We also tested other weightings, using a robust parameter ranging from -2 to 2 in steps of 0.5. However, prioritizing angular resolution over sensitivity resulted in very noisy images.}. This produced a continuum image with a synthesized beam of size of  $1.321\arcsec \times0.950\arcsec$ ($PA=-86.729\dego$), and achieved a noise level of $17~\mu$Jy/beam.
The clear detection of circumstellar material, which is  unresolved, is shown in Fig.~\ref{fig:im_contours_ti_polar_alma} (right image).  There is no evidence of any offset from the expected stellar position.
 
Our measurement of the flux of the debris disk is achieved by fitting a Gaussian in the uv-plane, giving an integrated flux value of $561 \pm 20~\mu$Jy in the continuum at $1.3~\mm$. In addition, we derived an upper limit of the $^{12}$CO value of $4.8~\mathrm{Jy\cdot km/s}$. This corresponds to three times the rms noise level measured over an area equal to the size of the synthesized beam.

\begin{figure*}[h]
    \centering
    \includegraphics[width=\linewidth]{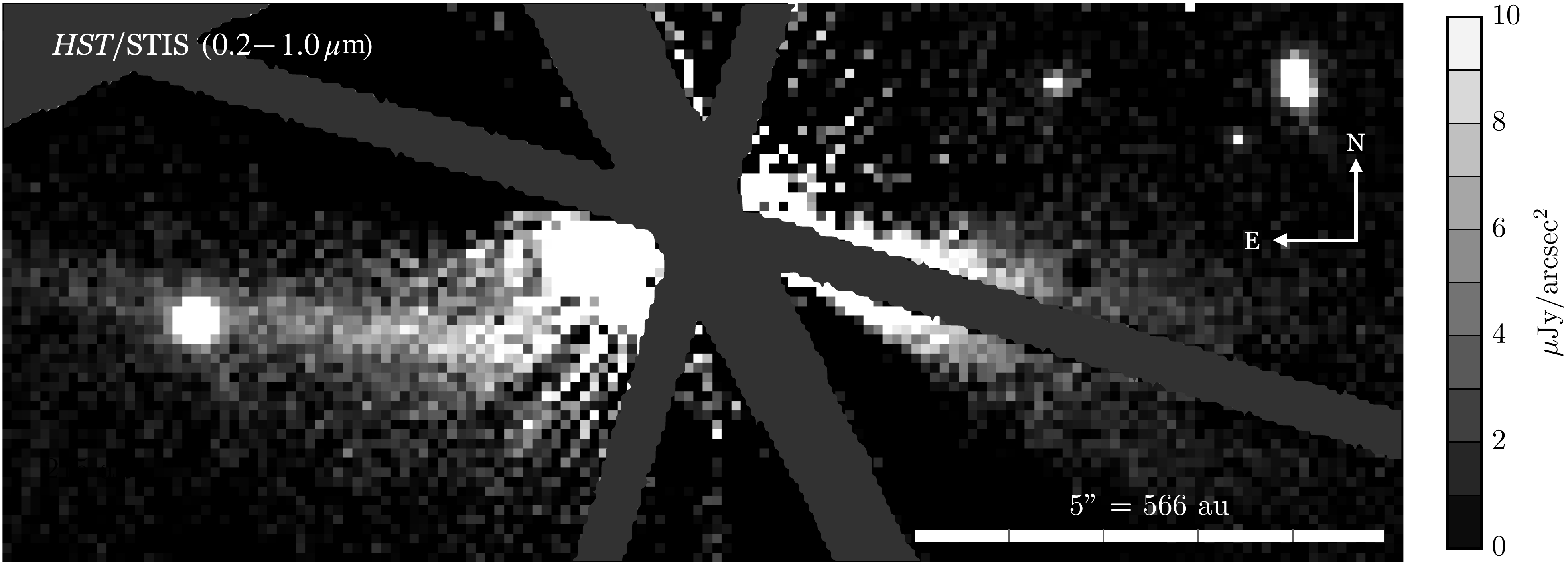} 
    \caption{Optical coronagraphic HST/STIS observations of HD\,120326 centered behind the WedgeA1.0''. We applied a spatial binning $2\times2$\,pixels$^2\rightarrow1\times1$\,pixel$^2$. The flux unit is $\mu$Jy/arcsec$^2$. The areas hidden by the wedges (WedgeA and WedgeB) and the diffraction spikes (the cross-pattern) are masked in gray.}
    \label{fig:im_HST_flux}
\end{figure*}

\subsection{HST/STIS optical observations \label{sec:obs_hst}}

The archival HST/STIS observation of HD\,120326 were already published in \citet{padgett_warm_2016} and \citet{bonnefoy_belts_2017}. These coronagraphic observations were acquired on 2013-06-20 with the star centered behind the WedgeA1.0 at two different rolling angles of separation $30\dego$ and in the broad band filter centered at $0.59~\mic$ ($d\lambda=0.44~\mic$).  Both papers published reduced images featuring an extended halo around HD\,120326, and strong stellar artifacts at close separations, without providing a flux calibration or deriving a $S/N$ map.

We postprocessed  the dataset again with a bespoke routine to remove most of the stellar light. We used as
inputs the two calibrated images acquired at ORIENT1 ($28\dego$) and ORIENT2 ($58\dego$), performed a roll-subtraction (IM1-IM2) and de-rotated the image, resulting in north being up and east on the left. The corresponding results are shown in Fig.~\ref{fig:im_HST_flux}. This image shows fewer instrumental artifacts than that of \citet{bonnefoy_belts_2017}, because we avoided duplicating the cross-pattern feature during postprocessing (see their Fig.~5). As seen here, we successfully recover the disk, which is impacted by self-subtraction (see black negatives feature at the bottom right of the wedge in Fig.~\ref{fig:im_HST_flux}).
We derived the flux map (Fig.~\ref{fig:im_HST_flux}) by converting the ADU in Jy/arcsec$^2$, following the procedure described on the STScI website\footnote{\url{https://www.stsci.edu/documents/dhb/web/c03_stsdas.fm3.html}}. At large separations ($>2.5\arcsec$), the disk has a flux density of a few $\mu$Jy/arcsec$^2$.

\begin{figure*}[p] \centering 
    \includegraphics[width=0.9\linewidth]{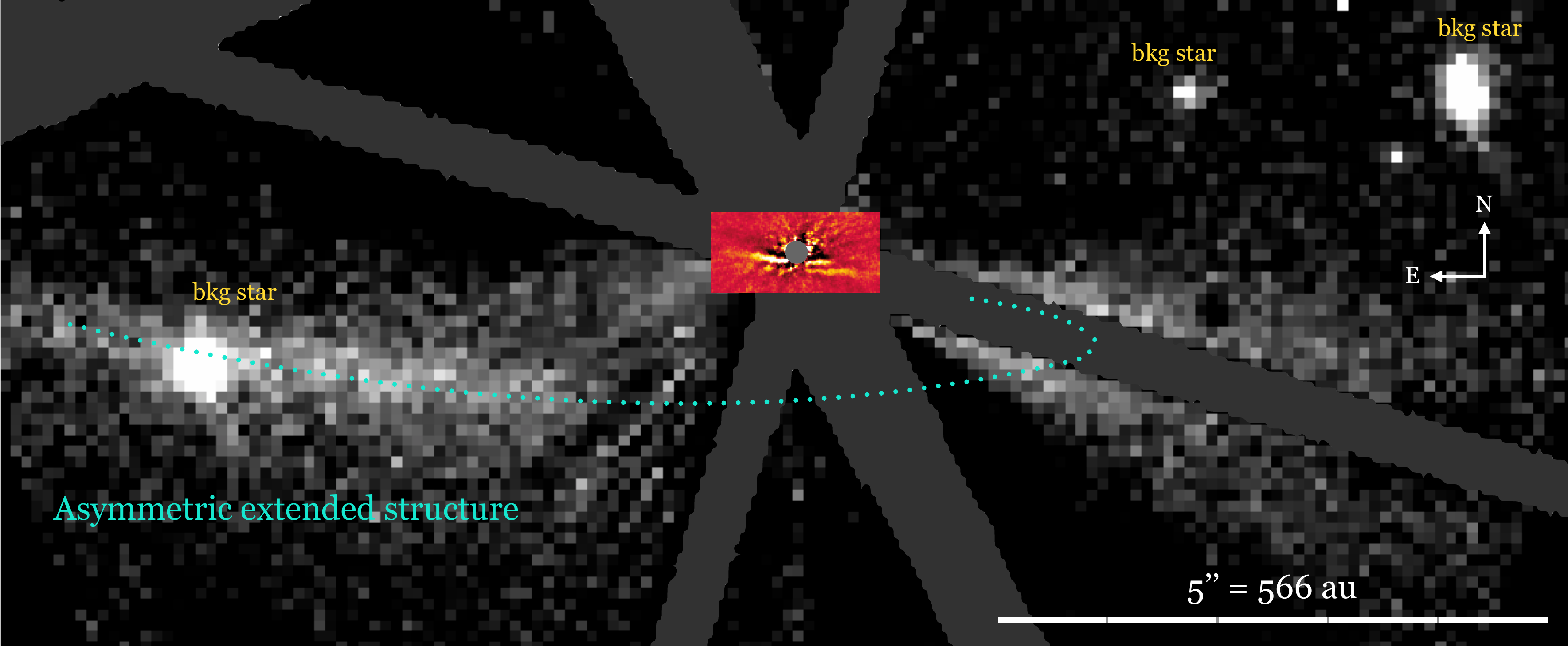}  \vspace{3mm}

    \includegraphics[width=0.9\linewidth]{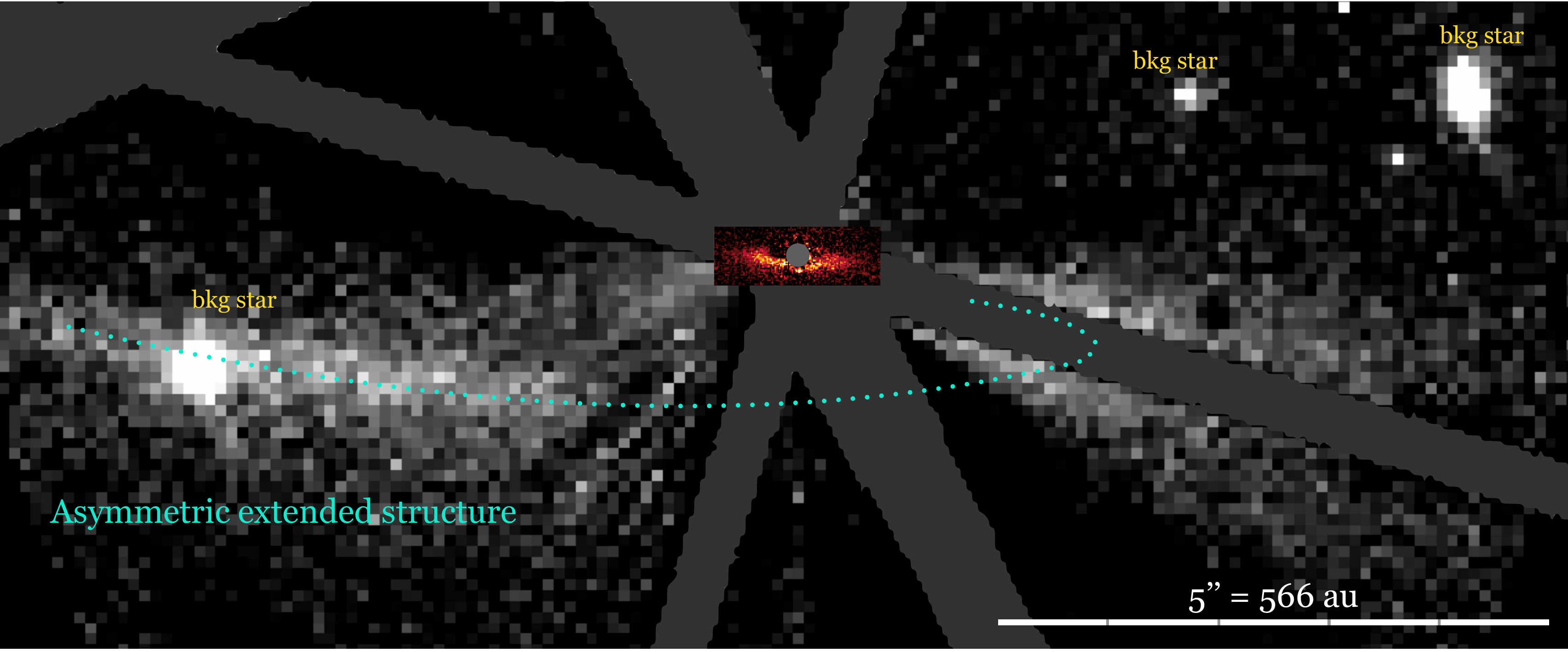}   \vspace{3mm}

    \includegraphics[width=0.9\linewidth]{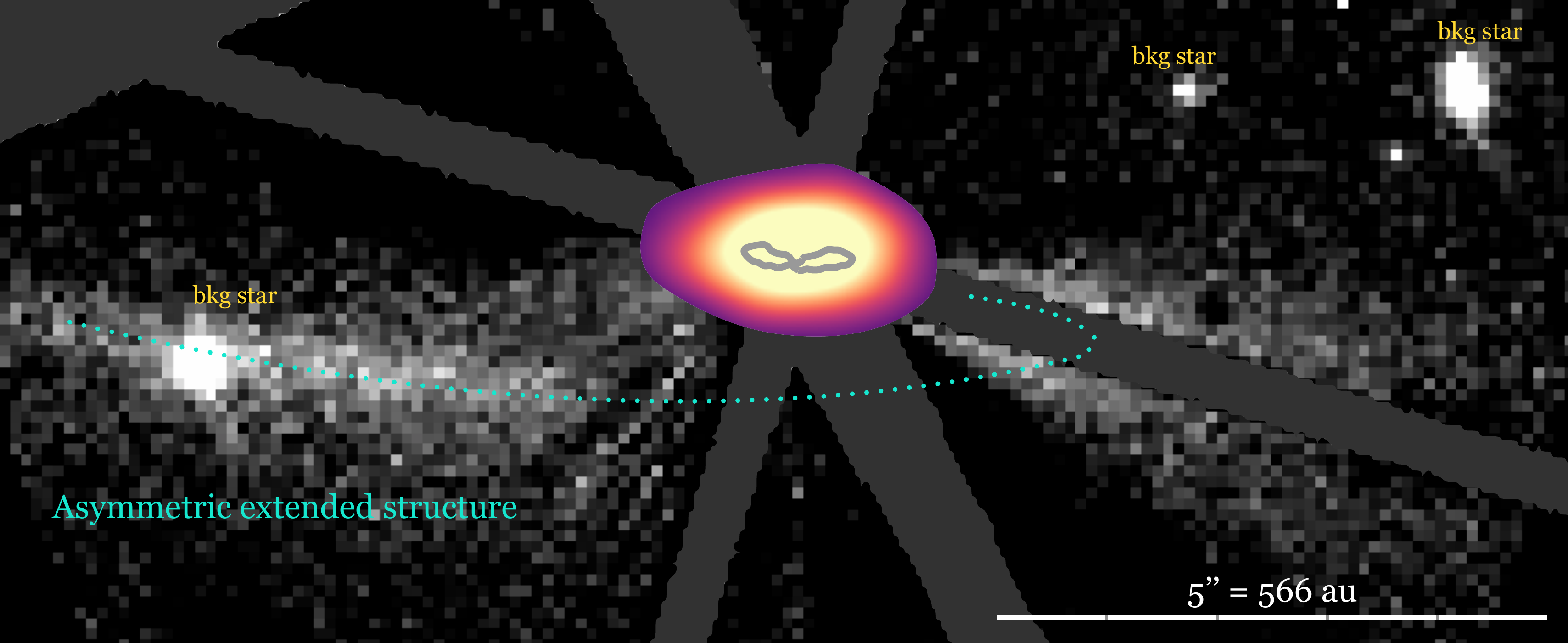}  \vspace{3mm}

    \includegraphics[width=0.5\linewidth]{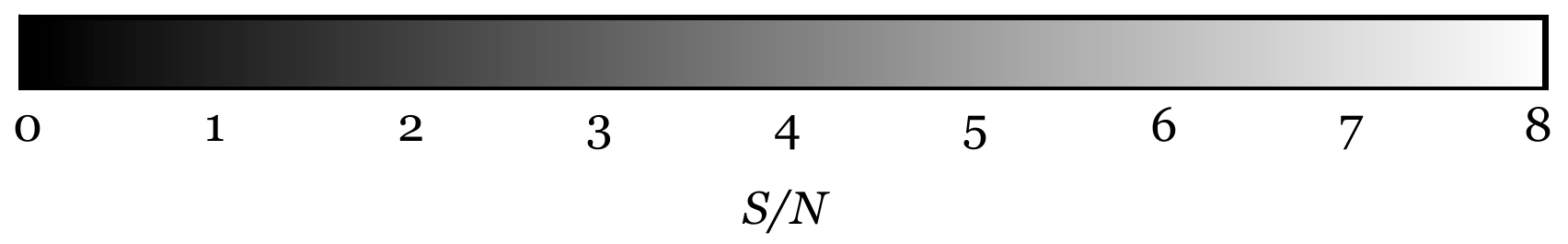}

    \caption{
    Overview of the young debris disk around HD\,120326. The black-to-white images are the $S/N$ map of the optical HST/STIS coronagraphic data, with a spatial binning $2\times2$\,pixels$^2\rightarrow1\times1$\,pixel$^2$. For reference, in the center of the images we display the inner structures detected with SPHERE in total intensity (top), in polarimetry (middle) and with ALMA (bottom).  The areas hidden by the coronagraphs  and the HST diffraction spikes are masked in dark gray.}
    \label{fig:HD120326_overview}
    \vspace{-0.5cm}
\end{figure*}

To better distinguish the disk features from the stellar residuals, we also computed the $S/N$ map. We defined the signal map as the reduced STIS image. Regarding the noise map, we computed radially the noise as the standard deviation in annuli. We did not consider the full image, but only the top-right quadrant, in which we masked the (small) regions defined by the location of the WedgeB1.0, one of the stellar spikes, a few background sources or suspicious, bright pixels. We applied a spatial binning of a factor four (two in height and two in width) to increase the $S/N$. We show the $S/N$ map in Fig.~\ref{fig:HD120326_overview}. A large asymmetric disk structure extends up to a projected separation of about $700~\au$. Its $S/N$ per pixel is between $5$ and $8$. This spiral-like feature is discussed in Sect.~\ref{sec:discussion_global_archi}.
Some signal is still visible relatively close to the star, at the eastern edge of the WedgeA1.0 which has a width of $1$''. For scale comparison, in Fig.~\ref{fig:HD120326_overview}, we added the image of the inner structures detected with SPHERE in total intensity (top) and in polarized intensity (middle), and with ALMA (bottom). In Sect.~\ref{sec:discussion_global_archi} we examine the global architecture of the debris disk around HD\,120326.

\medskip

In the next two sections we present our constraints on the properties of the inner disk structures of HD\,120326, which are detected in the NIR with VLT/SPHERE (Sect.~\ref{sec:modelling_results_sphere}) and in the millimeter with ALMA (Sect.~\ref{sec:modelling_results_alma}).

\section{Morphology, photometry, and polarization of the disk structures from SPHERE NIR data \label{sec:modelling_results_sphere}}

In Sects. \ref{sec:disk_modelling}--\ref{sec:disk_reflectance}, we describe how we constrained the morphology, polarization, and reflectance spectrum of the inner dust  belt of HD\,120326. To this aim, we modeled the VLT/SPHERE NIR polarimetric and total intensity data. 
In Sect.~\ref{sec:second_structure_nature}, we examine the leftover circumstellar signal in the residuals (indicated by the white arrow in Fig.~\ref{fig:im_results_mcmc}, bottom-right) and investigate its origin.

In particular, to constrain the disk morphology and photometry, we developed the open-access Python-based tool Modeling Disks in Scattered light\footnote{\url{https://github.com/cdesgrange/MoDiSc}} (\texttt{MoDiSc}) to constrain the morphology and photometry of a disk seen in scattered light. The tool \texttt{MoDiSc} was originally inspired by the code \texttt{DiskFM} \citep{mazoyer_diskfm_2020}. The tool  \texttt{MoDiSc}  takes as input a configuration file setting the paths to the datasets of the observations and defining the variables used in the simulations. By exploring the parameter space with a \texttt{MCMC} algorithm \citep[\texttt{emcee} package from][]{foreman-mackey_emcee_2013}, \texttt{MoDiSc} looks for the disk model matching at best the observations. To initialize the values of the \texttt{MCMC} simulation, a first satisfying guess can be determined by running a \texttt{Nelder-Mead} minimization \citep{nelder_simplex_1965}, which is implemented in \texttt{MoDiSc}.
The disk model can be composed of one or several belts, each of them generated with the module \texttt{fm.scattered\_light\_disk} from \texttt{VIP} \citep{gomezgonzalez_vip_2017,christiaens_vip_2023}, the module being based on the radiative transfer code \texttt{GRaTeR} \citep{augereau_hr_1999}. One or several observations in total and/or polarized intensity can be considered simultaneously, from one or several instruments.

\subsection{Disk modeling with one belt \label{sec:disk_modelling}}

To derive the disk morphology, we modeled both jointly and independently the SPHERE/IRDIS total and polarized intensity observations (2019-07-09 and 2018-06-01, respectively) by using the open-access code \texttt{MoDiSc}. We first describe our methodology and results for our joint modeling, and will then briefly also mention the results obtained for the independent modeling following a similar methodology.

In our simulations, we constrained the seven following parameters related to the belt: its reference radius $r_0$, its position angle $PA$, its inclination $i$, its scattering anisotropy parameter $g$, its outer slope $\alpha_\text{out}$ and  two scaling factors, one to match the flux of the total intensity observations ($\text{scaling}_{I}$) and the other one to match the flux of  the polarized intensity observations ($\text{scaling}_{pI}$).
To limit the number of free parameters, we fixed the eccentricity, ascending node and argument of periastron of the ring to zero, and a steep inner slope to $\alpha_\textnormal{in} = 10$. 
We used a gaussian vertical profile ($\gamma=2$), a linearly flared ring ($\beta=1$) and set the scale height $\xi_0$ to $1.5~\au$ at the reference radius $r_0$ (aspect ratio of $\sim$$3.75\%$), see Eqs.~\ref{eq:DD_dust_density_distri_1}-~\ref{eq:DD_scale_height} and \citet{augereau_hr_1999} for a thorough description of the model for the dust density distribution. To model the scattering phase function (SPF) of the disk in total intensity, we used the Henyey-Greenstein (HG) function \citep[][see Eq.~\eqref{eq:HG_1compo_ti}]{henyey_diffuse_1941} parametrized by the scattering anisotropy parameter $g$. The same goes for the polarized intensity, which is modeled by this HG function multiplied by the Rayleigh scattering function, see Eq.~\eqref{eq:HG_1compo_polar}.

Our simulations aimed to minimize the reduced $\chi_{r,I+pI}^2$ defined as
\begin{align}
    \chi^2_{r,I+pI} &\;=\; \chi^2_{r, I}  \;+\; \chi^2_{r, pI} ,\\
    &\;=\; \frac{\chi^2_{I}}{N_{\text{resol},I}-N_{\text{param},I}-1} \;+\; \frac{\chi^2_{pI}}{N_{\text{resol}, pI}-N_{\text{param},pI}-1}, 
    \label{eq:chi_r}
\end{align}
where $N_{\text{resol},I}$ and $N_{\text{resol},pI}$ are the number of pixels considered within the regions used to model the inner belt in the total and polarized intensity images (see Fig.~\ref{fig:im_disk_modelling_mask}), and $N_{\text{param},I}$ and $N_{\text{param},pI}$ are the number of free parameters used to model the total and polarized intensity images, respectively. In our case,  $N_{\text{resol},I} =  N_{\text{resol},pI} = 1968$ and $N_{\text{param},I} = N_{\text{param},pI} = 6$.

The $\chi^2_{I}$ and $\chi^2_{pI}$ are expressed as follows
\begin{align}
\label{eq:chi_I}
    \chi^2_{I} &\;=\; \sum_\text{pixel} \, \pac{\frac{\texttt{PCA(}\,\text{obs}_I\;-\;M_{I}\,(r_0,\, PA,\, i,\, \alpha_\text{out}, \,\text{scaling}_{I}) \,\texttt{)}}{\sigma_{I}}}^2 ,\\
    \chi^2_{pI} &\;=\; \sum_\text{pixel} \, \pac{\frac{Q_\phi\;-\;M_{pI}\,(r_0,\, PA,\, i,\, \alpha_\text{out}, \,\text{scaling}_{pI})}{\sigma_{pI}}}^2,
    \label{eq:chi_pI}
\end{align}
where $\text{obs}_I$ is the pre-processed total intensity science coronagraphic cube before applying PCA ADI, $Q_\phi$ is the IRDAP-processed polarized intensity image, $M_I$ and $M_{Q_\phi}$ are the convolved disk models in total ($I$) and polarized ($pI$) intensity, and $\sigma_{I}$ and $\sigma_{pI}$ are the noise maps of the observation in total and polarized intensity, respectively.
Both the total and polarized intensity ($I$ and $pI$, respectively) terms have similar values, they therefore give an equal contribution in this merit function. 

To explore the parameter space, we first ran simulations using a \texttt{Nelder-Mead} minimization \citep{nelder_simplex_1965}, to converge quickly to a minimum, before running a Markov Chain Monte Carlo (\texttt{MCMC}) algorithm with \texttt{emcee} \citep{foreman-mackey_emcee_2013} to explore the space parameter more extensively. The walkers were initialized by drawing random values in the interval set by the best parameters found with \texttt{Nelder-Mead} plus or minus $10\%$. The random draw was linearly uniform for the reference radius, the position angle, the scattering anisotropy parameter, the outer slope, the cosine of the inclination and the logarithm in base ten of the scaling flux factor. 
In total, we used $100$ walkers, with $3\,000$ iterations for each parameter. 
For each iteration, we generated a disk model with the radiative transfer code \texttt{GRaTeR} \citep{augereau_hr_1999}, available in the open-access library $\texttt{VIP}$ \citep{gomezgonzalez_vip_2017,christiaens_vip_2023}. 

\medskip

In the case of the total intensity data, we convolved directly the disk model $M_I$ with the SPHERE/IRDIS non-coronagraphic observation of the star, corresponding to the PSF of the instrument. Then, we rotated the convolved synthetic disk model for each parallactic angle of the sequence of observation, and subtracted it from the science coronagraphic cube. We processed this new cube by applying PCA using ten modes. We divided the cube by the noise map $\sigma_I$ (computed in Sect.~\ref{sec:obs_ti_only}), producing a normalized residual map. 
If the disk model matches the observations, no circumstellar signal should remain in this normalized residuals map, that should only represent noise. 

On the other hand, in the case of polarimetric observations, we matched the disk model with the $Q_\phi$ image obtained by processing with the IRDAP pipeline (see Sect.~\ref{sec:obs_polar_only}). To convolve robustly the disk model, we constructed the synthetic Stokes $Q$ and $U$ images from the disk model generated with $\texttt{VIP}$. We convolved those $Q$ and $U$ images with the PSF of the star and re-constructed the $Q_\phi$ image. Applying robustly the convolution on the reconstructed Stokes $Q$ and $U$ images is important for disks located at small separations, to account for positive and negative polarization signals that cancel each other when observed with a limited angular resolution \citep[e.g.,][]{engler_detection_2018,heikamp_polarimetric_2019}. We obtained the residual map by subtracting the synthetic image $M_{pI}$ from the processed image $Q_\phi$, and normalized it by the noise map $\sigma_{pI}$ (computed in Sect.~\ref{sec:obs_polar_only}).

By considering only pixels within the region of interest (where the disk signal may be located), we summed the square value of these pixels for the total and polarized intensity, to obtain $\chi^2_I$ and $\chi^2_{pI}$, respectively, and then $\chi^2_{r,I+pI}$ following Eq.~\eqref{eq:chi_r}. By minimizing $\chi^2_{r,I\,+\,pI}$, the \texttt{MCMC} exploration maximizes the likelihood directly related to it.

\medskip

\begin{table}[h]
    \setlength{\tabcolsep}{4pt}
    \centering \small
    \caption{Morphology of the inner dust belt  based on our \texttt{MCMC} exploration jointly fitting the SPHERE total and polarized intensity data. } 
    \begin{tabular}{cCCC} 
    \hline \hline  \noalign{\smallskip}
        Parameter & \text{Range} & \text{Best fit}~\chi^2_\text{r,min} & \text{Median} \pm 1\sigma \\
    \noalign{\smallskip} \hline \hline  \noalign{\smallskip}
        $r_0$ (au) & [10, 130] & 39.9 & 39.7\pm0.6 \\
        $PA$ ($\dego$) & [-120, -60] & -94.1 & -94.1\pm0.1 \\
        $i$ ($\dego$) & [0, 90] & 78.0 & 77.9\pm0.4 \\
        $g$ & [0.05, 0.999] & 0.78 & 0.77\pm0.01 \\
        scaling ($I$) & [100, 10^8] & 6.8 \times 10^3 & (6.8\pm0.3) \times 10^3 \\
        scaling ($pI$) & [100, 10^8] & 9.6 \times 10^4 & (9.3\pm0.6) \times 10^4 \\
        $\alpha_\text{out}$ & [-20, -1.1] & -2.17 & -2.18\pm0.9 \\
        \noalign{\smallskip}
        \hline 
        \noalign{\smallskip}
        $\text{SB}_\text{disk,\,I}/F_{\star,\, \text{I}}$  & & 2.0\times10^{-3} & (2.0\pm0.2)\times10^{-3} \\
        $\text{SB}_\text{disk,\,pI}/F_{\star,\,\text{pI}}$ & & 1.2\times10^{-4} & (1.2\pm0.2)\times10^{-4}   \\
     \noalign{\smallskip} \hline \hline  \noalign{\smallskip}
    \end{tabular}
     
    \tablefoot{The two last lines correspond to the total surface brightness of the disk in total and polarized intensity, respectively, expressed in terms of contrast, i.e., divided by the total stellar flux. The ranges correspond to the authorized values.
    The disk models and residuals are shown in Fig.~\ref{fig:im_results_mcmc}, and the corner plot of the posteriors in Fig.~\ref{fig:cornerplot_morpho_sphere}. One may wish to compare these results with those from the \texttt{MCMC} exploration independently fitting either the total or polarized intensity, see Tables~\ref{tab:mcmc_results_nir_indpt_I} and~\ref{tab:mcmc_results_nir_indpt_pI}, respectively.
    }
    \label{tab:mcmc_results_nir}
\end{table}

\begin{figure}[h] \centering
    \includegraphics[height=4.2cm]{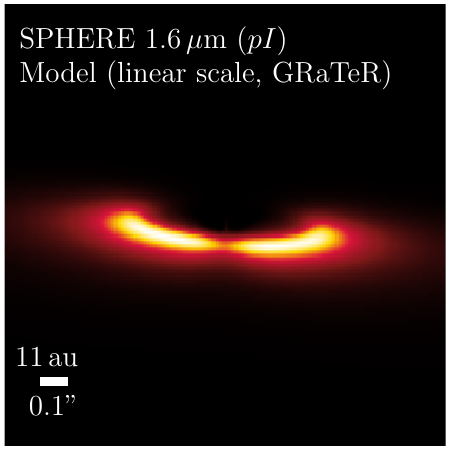}
    \includegraphics[height=4.2cm]{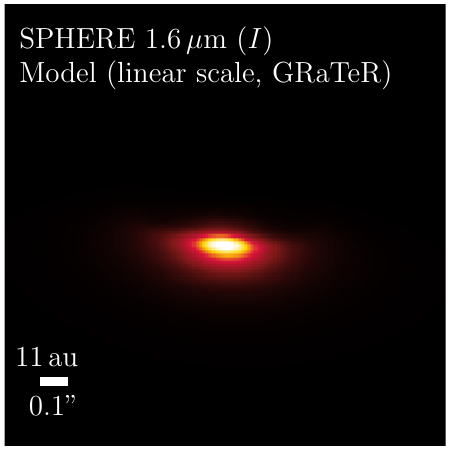} 
    \includegraphics[height=4.2cm]{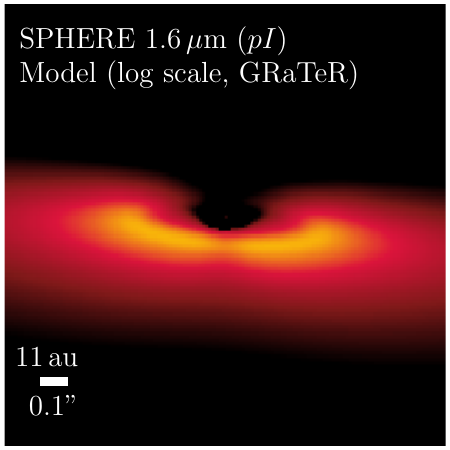} 
    \includegraphics[height=4.2cm]{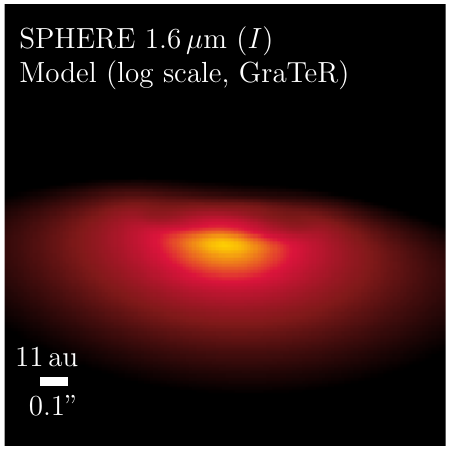} 
    \includegraphics[height=4.2cm]{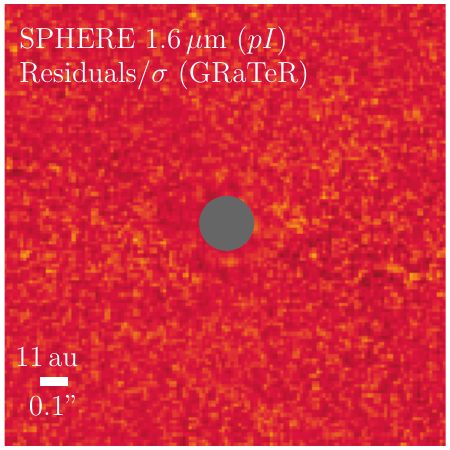} 
    \includegraphics[height=4.2cm]{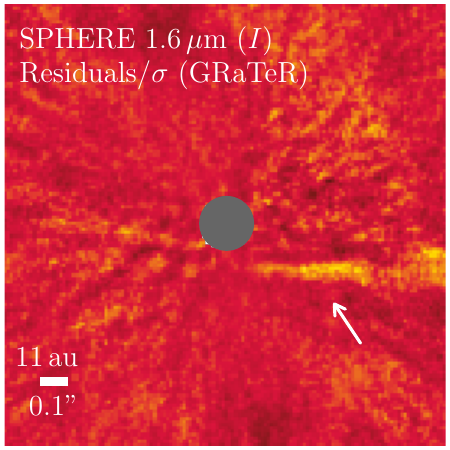} 
    \caption{  Results of our \texttt{MCMC} simulations, using the Pythonized version of the \texttt{GRaTeR} tool to generate the belt model.
    We show the best models in linear or logarithmic scales (top and middle), and the associated residuals normalized by the noise (“$\sigma$”,  bottom). Scattered light images for the polarized intensity images are on the left and those for total intensity images on the right. Some signal is still visible in the total intensity residuals, and is indicated by the white arrow.
    Unlike the disk model in total intensity, the disk model in polarized intensity accounts for Rayleigh scattering, hence visual brightness differences between these disk models depending on the scattering angle. The best-fitting parameters  are given in Table~\ref{tab:mcmc_results_nir}. As a reminder, since the disk inclination found is $78\dego$,  the scattering angle is $12\dego$ at the closest location of the belt, $90\dego$ in the anses, and $168\dego$ at its furthest location \citep[see also Fig.~5 from][to have an insight of the scattering angles around a dust belt]{perrin_polarimetry_2015}.
    }
    \label{fig:im_results_mcmc}
\end{figure}

We list the best parameters derived for the dust belt based on our disk modeling in Table~\ref{tab:mcmc_results_nir}.  The total surface brightness of the disk is determined to be $(1.2\pm0.2)\times10^{-4}$ in polarized intensity and $(2.0\pm0.2)\times10^{-3}$ in total intensity, corresponding to $0.12\pm0.2~\mu$Jy and $2.0\pm0.2~\mu$Jy, respectively, by using a stellar flux of $0.975~\mjy$ at $1.6~\mic$ \citep{ofek_calibrated_2008}.
Figure~\ref{fig:im_results_mcmc} shows  the synthetic best disk models and residuals for the observations in polarized and total intensity at $1.6~\mic$.

The posteriors on the disk parameters are reported in Fig.~\ref{fig:cornerplot_morpho_sphere}, for which we removed the first $2\,000$ iterations for each walker, considered as the burn-in phase. 
We can see that the reference radius is correlated with the inclination of the belt and the outer slope. 
The reference radius $r_0$ of $39.9^{+0.4}_{-0.8}~\au$ corresponds to a semi-major axis at maximum dust surface density in the mid plane of   $\,43.7^{+0.6}_{-1.0}$~au with a full-width at half maximum of $24.6$$~\au$ and a shallow outer edge, thus decreasing slowly in density for higher radii. 

As the inner and outer slopes used to model the disk in our work ($\alpha_\text{in,fixed}=10$ and $\alpha_\text{out,fitted}=-2.17^{+1.0}_{-0.8}$ ) differ from those used in \citet{bonnefoy_belts_2017} and \citet{olofsson_vertical_2022}, respectively $\alpha_\text{in}=10$ and $\alpha_\text{out}=-5$ and $\alpha_\text{in}=12.0\pm3.9$ and $\alpha_\text{out}=-1.9\pm0.1$, it is more relevant to compare the maximum dust surface density between our studies than the reference radius. The disk parameters of  \citet{bonnefoy_belts_2017} and \citet{olofsson_vertical_2022} correspond to a maximum dust surface density peaking at $64.0\pm3.3$ and $37.6\pm4.0$~au, respectively, using Eq.~(1) from \citet{augereau_hr_1999}. Therefore, our value ($\,43.7$~au) is much closer to that of \citet{olofsson_vertical_2022}. Potential factors that could explain the difference between the values are, first, the type of data used (polarized or total intensity), which may probe different dust grains (e.g., depending on their scattering angles, see Fig.~\ref{fig:SPFs}).  Second, for a given data type, the epoch of observation considered plays a role, as it may be more or less sensitive to the inner belt (Sect.~\ref{sec:obs_ti_only}). 
Third, the inclination and the distance of the belt to the star are correlated (see Fig.~\ref{fig:cornerplot_morpho_sphere}). For instance, decreasing the inclination by one degree (for inclination values around $78\degr$) results in a reduction of about $2~\au$ in both the reference radius and the maximum dust surface density radius. The inclinations found by \citet{bonnefoy_belts_2017} and \citet{olofsson_halo_2022} are  $80\dego\pm1\dego$ and $76.7\dego\pm0.6\dego$, respectively.  This may explain part of the differences, as the smallest inclination and smallest maximum dust surface density radius first correspond  to the values from \citet{olofsson_vertical_2022}, then to those from our work, and, finally, to those from \citet{bonnefoy_belts_2017}.

For completeness, we also ran additional simulations to model independently the debris disk seen either in the SPHERE/IRDIS total intensity or polarized intensity. We display the best-fitting parameters obtained in each case in Tables~\ref{tab:mcmc_results_nir_indpt_I} and~\ref{tab:mcmc_results_nir_indpt_pI}, respectively. The results based on total intensity give higher values for the inclination, the reference radius and the anisotropic parameter compared to the results based on polarized intensity only: $i_I=79.4\dego\pm0.3\dego$ and $i_{pI} = 77.1^{+0.6}_{-0.5}\,\dego$, $r_{0,I}=43.2^{+1.0}_{-0.9}~\au$ and $r_{0,pI}=38.5^{+1.1}_{-0.7}~\au$  (corresponding to maximum dust surface density at $46.7^{+1.6}_{-0.8}$ and $42.3^{+1.7}_{-0.8}~\au$, respectively), $g_I=0.81\pm0.01$ and $g_{pI}=0.74^{+0.03}_{-0.02}$. Results are consistent with each other at $2\sigma$.
The uncertainties provided on the disk parameters are statistical uncertainties derived from the \texttt{MCMC} exploration process (see Fig.~\ref{fig:cornerplot_morpho_sphere}). They are likely underestimated, which is common in morphological simulations of disks using \texttt{MCMC} exploration \citep[see][]{mazoyer_diskfm_2020}. 
We can see that the best parameters derived by jointly fitting the total and polarized intensity are roughly an average of the best parameters derived by independently fitting these observations (Tables~\ref{tab:mcmc_results_nir}, ~\ref{tab:mcmc_results_nir_indpt_I} and~\ref{tab:mcmc_results_nir_indpt_pI}).  Compared to the inclination $i$ we determined from jointly fitting the total and polarized intensity data, the inclinations $i_I$ and $i_{pI}$ are closer to those found by \citet{bonnefoy_belts_2017} and \citet{olofsson_vertical_2022}. To some extent the same applies to the maximum dust surface density radius. Nonetheless, the maximum dust surface density radius determined by our modeling of the total intensity data alone still differs significantly from that of determined by \citet{bonnefoy_belts_2017}. This could be related to the fact it is difficult to detect in total intensity the ansea of the belt around HD~120326, thus to constrain its semi-major axis and so its maximum dust surface density radius (see Fig.~\ref{fig:im_snr_red_ti}).

To summarize, our best model fitting jointly the observations acquired in total and polarized intensity seems to reproduce well the observed disk signals, excepting the signal in the residuals indicated by the white arrow (Fig.~\ref{fig:im_results_mcmc}, bottom-right). This signal corresponds to the second, outer structure already mentioned in Sect.~\ref{sec:obs_ti_only} (see also Fig.~\ref{fig:im_snr_red_ti}), and we will investigate its nature in Sect.~\ref{sec:second_structure_nature}. Since most of the signal is reproduced by modeling with one dust belt jointly the polarized and total intensity observations, the difference between both images can be mostly assigned to processing effects (Sect.~\ref{sec:obs_sphere}) and the scattering phase function (Sect.~\ref{sec:SPF_DP}).

\subsection{Scattering phase function and degree of polarization \label{sec:SPF_DP}}

\begin{figure}[h] \centering 
    \includegraphics[width=\linewidth]{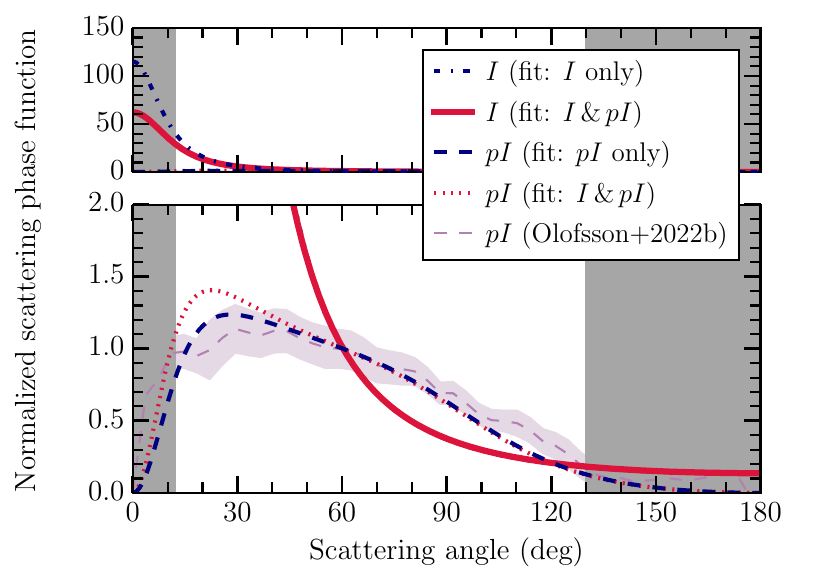}
    \caption{Scattering phase function for our best model of the polarized and total intensity observation, fit independently or jointly. For comparison, we added the model retrieved by \citet{olofsson_vertical_2022} for the polarized intensity data.
    We hide the part below $12\dego$ and above $130\dego$, because we do not detect the disk for such scattering angles, neither in polarized or total intensity light (see text).
    }
    \label{fig:SPFs}
\end{figure}

In Fig.~\ref{fig:SPFs}, we compare the scattering phase function of the joint and independent modeling of the polarized and total intensity observations derived from the disk modeling described above. Using the ad hoc formalism of Henyey-Greenstein (see Eqs.~\eqref{eq:HG_1compo_ti} and~\eqref{eq:HG_1compo_polar}), the scattering phase function is  parameterized by one scattering anisotropy parameter $g$, which is the same for the polarized and total intensity data in the case of the joint modeling.

\begin{figure*}[h!] \centering
    \includegraphics[height=6.4cm]{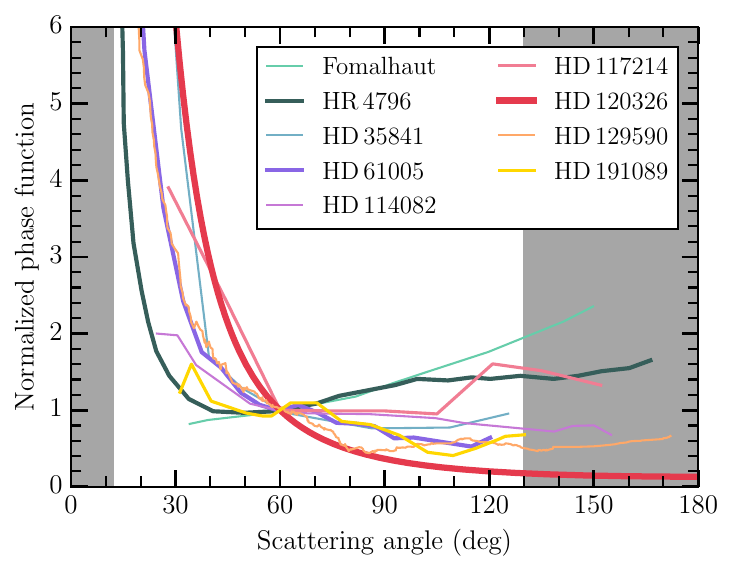} \quad
    \includegraphics[height=6.4cm]{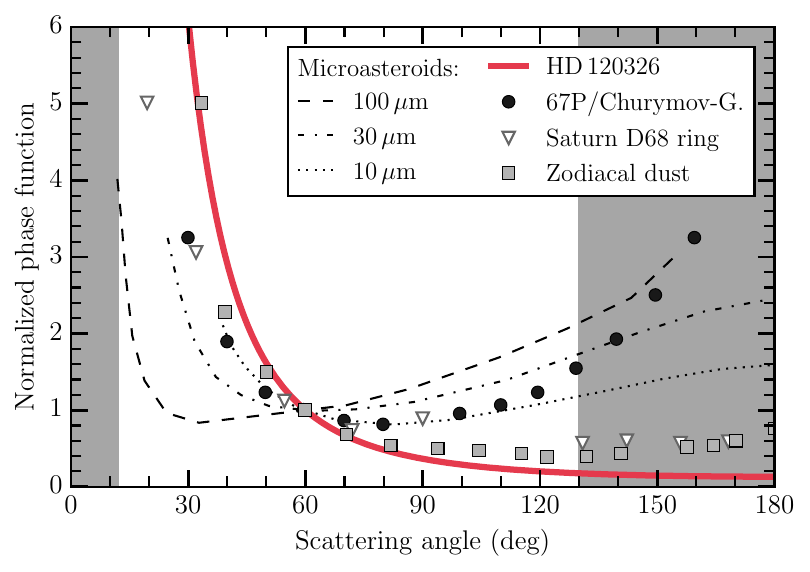}
    \caption{ Scattering phase function of the inner belt of HD\,120326 seen in total intensity compared to other extrasolar debris disks (left) and different dust populations in the Solar System (right).}
    \label{fig:SPF_I_literature}
\end{figure*}

Both polarized SPF mostly agree with each other, with some differences at low scattering angles ($10$--$30\dego$) that correspond to a region of the disk close to or below the inner working angle of the coronagraph. 
In Fig.~\ref{fig:SPFs}, we also compare our polarized SPF with the one derived by \citet{olofsson_vertical_2022}. They are consistent with each other at $2\sigma$, and mostly at $1\sigma$, with differences for scattering angles between $105\dego$ and $125\dego$, corresponding to backward scattering from the disk. We hide in the Fig.~\ref{fig:SPFs} (and Fig.~\ref{fig:SPF_I_literature}) the parts of the SPFs for which we do not have access to the scattering angles, that is to say, below $12\dego$ (i.e., $90\dego$$-$\,inclination) and above $178\dego$ (i.e., $90\dego$$+$\,inclination) due to geometrical effects, and also above $130\dego$, because no disk signal is detected in either polarized or total intensity data.

As for total intensity, the scattering anisotropy parameter $g$ increases when modeling independently the total intensity data, compared to modeling it jointly with polarimetric observations (see Tables~\ref{tab:mcmc_results_nir} and~\ref{tab:mcmc_results_nir_indpt_I}). This results in a disk which seems even more strongly forward scattering, in other words, an SPF that is steeper at low scattering angles (Fig.~\ref{fig:SPFs}). Nonetheless, the height of the peak must be taken with caution, first because it cannot be constrained below $12\dego$ (geometrical effects).
Second, the inner belt of HD\,120326 at low scattering angles ($10$--$30\dego$)  corresponds to a projected separation below or at the edge of the inner working angle of the coronagraph. Moreover, at small separations, the self-subtraction effects caused by ADI removes most of the light \citep{milli_impact_2012}. Estimating the SPF of the inner belt of HD\,120326 is therefore particularly challenging at low scattering angles.

Figure~\ref{fig:SPF_I_literature} shows the comparison of the SPF of HD\,120326 in total intensity at $1.6~\mic$ with that of other extrasolar debris disks and dust populations in the Solar System. 
The dust populations in the Solar System include the zodiacal dust \citep{leinert_interpretation_1976}, the ring D68 of Saturn \citep{hedman_saturns_2015}, the comet 67/Churyumov-Gerasimenko \citep[using the dataset MTP020 acquired on 2015-08-28;][]{bertini_scattering_2017}, and microasteroids of diameter $10~\mic$, $30~\mic$, and $100~\mic$ \citep{min_lunar_2010}. The SPFs of these microasteroids, corresponding to dust grains covered by small regolith particles, were derived theoretically using Fraunhofer diffraction model and Hapke reflectance theory \citep{hapke_bidirectional_1981} by \citet{min_lunar_2010}. All SPFs are normalized at $60\dego$ for easier comparison. 

Based on Fig.~\ref{fig:SPF_I_literature}, one can see that the SPF of HD\,120326 is somehow similar to the SPFs of HD\,35841 \citep{esposito_direct_2018}, HD\,61005 \citep{olofsson_azimuthal_2016} and HD\,114082 \citep{engler_highalbedo_2023}, with a significant forward scattering peak and a continuous decline as the scattering angle increases. Nonetheless, we note that the derived SPF of HD\,120326 is limited by the one-component HG function we used in our modeling, which prevented us from obtaining both forward and backward scattering contributions for the inner belt of HD\,120326. This can be observed, for instance, in the cases of HR\,4796 \citep{milli_nearinfrared_2017} and HD\,11721 \citep{engler_hd_2020}. Even though we did not observe any backward scattering for HD\,120326 in the images, it may be present at a lower level that our observations might not be sensitive to. Regarding the dust populations in the Solar System, the SPF of HD\,120326 is very similar to that of the zodiacal dust, as already reported for other debris disks in \citet{hughes_debris_2018}.

\subsection{Maximum degree of linear polarization \label{sec:max_dolp}}

We here derived the maximum polarization fraction of the inner belt, defined as the maximum of the linear degree of polarization over the scattering angles. We divided the scattering phase function of the polarized intensity data by that of the total intensity data for our best synthetic models, using the scaling flux found in our modeling to express them in the same unit. This led us to obtain the linear degree of polarization for our best disk model.

The  maximum polarization fraction occurs at a scattering angle of $90\dego$, because of the Rayleigh scattering assumed in our formalism to model the polarized SPF (Eq.~\ref{eq:HG_1compo_polar}), and the fixed scattering anisotropy parameter $g$ for both total and polarized intensity models. This sets the bell shape of the linear degree of polarization of the inner belt. 

We found that the inner belt of HD\,120326 has a maximum polarization fraction of $51\pm6\,\%$. The $1\sigma$ uncertainty accounts for the error of the flux in total and polarized intensity (Table~\ref{tab:mcmc_results_nir}). 
This maximum polarization fraction is similar to HR\,4796\,A \citep[$50\,\%\pm3\,\%$ at $1.9$--$2.19~\mic$;][]{perrin_polarimetry_2015}, and higher than other debris disks \citep[$10$--$20\,\%$ in the NIR, e.g., Beta~Pictoris, HD\,15115, HD\,32297, HD\,114082, see][]{tamura_first_2006,asensio-torres_polarimetry_2016,bhowmik_spatially_2019,engler_highalbedo_2023}. A high maximum polarization fraction can for instance be associated with a low albedo of dust particles \citep{umov_chromatische_1905}, or also to small monomers in dust grain aggregates \citep{tobonvalencia_scattering_2022}.

\subsection{Reflectance of the inner belt \label{sec:disk_reflectance}}

We derived the reflectance of the inner ring based on the spectroscopic SPHERE/IFS (YJ bands) and photometric SPHERE/IRDIS (broad band H) data acquired simultaneous during the best epoch of observation (2019-07-09) in total intensity, and the additional photometric points acquired in 2016-04-05 in the narrow bands H2 and H3. 
The reflectance represents the surface brightness of the disk ($S\hspace{-0.5mm}B_\text{disk}$) divided by the total stellar flux ($F_\star$). 

\begin{figure} \centering
    \includegraphics[width=0.95\linewidth]{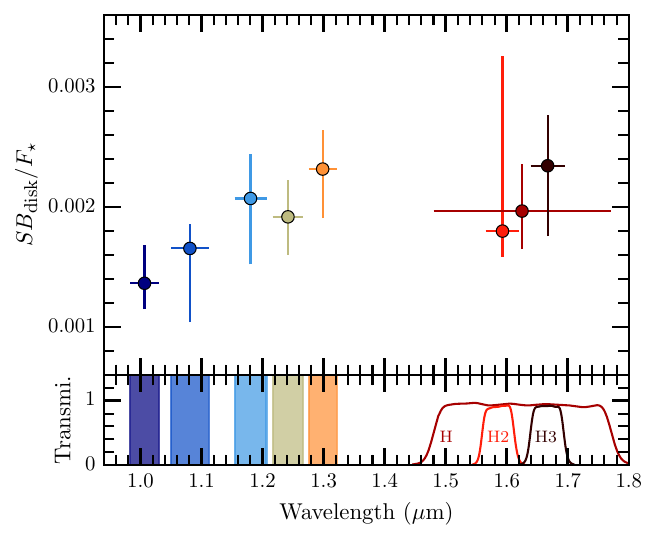}
    \caption{Reflectance spectrum of the inner belt of HD\,120326.
    For each point, we derive the surface brightness of the disk ($S\hspace{-0.5mm}B_\text{disk}$) and the stellar flux ($F_\star$) over a spectral range shown by the horizontal line around each point, and also by the colored areas (IFS) or the filter transmission (IRDIS) at the bottom. 
    }
    \label{fig:reflectance}
\end{figure}
In practice, we ran new \texttt{MCMC} simulations for the five IFS images (Fig.~\ref{fig:im_red_ti_IFS}) to estimate the surface brightness of the inner belt by using the open-access code \texttt{MoDiSc}. In the \texttt{MCMC} simulations, one free parameter (the flux scaling factor) was used, and the five others ($r_0$, $PA$, $i$, $g$, $\alpha_\text{out}$) were fixed to the best fitting values (Table~\ref{tab:mcmc_results_nir}) found previously (Sect.~\ref{sec:disk_modelling}) based on our modeling of the joint total and polarized intensity observations in the broad band H. 
We determined the surface brightness of the inner belt as the total flux of the best synthetic belt model. Then, we divided the surface brightness of the belt by the stellar flux estimated within a large circular region of diameter $12$~FWHM, representing between $20$ and $27$ pixels depending on the wavelength and plate scale of the instrument (IFS or IRDIS). 

We show the reflectance spectrum of the inner ring between $1.0$ and $1.8~\mic$ in Fig.~\ref{fig:reflectance}. Based on the SPHERE data, we conclude that the inner belt seems to have a red color at $1\sigma$ between $1.0$ and $1.3~\mic$, and a gray color between $1.3$ and $1.8~\mic$. 
Although there are large uncertainties on the $1.6~\mic$ measurement, there might be a tentative evidence of a break in the reflectance spectrum between $1.3$ and $1.5~\mic$. Such a break could be caused by an absorption band around $1.5~\mic$, possibly due to H$_2$O in water ice \citep{ciarniello_visir_2021} or in minerals \citep[e.g., phyllosilicates;][]{hu_theoretical_2012,lane_reflectance_2024}. The red spectral slope from $1.3$ to $1.5~\mic$ would be due to additional compound(s) mixed with H$_2$O ice or H$_2$O-bearing minerals, possibly opaque minerals or organic matter \citep[e.g., mixture of water ice and kerite in ][]{ciarniello_visir_2021}.

Regarding the uncertainties on $S\hspace{-0.5mm}B_\text{disk}$/$F_\star$, we derived them by including the errors on the estimations of the surface brightness of the disk and the stellar flux. 
Concerning the surface brightness, we estimated the error via the posteriors of the \texttt{MCMC} simulations. The lower and upper errors are defined to cover an interval of $68\%$ of the posteriors, centered on their median value (e.g., Fig.~\ref{fig:cornerplot_morpho_sphere}). As for the stellar flux, we considered a conservative $5\%$ variation during the sequence of observation, based on empirical experience.

\subsection{Nature of the second component of the disk (seen in the SPHERE total intensity data) \label{sec:second_structure_nature}}

By examining the residual image in Fig.~\ref{fig:im_results_mcmc} (bottom-right), one can see that there is still some signal in the southwestern region of the disk, indicated by the white arrow. Such a signal is located beyond the inner belt seen in polarimetry (Fig.~\ref{fig:reduced_images_polar_Jy_snr}), and is unlikely to be caused by an effect of the ADI postprocessing on the inner belt. Indeed, the residual image is obtained by processing a inner-belt-free cube, since we subtracted the disk model to the cube of observation, before processing it with PCA~ADI (Sect.~\ref{sec:disk_modelling}). In addition, this second structure is imaged at different epochs and with different algorithms (Figs.~\ref{fig:im_snr_red_ti} and \ref{fig:im_ipca}).

Therefore, we firmly confirmed this second disk structure, whose exact nature remains to be determined. Previously, \citet{bonnefoy_belts_2017} considered whether it could be a halo of dust grains, or, a second belt. We revisit both options below, by using in particular the fact that the second structure is not detected in the polarimetric data \citep[][and this work]{olofsson_vertical_2022}.

\subsubsection{Hypothesis: Second belt \label{sec:2nd_ring} }

By assuming that the second structure is an additional belt, \citet{bonnefoy_belts_2017} found  that its reference radius would be $137\pm9~\au$\footnote{To be consistent, we updated their original value $130\pm8~\au$ to $137\pm9~\au$ using the most recent distance estimation of $113.4~\pc$ (and not $107.4~\pc$ as in \citet[][]{bonnefoy_belts_2017}).} and its flux ratio relative to the inner one would be $0.08\pm0.03$  at the scattering angle for which the flux is maximum in total intensity. By using their disk models, we checked that this value of flux ratio is the same at scattering angles of $\sim$$30\dego$, namely, the maximum of the polarization intensity peak based on Fig.~\ref{fig:SPFs}, and $\sim70\dego$, which still has a high level of polarization intensity and at which we do not detect the second structure in Fig.~\ref{fig:reduced_images_polar_Jy_snr}. 

Thus, by assuming that the dust particles of the second belt have the same polarimetric properties as those of the inner disk, we expect the flux of the second belt to be $8\pm3\%$ of the flux of the inner belt, that is, $12.5$~times smaller. However, such a flux is too faint to be recovered in our polarimetric data, because we only detect the inner belt with a $S/N$ of $3$--$7$ per pixel (see Fig.~\ref{fig:reduced_images_polar_Jy_snr}), so a flux $12.5$~times smaller results in a $S/N$ below one per pixel. Therefore, the second structure seen in total intensity could be a second ring, that remains undetected in our SPHERE polarimetric (2018-06-01) observations due to a lack of sensitivity.

\subsubsection{Hypothesis: Apocenter pile-up of small particles originating from the inner belt \label{sec:betadisk}}

\begin{figure}[h!] \centering
    \includegraphics[height=4.2cm]{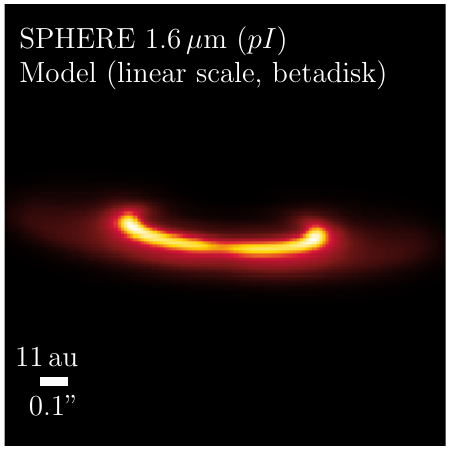} 
    \includegraphics[height=4.2cm]{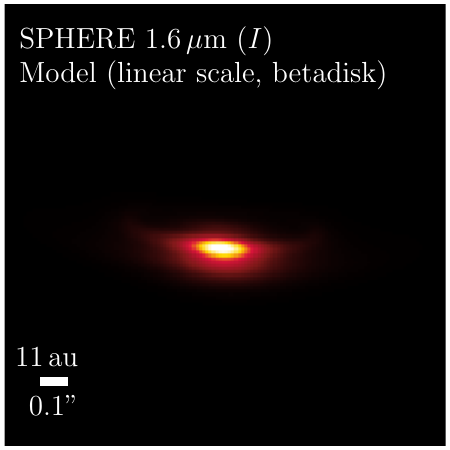} 

    \includegraphics[height=4.2cm]{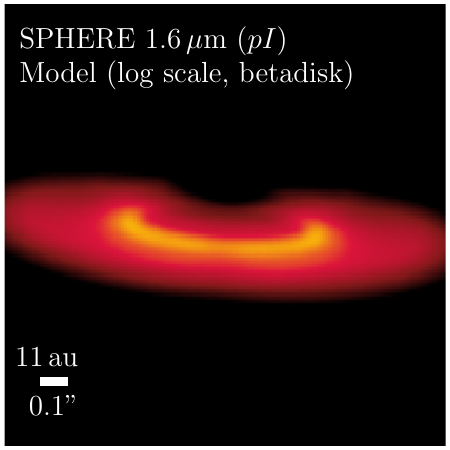} 
    \includegraphics[height=4.2cm]{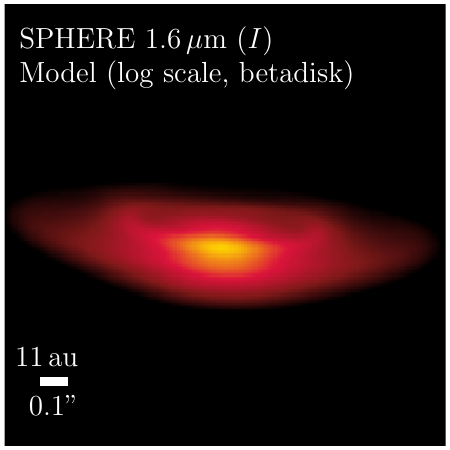}

    \includegraphics[height=4.2cm]{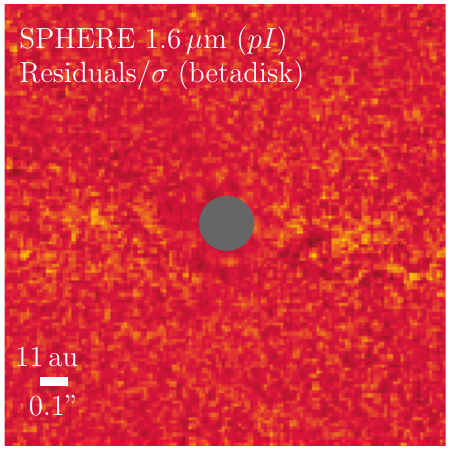} 
    \includegraphics[height=4.2cm]{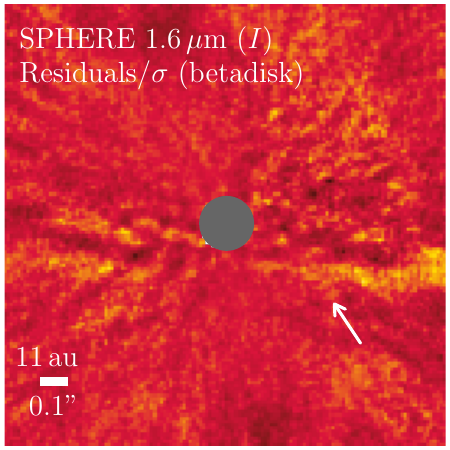} 
    \caption{Scattered light images of one dust belt generated using the code \texttt{betadisk}, with the associated residual map normalized by the noise (“$\sigma$”), in polarized and total intensity (left and right, respectively). The second structure seen with SPHERE (indicated by the white arrow) is somehow more removed, compared to the \texttt{GRaTeR} modeling (Fig.~\ref{fig:im_results_mcmc}, bottom-right).}
    \label{fig:im_betadisk}
\end{figure}
Alternatively, the second structure could be a halo of small particles related to the inner belt. Indeed, these small particles could originate from the collisions of larger bodies within the inner belt \citep[“birth ring”,][see also \citet{lecavelierdesetangs_dust_1996,augereau_dynamical_2001}]{strubbe_dust_2006} similarly to the debris disk around HD\,129590 \citep{olofsson_apocenter_2023}. This halo would correspond to the eccentric orbits of small particles, whose size sets their eccentricity. Instead of a smooth continuous halo, it could have a visual gap between the birth belt and the second disk structure. This could be explained by the apocenter pile-up model \citep{olofsson_apocenter_2023}, that uses the fact that the small, eccentric particles spend most of the time at their apocenter. By considering many small particles in a narrow range of sizes, their eccentricities would be similar, and so too their apocenter distances (but with arguments of pericenter uniformly distributed between [$0$, $2\pi$]). This results in a visual arc structure located at their apocenter distance for scattering angles giving a bright enough SPF. This arc structure could be enhanced by scattering efficiencies that depend on the size (and composition, porosity) of dust particles \citep{olofsson_apocenter_2023} and the wavelength of observation. We stress that the apocenter pile-up model introduces a visual pile-up (not a dynamical one), and that it is different to the apocenter glow, which requires an asymmetric disk \citep[][ see also \citet{wyatt_how_1999}]{pan_apocenter_2016}. 

To investigate this, we used the open-access \texttt{betadisk} tool\footnote{Accessible here \url{https://github.com/joolof/betadisk}} \citep{olofsson_halo_2022}. This code can compute scattered light images for different grain sizes, parametrized by the parameter $\beta$.  This parameter is the ratio between the stellar radiation pressure and gravitational forces. It is inversely proportional to the particle size $s$ in the case of stars bright enough,  $\beta<0.5$ and  $s\gtrsim0.5~\mic$ (e.g.,
\citet{burns_radiation_1979}, Figs. 1 and 2 from \citet{artymowicz_radiation_1988}, and Fig. 7 from \citet{lohne_modelling_2012}).

By assuming the disk parameters determined in Sect.~\ref{sec:disk_modelling} (disk inclination, position angle, reference radius, and the scattering anisotropy parameter), we modeled the inner belt with \texttt{betadisk}.  This synthetic disk is a linear combination of $14$ scattered light images computed for different grain sizes, corresponding to $\beta$ from $0.01$ to $0.4$. 
We found out that a combination of large particles, mainly in the parent bodies corresponding to the inner belt, and eccentric, small particles could better model the disk structures than the previous model generated with the radiative transfer \texttt{GRaTeR} code, for which we used its fast-computing version. This version does not assume any size distribution, but an overall scattering efficiency for the dust without any underlying assumption on the grains. 
We can indeed see that the residual map corresponding to the \texttt{betadisk} model (Fig.~\ref{fig:im_betadisk}) is to some extent cleaner than the one corresponding to the \texttt{GRaTeR} model (Fig.~\ref{fig:im_results_mcmc}). The apocenter pile-up model may therefore explain some signal of the second structure. 

Yet, this second structure is asymmetric, with a higher flux and larger extension in the west than in the east. In addition, this asymmetry is reversed for the inner belt, which is brighter and more extended in the east than in the west, as also seen in Fig.~\ref{fig:im_snr_red_ti} here and  Fig.~2 from \citet{bonnefoy_belts_2017}. This aspect still remains to be explained.

\section{Morphology and photometry of the dust belt seen with ALMA millimeter observations \label{sec:modelling_results_alma}}

In Sect.~\ref{sec:alma_morpho}, we constrain the morphology of the disk detected with ALMA (Fig.~\ref{fig:im_contours_ti_polar_alma}). Then, in Sect.~\ref{sec:sed}, we describe our modeling of the SED, taking into account the new ALMA measurement. Last, in Sect.~\ref{sec:dust_mass}, we explain how we derived the dust mass of HD\,120326.

\subsection{Morphology of the disk \label{sec:alma_morpho}}

The ALMA observation does not resolve the disk (or only very marginally, see Fig.~\ref{fig:im_contours_ti_polar_alma}), so we used parametric modeling of the visibilities to obtain constraints on the disk parameters. We used an optically thin 3D model that takes line of sight effects into account, which was previously used to model debris disks \citep[e.g.,][]{kennedy_unexpected_2020}. To obtain constraints on parameters, we modeled the data using \texttt{galario} \citep{tazzari_galario_2018} to Fourier transform the model images and compare them to the observed visibilities, and \texttt{emcee} to perform Markov-Chain Monte-Carlo sampling and extract posterior distributions for parameters. We assumed a Gaussian torus model, for which the model parameters are $x/y$ offsets from the observation phase center, the disk inclination and position angle, the disk flux, the ring radius and width. The scale height ($H/r$) was restricted to be small ($<0.05$). We did not include the stellar flux as it is $\sim$100 times fainter than the disk at $1~\mm$. We ran the \texttt{MCMC} until all parameters  converged, and then continued sampling to build up the posterior distributions.

Posterior distributions from the modeling are shown in Fig.~\ref{fig:cornerplot_morpho_alma}. The inclination ($70\pm5\dego$) and position angle ($-91.7\pm4.5\dego$) are consistent at $2\sigma$ with the scattered light results. 
While the resolution clearly limits the conclusions that can be drawn, the disk detected with ALMA peaks at radii of $35\pm25~\au$ at $1\sigma$ based on our modeling.
At $1\sigma$, the outer edge of the disk in the millimeter is $75~\au$ at most, thus closer to the star than the second structure seen with SPHERE, which peaks at radii larger than $100~\au$ (Sect.~\ref{sec:second_structure_nature}).
 
In particular, the millimeter-wave emission peaks at a location consistent with that of the inner belt seen in the NIR, for which scattered light models give disk radii between $30$ and $65~\au$ at $1\sigma$, depending on the modeling and the SPHERE data used (see Sect.~\ref{sec:disk_modelling}). By assuming millimeter dust traces the presence of planetesimals, we would expect that the inner belt resolved with SPHERE is a planetesimal belt. 
To be more confident would require higher resolution ALMA observations.

\subsection{Modeling of the spectral energy distribution \label{sec:sed}}

\begin{figure}[h!] \centering
    \includegraphics[width=\linewidth]{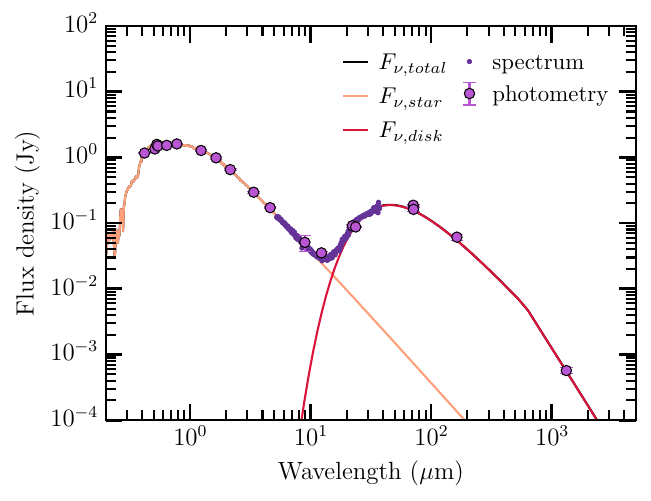}
    \caption{Spectral energy distribution of HD\,120326 based on archival data (see main text) and the new ALMA measurement at $1.3~\mm$.}
    \label{fig:sed}
\end{figure}

To construct a spectral energy distribution, we collected photometry from various sources, for instance: $UBV$, Stromgren, \textsc{Hipparcos}, \textit{Gaia}, 2MASS, WISE, \textit{Spitzer}, and \textit{Herschel}. We also used the new ALMA photometric point derived in this work, and a \textit{Spitzer} IRS spectrum. These data were fit with stellar and blackbody disk models as described previously by \citet{yelverton_statistically_2019}. The best fitting model is shown in Fig.~\ref{fig:sed}. The disk model used is a single component modified blackbody, as we found that this was sufficient to explain the data. We determined that the star has an effective temperature of $6\,940 \pm 100\,\kelvin$, and a luminosity of $4.5 \pm 0.1$\,$L_\odot$. 
We derived for the disk a temperature of $110 \pm 1$\,K and a fractional luminosity of $1.8 \times 10^{-3}$. The ALMA photometry is somewhat lower than expected for a pure blackbody, but this is expected for debris disks where much of the emission comes from bodies that are significantly smaller than 1\,mm.

Assuming blackbody absorption and emission, the dust temperature implies a dust population at 13\,au from the star, which is clearly closer than it is observed. However, the small dust sizes also mean that their temperatures are higher than a blackbody at the same stellocentric distance, meaning that the 13\,au distance is a lower limit. Empirically, \citet{pawellek_dust_2015} found that the difference in disk radii at 4.5\,$L_\odot$ is about a factor of four, which would make the SED estimate consistent with the semi-major axis derived for the SPHERE or ALMA data.

\subsection{ Dust mass \label{sec:dust_mass}}

Based on the results from our SED modeling, we derived the dust mass,  which consists of the mass of dust particles of size below a few millimeters. This is a significant underestimation of the total mass budget of the debris disk, as most of the mass is represented by the largest bodies (e.g., planetary embryos or kilometer-size planetesimals). The dust mass is regularly derived in the literature, expressed as
\begin{equation}
    M_\text{dust} \;=\; \frac{F\nu\,d^2}{\kappa\,B_\nu(T_\text{dust})}\,,
\end{equation}
where $F_\nu$ is the flux density of the dust belt in W.m$^{-2}$.Hz$^{-1}$, $d$ is the distance of the star hosting the dust belt, $\kappa_\nu$ is the dust opacity assumed to be $1.7~\mathrm{cm^2 g^{-1}}$ ($850~\mic/\lambda_\text{obs}$) from \citet{beckwith_survey_1990} (also assumed in \citet{krivov_solution_2020}), $B_\nu(T_\text{dust})$ is the Planck function, and $T_\text{dust}$ is the temperature of the dust grains chosen to be consistent with the observed SED ($110$~K, see Sect.~\ref{sec:sed}).

\begin{figure} \centering
    \includegraphics[width=\linewidth]{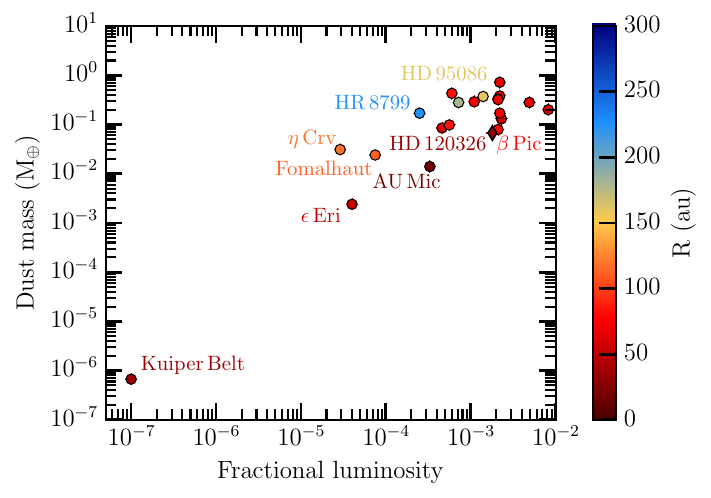}
    \caption{ Dust mass and fractional luminosity of the debris disk around HD\,120326 (diamond marker) compared to those of the sample used in \citet{krivov_solution_2020} (circle markers).  The dust mass and fractional luminosity of HD\,120326 is in particular very similar to $\beta$\,Pic. The color represents the location of the dust belt relative to its host star.}
    \label{fig:dust_mass}
\end{figure}

We obtained a dust mass for HD\,120326 of $0.067~\mearth$. In Fig.~\ref{fig:dust_mass}, we compare this dust mass to the sample from \citet{krivov_solution_2020}, which is restricted to the best-quality ALMA data from \citet{matra_empirical_2018}, observed either in band 6 ($\lambda\sim1.3~\mm$) or 7 ($\lambda\sim850~\mic$). From Fig.~\ref{fig:dust_mass}, HD\,120326 has a fractional luminosity and dust mass very similar to $\beta$\,Pictoris ($2.1\times10^{-3}$ and $0.079~\mearth$; A6V-type star), and for the two next closest matches, to HD\,61005 ($2.3\times10^{-3}$ and $0.13~\mearth$; G8Vk star) and HD\,146181 ($2.2\times10^{-3}$ and $0.17~\mearth$; F6V star), based on the values derived in \citet{krivov_solution_2020} (dust mass) and \citet{matra_empirical_2018} (fractional luminosity).

\section{Discussion \label{sec:discussion}}

We discuss  our results on the inner belt (Sect.~\ref{sec:discussion_inner_belt}) below. We also investigate the global morphology of the debris disk around HD\,120326 (Sect.~\ref{sec:discussion_global_archi}).

\subsection{Inner belt \label{sec:discussion_inner_belt}}

In Sects.~\ref{sec:disk_modelling}-\ref{sec:disk_reflectance}, we discuss the constraints on different properties of the inner belt based on the VLT/SPHERE data, which we then identified as a planetesimal belt based on the ALMA data (Sect.~\ref{sec:alma_morpho}). To summarize the results based on the SPHERE data, the inner belt has a semi-major axis of $42.5~\au$, with a FWHM of $24.3~\au$ (fractional width of 0.57). Its dust grains are shown to be particularly forward scattering in terms of the total intensity. Assuming a Rayleigh scattering for the SPF in polarized intensity data does match  the polarized intensity data well. Their relatively high maximum degree of polarization ($51\%\pm6\%$) could be related to various physical properties (e.g., low albedo or small monomers if particles are structured in dust aggregates; Sect.~\ref{sec:max_dolp}). The red ($1.0$--$1.3~\mic$) and then gray  ($1.5$--$1.8~\mic$) slopes of the dust particles can be  interpreted as either related to the size or the composition of the dust grains. In the former case, dust particles would rather have a size ($s$) equal or larger than the one defined by the size parameter size ($2\pi s/\lambda$$\,\sim\,$$1$) at the turnover wavelength  $1.3~\mic$ (see Fig.~\ref{fig:reflectance}).
On the other hand, in the latter case, the slope and tentative break in the reflectance spectrum could be caused by absorption related to the presence of H$_2$O (see Sect.~\ref{sec:disk_reflectance}). 
Increasing the wavelength coverage and/or the spectral resolution could help leverage the degeneracies between different dust grain properties.

\begin{figure}[h] \centering
    \includegraphics[width=\linewidth]{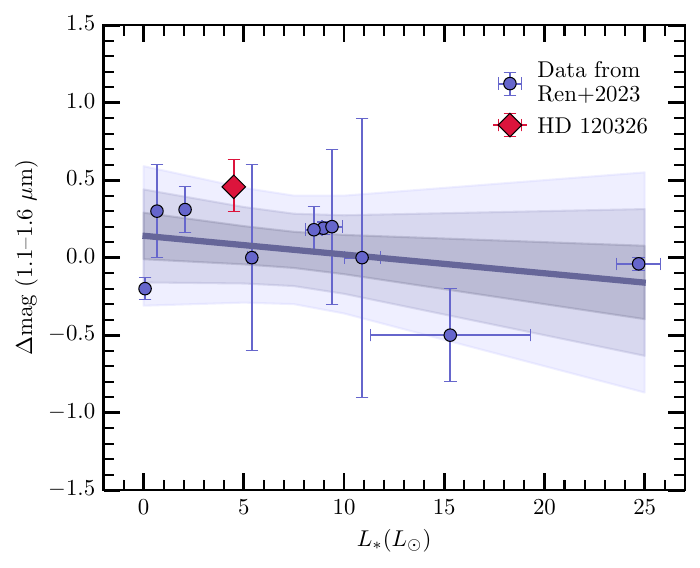}
    \caption{Dust color at scattering angles between $80$ and $100\dego$ for the debris disk around HD\,120326 compared to other debris disks \citep{ren_debris_2023}. The trend and shaded areas corresponding to the uncertainties of the trend at $1\sigma$, $2\sigma$, and $3\sigma$ are from \citet{ren_debris_2023} based on their analysis (see their Sect. 3.2).}
    \label{fig:color_ren+2023}
\end{figure}
We compared the color $1.1$$-$$1.6~\mic$ of the inner belt of HD\,120326 with other debris disks \citep{ren_debris_2023} in Fig.~\ref{fig:color_ren+2023}. The inner belt of HD\,120326 has a color redder than $0.36$ mag compared to what is predicted from \citet{ren_debris_2023} for a star of $4.5~\lsun$, lying at $3\sigma$ of their confidence region. Nonetheless, as  they highlighted in their paper, the trend is based on a relatively small sample (ten disks), and some systematics could occur, as their data are from a different instrument, HST/NICMOS (but with a similar spectral coverage, with the filters F110W and F160W).

We point out that it would be particularly interesting to image the inner belt of HD\,120326  with HST/STIS, which has a wavelength range centered at $0.59~\mic$ ($d\lambda=0.44~\mic$), using a mask with a smaller inner working angle (e.g., the Cbar5). 
Indeed, \citet{ren_debris_2023} derived the color $0.59-1.1~\mic$ for  twenty debris disks, and reported that debris disks with stellar luminosity of $4.5~\lsun$ have a bluer color of $0.7$~mag at these spectral ranges. If HD\,120326 would also have a blue color $0.59-1.1~\mic$ and assuming the color is mainly related to the size of dust grains rather than their  composition, this could hint at two different dust populations \citep[e.g., Fig.~2 from][]{thebault_there_2019}. Very small grains would be responsible for a blue color between $0.59$ and $1.1~\mic$, and larger ones would be responsible for a red color between $1.1$ and $1.6~\mic$.

\subsection{Overall picture of the debris disk around HD\,120326 \label{sec:discussion_global_archi}}

The complex, global morphology of the debris disk around HD\,120326 is illustrated in Fig.~\ref{fig:HD120326_overview}. Two key questions are, first, what is the nature of the extended asymmetric structure seen with HST/STIS at 100s of au and the second one seen with SPHERE in total intensity at smaller distances, and second, to what extent the inner structures detected with SPHERE and ALMA connect with the outer one seen with STIS.

Regarding the first question, the extended, asymmetric structure could be an eccentric or misaligned ring, or a spiral. The onset of an eccentric or misaligned disk is preceded by the appearance of spirals \citep[][assuming massless disks]{wyatt_spiral_2005}, and requires timescales that are an order of magnitude greater than the timescale governing the appearance of spirals, amounting to typically several $10$--$100~\myr$ at the edges ($\gtrsim100~\au$) of planetary systems \citep[e.g.,][]{pearce_dynamical_2014}. Thus, dynamical timescales are shorter with respect to the creation of a spiral, as opposed to  an eccentric or misaligned disk.  
Therefore, considering the young age \citep[$16~\myr$;][]{mamajek_post_2002} of the HD\,120326 system, the asymmetric structure could is more likely to be a spiral than an eccentric or misaligned disk.

While spirals are regularly observed in gas rich protoplanetary disks, their detection is much rarer in debris disks. The reason for this is still unclear. The argument usually advanced is that they are only transient \citep{wyatt_spiral_2005,farhat_case_2023}. However, theoretical studies  of planet debris disk interactions indicate that these features can be more persistent depending on the importance of disk self-gravity \citep{hahn_secular_2003,ward_dynamics_1998,jalali_density_2012,sefilian_secular_2022}.
Notably, HD\,120326 is one of the very rare debris disks with an imaged spiral-like feature. The other ones are the debris disks around TWA~7 \citep{olofsson_resolving_2018}, and possibly HD~106906 \citep{rodet_origin_2017,farhat_case_2023}, and the gas-rich debris disk around HD~141569 \citep[][but this one could be considered as a transition disk as with an age of $5~\myr$]{konishi_discovery_2016}. 
Among all of them, HD\,120326 is the only one with a spiral-like feature spanning over 100s of au, up to $1\,000~\au$ if co-planar with the inner belt. 
By assuming it is indeed a spiral, different mechanisms of dynamical interactions could be outlined to shape it.
Stellar flybys can generate spirals \citep[e.g., HD\,141569;][]{reche_investigating_2009}, but this scenario remains very unlikely for HD\,120326 based on the extensive search for such a dynamical perturber by \citet{bertini_flybys_2023}.

As a result, the likely remaining possible mechanisms are, first, secular dynamical interactions with a planet \citep{wyatt_spiral_2005,farhat_case_2023} 
possibly combined to disk self-gravity \citep{hahn_secular_2003,ward_dynamics_1998,sefilian_secular_2022,sefilian_formation_2021,sefilian_formation_2023}, or considering disk self-gravity alone \citep{jalali_density_2012}. It could also be collisional dust avalanches following the breakup of a large planetesimal-like object \citep{artymowicz_beta_1997,grigorieva_collisional_2007},  giant impacts \citep{jones_giant_2023},
interstellar medium wind \citep[as for HD\,61005;][]{maness_hubble_2009}, or  it could be potentially inherited from the protoplanetary disk phase \citep[e.g.,][]{najita_pebbles_2022}. 

\begin{figure}
    \centering
    \includegraphics[width=0.95\linewidth]{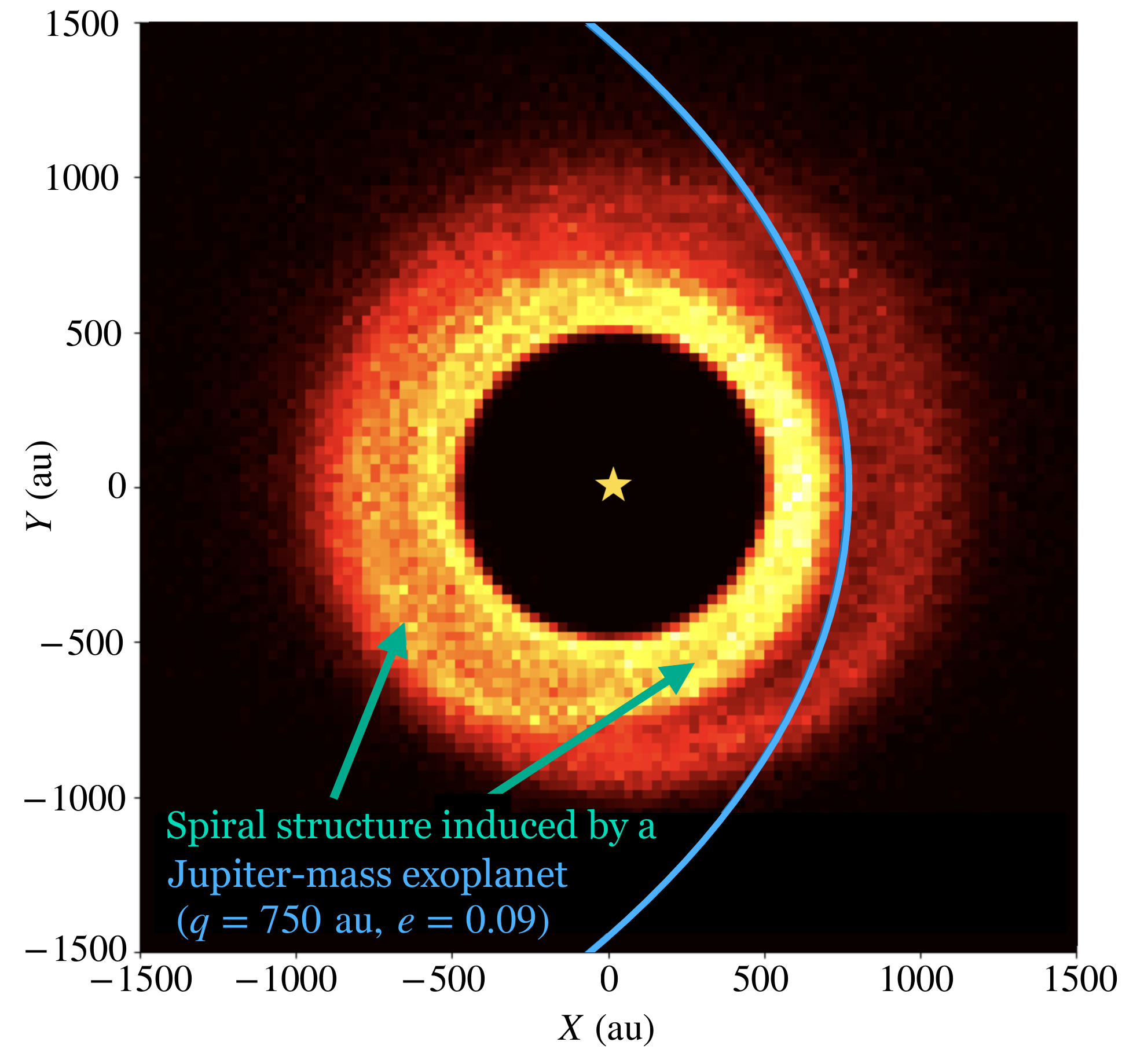}
    \caption{ Very eccentric, wide-separation Jupiter-mass exoplanet that could generate a spiral-like feature spanning 100 of au, during the timescale of the age of the system. The blue curve shows part of the planetary orbit used in our N-body simulations.} 
    \label{fig:simu_Nbody}
\end{figure}

To further investigate the hypothesis of dynamical interactions with a putative planet, we ran N-body simulations using the symplectic code \texttt{SWIFT-RMVS3} \citep{levison_long-term_1994}, assuming a massless disk and a giant planet with various orbital configurations, to assess whether an extended asymmetric structure spanning 100 of au could be generated.
We successfully reproduced a spiral structure extending over 100s of au within $16~\myr$ (i.e., the age of the system), by considering a giant exoplanet of  $1~\mj$ on a highly eccentric ($e = 0.9$) orbit with a periastron of $750~\au$ (see Fig.~\ref{fig:simu_Nbody}). A Jupiter-mass exoplanet remains below the detection limits achievable for this system using SPHERE \citep[see Fig.~B.1 from][]{bonnefoy_belts_2017}. This highly eccentric, wide-orbit Jupiter-mass scenario is possible under the assumption of a massless disk. However, other scenarios involving exoplanet(s) --whether accounting for disk self-gravity or not-- may also be possible \citep{sefilian_secular_2022}. A comprehensive investigation into the physical and orbital constraints required for an exoplanet to generate the extended spiral-like feature is beyond the scope of this paper. In addition, based on our N-body simulations, this giant exoplanet alone would not explain the asymmetric structure observed at 100-150 au, which likely has a separate origin.   

Alternatively, we point out that the location of a bright point-like source overlays with the southeastern end of the spiral-like feature (see Figs.~\ref{fig:im_HST_flux} and~\ref{fig:HD120326_overview}). This bright source was determined to be a background star (see Appendix~\ref{app:PMD} and in particular Fig.~\ref{fig:PMD}). We investigated whether the asymmetry of this extended disk structure could be caused by a fortuitous alignment of the background star with dust grains from the halo of dust grains around HD\,120326. In such cases, the particularly forward-scattering dust grains could scatter some of the light from the background star. This could result in more scattered light on the southeastern grains than on the southwestern ones, because there is no bright background star there. This would mean that the spiral-like feature seen might actually be a fraction of an outer belt. Initially, we postulated that this candidate outer belt could be illuminated by the background star. Because of its location and strong forward scattering, the fact that the background star is located on the west side could explain the west-east asymmetry. However, as further described in Appendix~\ref{app:PMD}, the difference in scattering angles between the eastern and western sides is not sufficient to explain such a strong asymmetry.

To determine what is the true morphology of the extended debris disk around HD\,120326, and ultimately identify the mechanism causing it, new HST/STIS data are required. Indeed, large areas where disk signal is expected are masked by the WedgeA1.0 or the cross-pattern of the spiders, or are affected by self-subtraction effects from the data processing (see Sect.~\ref{sec:obs_hst} and Fig~\ref{fig:HD120326_overview}). 
The new STIS data should be acquired with an optimized strategy of observation, using a color-matched reference star, more angular diversity (e.g., with three rolling angles instead of two), and an appropriate position of the wedges relatively to the disk, to probe a larger spatial coverage \citep[e.g.,][]{schneider_stis_2009}.

Concerning the second, enigmatic structure seen with SPHERE in total intensity at smaller radii (e.g., Sect.~\ref{sec:obs_ti_only}), we investigate  its nature in Sect.~\ref{sec:second_structure_nature}. 
Analysing this faint structure requires pushing SPHERE to its limits in terms of sensitivity, so we could not draw decisive conclusions about its true nature. We confirmed that the second structure could be a second ring, which was previously hypothesized by \citet{bonnefoy_belts_2017}, and we showed that due to its faintness, it is not expected to be detected in the SPHERE polarimetric observations (see Sect.~\ref{sec:2nd_ring}). Alternatively, instead of a second ring, it could also be a halo of small particles that originate from the birth ring (see Sect.~\ref{sec:betadisk}), identified as a planetesimal belt (see Sect.~\ref{sec:alma_morpho}). This is supported by the apocenter pile-up model \citep{olofsson_apocenter_2023}.

One way to distinguish between the belt and halo hypotheses is through a color analysis of the dust grains in the second structure. This would help determine the grain size and, consequently, the nature of the structure. If the grains have a blue color, this could indicate they are small, and smaller than those in the inner belt, given the red-gray reflectance spectra of the dust grains in the inner belt (Sect.~\ref{sec:disk_reflectance}). This would support a halo nature for the second structure.
Unfortunately, we do not detect well the second structure at wavelengths shorter than those of the broad band H ($1.48$--$1.78~\mic$)  in the SPHERE/IFS data (see Fig~\ref{fig:im_red_ti_IFS}). In these data, we only faintly detect the western part of the disk when combining 30 spectral channels ($0.98$--$1.32~\mic$) and when applying a strong spatial binning (Fig.~\ref{fig:im_red_ti_IFS}, bottom-left image). In addition, this western structure is located very close to a bright, extended signal of unclear origin, which could bias the flux measurement. As a result, only a handful of pixels capture a small fraction of the second structure. Therefore, highly sensitive HST/STIS observations using the Cbar5 mask would be highly valuable for detecting the second structure at short wavelengths and analyzing its color.

Furthermore, we stress that the second structure is asymmetric, with a higher flux and larger extension in the south-west region. This could be explained by a giant impact at the opposite direction in the birth ring, generating small, eccentric dust particles pushed away by the stellar radiation pressure \citep{jones_giant_2023}.

\medskip

As for the second question, regarding whether both the outer (imaged with STIS) and inner (detected with SPHERE and ALMA) structures connect with each other, tackling it also suffers from the lack of spatial coverage of STIS, and in particular at small separations, hidden by the too large mask WedgeA1.0. In addition, some parts of the spiders coincide with the spiral-like feature, in particular in the region where it seems it could connect with the inner ring, see Fig.~\ref{fig:HD120326_overview}. 

Concerning the SPHERE and ALMA images, in both images we do not detect a disk structure above a projected separation of $1.0$--$1.5\arcsec$, where it could overlay with the disk signal detected with STIS data.  However, focusing on the SPHERE NIR total intensity data, we cannot rule out the possibility that the second structure (or halo) extends further away. If such an extension exists, it may remain undetected in the SPHERE data due to sensitivity limitations. Indeed, the starlight reflected onto dust grains decreases with their distance from the star, following a $1/r^2$ dependence. Alternatively, in the outer regions of the halo, the dust grains are likely smaller, resulting in scattered light reduced in the NIR and increased in the optical. In any case, the second structure detected with SPHERE in total intensity is intriguing, and lies in the direction where the spiral-like feature begins. 

Therefore, the existing observations of HD\,120326 cannot distinguish whether the intermediate-scales structures seen with SPHERE and ALMA are connected with the large-scale structure detected with STIS. However, new STIS data using masks with a smaller inner working angle ,such as the Cbar5 or WedgeA0.6, may provide crucial constraints.

\section{Conclusion \label{sec:ccl}}

In this paper, we present a panchromatic study of the young ($16~\myr$) debris disk around HD\,120326 surrounding a F5V-type star and located in the Sco-Cen OB association.
We used optical (HST/STIS), NIR (VLT/SPHERE, both in total and polarized intensity), and millimeter (ALMA) data, with some of the data published for the first time in this work. These observations were sensitive either to the scattered starlight on dust grains (STIS, SPHERE) or their thermal emission (ALMA).

The debris disk around HD\,120326 shows substructures both at intermediate and large distances, imaged with SPHERE and ALMA ($30$--$150~\au$) and STIS ($\leq$$\,1\,000~\au$). 
From our analysis, we show that modeling the ALMA data with one belt results in a belt morphology and location that are consistent with that of the inner belt imaged with SPHERE. By assuming that millimeter dust particles trace the location of planetesimals, we identified the inner belt as a planetesimal belt. This belt has a semi-major axis at maximum dust volume density of $\sim$$\,43.7$~$\au$ and an inclination of $\sim$$\,78\dego$, based on our joint modeling  of the SPHERE NIR polarized and total intensity data using our open-access code \texttt{MoDiSc}. Constraints on the belt morphology using millimeter data are looser and, thus, higher resolution ALMA observations are needed to be more accurate. Moreover, we derived from the ALMA data an integrated flux value of $561\pm20~\microjy$ in the continuum at $1.3~\mm$ and an upper limit of  $^{12}$CO emission of $4.8~\jy\cdot\kms$.
Using the new photometric ALMA measurement, we modeled the SED and determined a fractional luminosity of $1.8\times10^{-3}$ and a dust mass of $0.067~\mearth$, with both values being very similar to the debris disk around $\beta$ Pictoris.

Based on the SPHERE data, still regarding the inner belt, we derived its scattering phase function both in total and polarized intensity and its maximum degree of polarization ($51\,\%\pm6\,\%$) at $1.6~\mic$. We derived its reflectance spectrum, which seems to show a red color between $1.0$ and $1.3~\mic$ and a gray color between $1.5$ and $1.8~\mic$. There might also be a tentative evidence of break in the reflectance between about $1.3$ and $1.5~\mic$. Such a break could be caused by various absorption features, for instance, water in hydrated minerals (e.g., phyllosilicates) or opaque mixtures of water ice and kerite.

Furthermore, we also explored the nature of the second faint structure ($\leq$$\,150~\au$), seen  in several SPHERE epochs in total intensity, with various processing algorithms. We confirmed it could be a second dust belt or a halo of small particles.  In the context of the halo hypothesis, the asymmetry structure, which is brighter and more extended in the southeast than in the southwestern part, may be explained by a giant impact  that could have occurred in the opposite direction in the inner ring (northeast) and that would generate small, eccentric dust particles pushed away by the stellar radiation pressure \citep[][]{jones_giant_2023}. 
We show that the non-detection in the polarized intensity data of the second structure is not contradictory with the halo or second belt hypotheses. 
A further investigation of both hypotheses could be achieved through a color analysis of the dust grains in the second structure. This would require new observations, such as STIS imaging using the Cbar5 mask. Such observations would also help determine whether the second structure is connected to the extended structure previously detected in archival STIS data acquired using the WedgeA1.0 mask.

As for the extended asymmetric-like feature ($\leq$$\,1\,000~\au$), we investigated  its morphology on the basis of  a new processing of the STIS archival data and theoretical considerations. New, optimized STIS observations are required to precisely constrain its morphology and the physical mechanisms that could generate and shape it. If the spiral feature is confirmed,  HD\,120326 would be one of the very rare debris disks with an imaged spiral feature so far and the only one with a spiral extending over hundreds of au, up to $1\,000~\au$. Yet, it is interesting to note that spiral features are a common outcome of planet-debris disk interactions, both in massless \citep[e.g.,][]{wyatt_insignificance_2005,pearce_dynamical_2014,farhat_case_2023} and massive disks \citep[with or without planets; e.g.,][]{ward_dynamics_1998,hahn_secular_2003,jalali_density_2012,sefilian_secular_2022}. 

To date, no planet has been discovered in HD\,120326. \citet{bonnefoy_belts_2017} ruled out the presence of $>2~\mj$ companions beyond projected separations of $50~\au$ (and below $750~\au$). Future observations with the upgrade of SPHERE \citep[SPHERE+;][]{boccaletti_sphere_2020}, as well as high-contrast imaging instruments on  extremely large telescopes (ELT and GMT) or on the JWST and \textit{Nancy Grace Roman} Space Telescope would be sensitive to less massive exoplanets in HD\,120326 and/or located closer to the host star. Such observations will provide major constraints on the global architecture of HD\,120326, including any potential disk-planet interactions that are sculpting the observed disk structures. Ultimately, these advancements would help to determine how unique the architecture of the young system HD\,120326 really is.

\begin{acknowledgements} 
    C.D. thanks for fruitful discussions \'Elodie Choquet regarding handling HST/STIS data, Gaspard Duchêne concerning scattering phase function and disk modeling, Veronica Roccatagliata concerning any potential flybys, Natalia Engler about albedo derivation for dust grains, and Anne-Marie Lagrange, regarding the presence of putative planets in HD\,120326. C.D. also thanks Karl Stapelfeldt, Elisabeth Matthews, Sophia Stasevic, Xie Chen, Bin Ren, and Oliver Absil for useful discussions. C.D. is also grateful to Anthony Boccaletti, Deborah Padgett, and Sasha Hinkley, who obtained as PI some of the datasets used in this papers.
    Last but not least, we thank the referee for their comments which improve the quality of this article.
    This work is based on observations collected at the European Southern Observatory at Paranal with SPHERE under ESO programmes 095.C-0607(A), 095.C-0487(A), 097.C-0060(A), 097.C-0949(A), 097.C-0865(F), 0101.C-0128(D), 0101.C-0016(A), with HST/STIS (program:~12998). In addition, this paper makes use of the following and also observations collected with ALMA (project ID: 2022.1.00968.S) andALMA data: ADS/JAO.ALMA\#2022.1.00968.1. ALMA is a partnership of ESO (representing its member states), NSF (USA) and NINS (Japan), together with NRC (Canada), MOST and ASIAA (Taiwan), and KASI (Republic of Korea), in cooperation with the Republic of Chile. The Joint ALMA Observatory is operated by ESO, AUI/NRAO and NAOJ. The National Radio Astronomy Observatory is a facility of the National Science Foundation operated under cooperative agreement by Associated Universities, Inc. 
    This work has made use of the High Contrast Data Centre, jointly operated by OSUG/IPAG (Grenoble), PYTHEAS/LAM/CeSAM (Marseille), OCA/Lagrange (Nice), Observatoire de Paris/LESIA (Paris), and Observatoire de Lyon/CRAL, and supported by a grant from Labex OSUG@2020 (Investissements d’avenir – ANR10 LABX56).
    C.D.  is part of Labex OSUG (ANR10 LABX56). C.D. acknowledges support from the European Research Council under the European Union’s Horizon 2020 research and innovation program under grant agreement No. 832428-Origins. 
    G.M.K. is supported by the Royal Society as a Royal Society University Research Fellow.
    A.A.S. is supported by the Heising-Simons Foundation through a 51 Pegasi b Fellowship. 
    T.D.P. is supported by a UKRI/EPSRC Stephen Hawking Fellowship.
    F.M. has received funding from the European Research Council (ERC) under the European Union's Horizon Europe research and innovation program (grant agreement No. 101053020, project Dust2Planets).
    J.M. acknowledges support from FONDECYT de Postdoctorado 2024 \#3240612
    V.F. acknowledges funding from the National Aeronautics and Space Administration through the Exoplanet Research Program under Grants No. 80NSSC21K0394 (PI: S. Ertel) and No 80NSSC23K0288 (PI: V. Faramaz).
    M.B. is supported by the European Union’s Horizon 2020 research and innovation programme under grant agreement no. 951815 (AtLAST).
\end{acknowledgements}

%\bibliographystyle{aa.bst}
%\bibliography{bib.bib}

\newpage

\begin{appendix}

\section{Log of the VLT/SPHERE observations of the system HD\,120326}

In Table~\ref{tab:obslog}, we list all the SPHERE observations on HD\,120326 to date.

\begin{table*}[h]
    \centering \small
    \caption{VLT/SPHERE observations of the system HD\,120326}
    \begin{tabular}{llCCCCCCCll}
\hline\hline\noalign{\smallskip} 
Epoch & Instr. Filt. & N_\text{exp} & \text{DIT} & \Delta\pi & \text{seeing} & \tau_0 & \text{airmass} & \text{strehl} & PI & ESO program \\
 &  &  & \text{(s)} & (\deg) &  (") & \text{(ms)} & & \text{ (RTC)} & &  \\
\noalign{\smallskip}  \hline\hline\noalign{\smallskip}
2015-04-09 & IRDIS H23 & 13 & 16 & 1.7 & \multirow{2}{*}{$-$} & \multirow{2}{*}{$-$} & 3.3 & - & Bonnefoy & 095.C-0607(A) \\
2015-04-09 & IFS YJ & 3 & 16 & 0.3 &  &  & 3.3 & - & Bonnefoy & 095.C-0607(A) \\
\noalign{\smallskip}\hline\noalign{\smallskip}
2015-08-01 & Zimpol IPRIM & 48^a & 28\,\times\,5 & - & 0.65\pm0.10 & 3.2\pm0.5 & 1.1 & - & Booth & 095.C-0487(A)\\ 
\noalign{\smallskip}\hline\noalign{\smallskip}
2016-04-05  & IRDIS H23 ($\dagger$) & 128 & 32 & \multirow{2}{*}{$37$} & \multirow{2}{*}{$1.06\pm0.13$} & \multirow{2}{*}{$3.4\pm0.6$} & \multirow{2}{*}{1.12} & 0.63 & Bonnefoy & 097.C-0060(A) \\
2016-04-05  & IFS YJ & 64 & 64 & &  &  &  & 0.63 & Bonnefoy & 097.C-0060(A) \\
\noalign{\smallskip}\hline\noalign{\smallskip}
2016-06-03  & IRDIS H23 ($\dagger$) & 80 & 32 & \multirow{2}{*}{$22$} & \multirow{2}{*}{$0.72\pm0.14$} & \multirow{2}{*}{$3.1\pm0.5$} & \multirow{2}{*}{$1.11$} & - & Hinkley & 097.C-0949(A) \\ 
2016-06-03  & IFS YJ & 40 & 64 &  &  &  & & - & Hinkley & 097.C-0949(A) \\ 
\noalign{\smallskip}\hline\noalign{\smallskip}
2016-06-13  & IRDIS J23 & 80 & 64 & 44.7 & 1.18\pm0.17 & 2.1\pm0.4 & 1.11 & 0.54 & GTO & 097.C-0865(F)\\
\noalign{\smallskip}\hline\noalign{\smallskip}
2018-06-01  & IRDIS Pol. BBH ($\ddagger$) & 80 & 64 & - & 0.51\pm0.06 & 4.4\pm0.8 & 1.11 & 0.69 & Boccaletti & 0101.C-0128(D) \\
\noalign{\smallskip}\hline\noalign{\smallskip}
2019-06-26 & IRDIS BBH & 56 & 32 & \multirow{2}{*}{$36$} & \multirow{2}{*}{$1.76\pm0.26$} & \multirow{2}{*}{$2.1\pm0.3$} & \multirow{2}{*}{$1.13$} & 0.53 & Bonnefoy  & 0101.C-0016(A)\\
2019-06-26 & IFS YJ & 28 & 64 &  &  &  &  & 0.53 & Bonnefoy  & 0101.C-0016(A)\\
\noalign{\smallskip}\hline\noalign{\smallskip}
2019-07-09 & IRDIS BBH ($\ddagger$) & 224 & 32 & \multirow{2}{*}{$58$} & \multirow{2}{*}{$0.63\pm0.07$} & \multirow{2}{*}{$3.1\pm0.4$} & \multirow{2}{*}{$1.13$} & 0.60 & Bonnefoy & 0101.C-0016(A)\\
2019-07-09 & IFS YJ ($\ddagger$) & 112 & 64 &  &  &  &  & 0.60 & Bonnefoy & 0101.C-0016(A)\\
\noalign{\smallskip}\hline\hline
    \end{tabular}
    \tablefoot{The results from this work are mainly based on the high-quality observations acquired on 2018-06-01 (polarized intensity) and on 2019-07-09 (total intensity), highlighted via ($\ddagger$). We also successfully detect the disk in the SPHERE/IRDIS total intensity observations of 2016-04-05 and 2016-06-03 ($\dagger$; see Fig.~\ref{fig:im_snr_red_ti}). The other datasets are of poorer quality  with for some no (2015-08-01) or almost no (2016-06-13, 2019-06-26) recovery of disk signal. The observation acquired in 2015-04-09 is not exploitable because the focal plane mask was missing during the observation sequence.  
    ($^a$) The $48$ exposures correspond to twelve HWP cycles, resulting in a total time on source of $N_\text{exp}\times\text{DIT}$. GTO stands for Guaranteed Time Observations (PI: Beuzit).}
    \label{tab:obslog}
\end{table*}

\section{NIR total intensity SPHERE observations \label{app:ti_obs}}

\begin{figure*}[h] \centering
    \includegraphics[height=4.2cm]{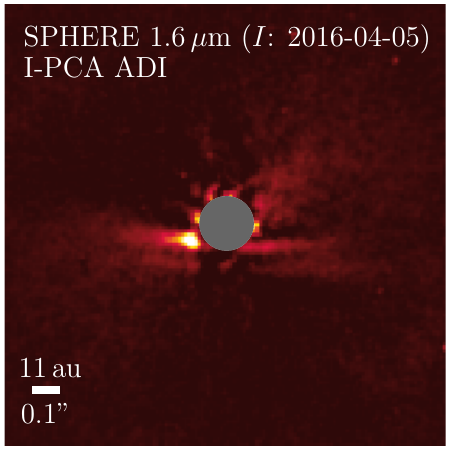} \quad
    \includegraphics[height=4.2cm]{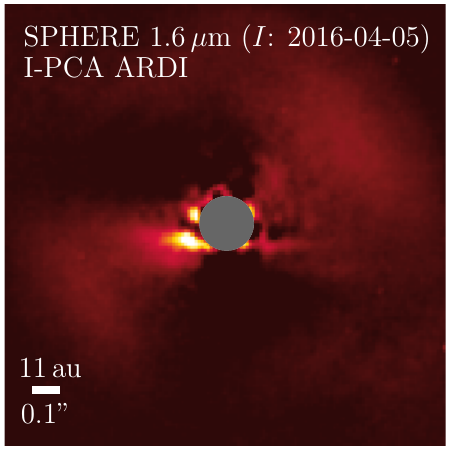} \quad
    \includegraphics[height=4.2cm]{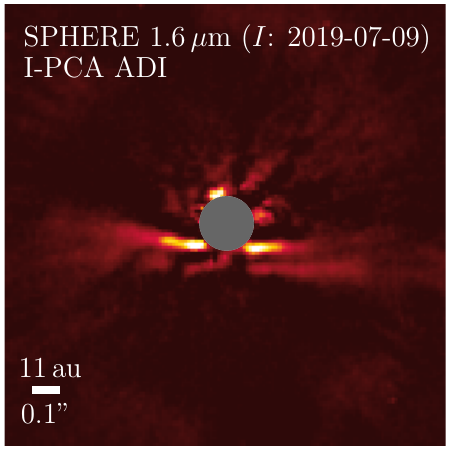} \quad
    \includegraphics[height=4.2cm]{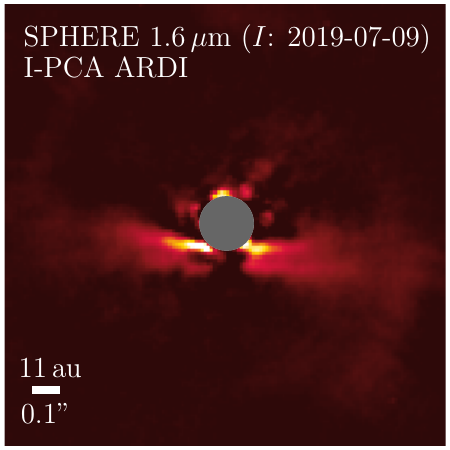} 
    \caption{Total intensity images in the broad band H and reduced using an iterative PCA~ADI or ADI+RDI algorithm. }
    \label{fig:im_ipca}
\end{figure*}

In the main text, we describe our PCA~ADI processing of the SPHERE/IRDIS data of the epochs acquired in total intensity in Fig.~\ref{fig:im_snr_red_ti}.
For the best epoch in the dual band H23 (2016-04-05) and the epoch in the broad band H (2019-07-09), we also processed the data with an iterative PCA, using ADI, RDI or ADI+RDI (ARDI) techniques. We only show the I-PCA~ADI and ARDI reductions in Fig.~\ref{fig:im_ipca}, because our I-PCA RDI processing is not able to retrieve the disk.

\begin{figure*}[h] \centering
    \includegraphics[height=4.2cm]{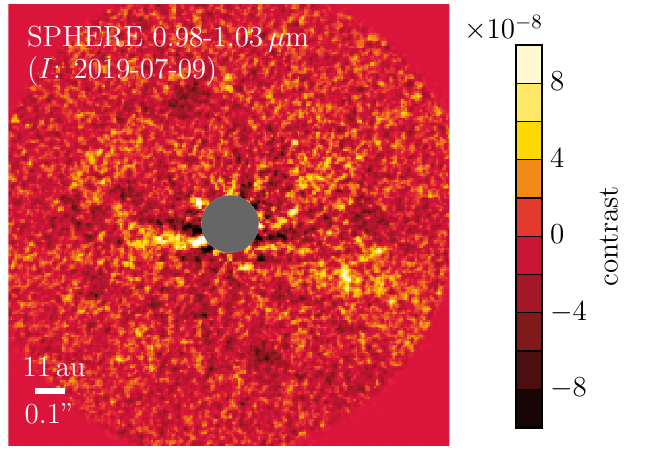} 
    \includegraphics[height=4.2cm]{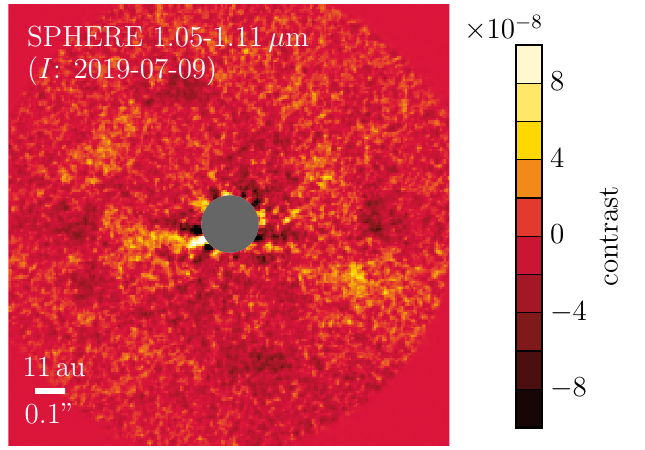} 
    \includegraphics[height=4.2cm]{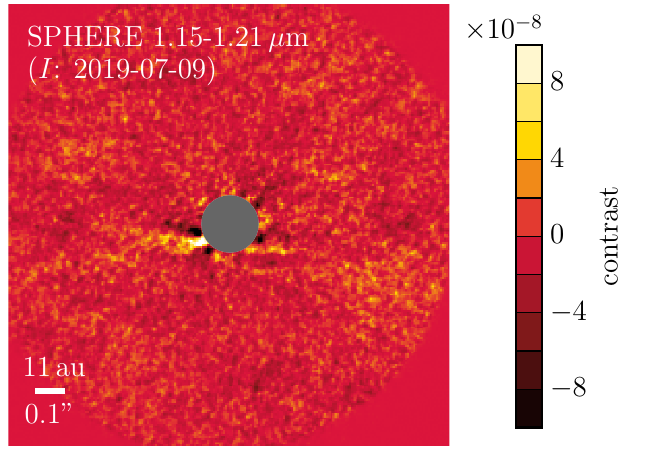} 
    
    \includegraphics[height=4.2cm]{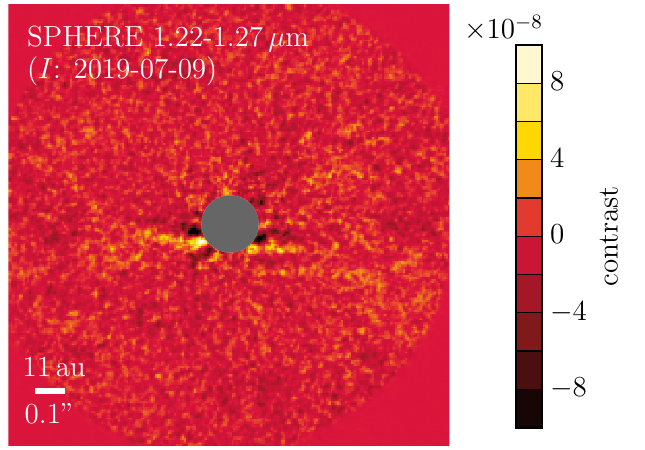}  
    \includegraphics[height=4.2cm]{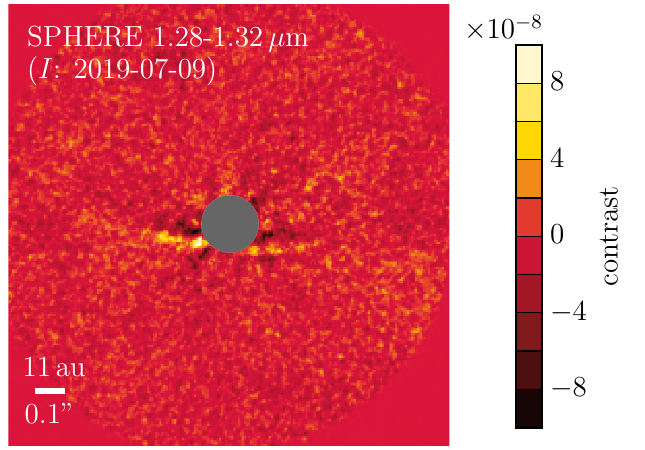}
    \includegraphics[height=4.2cm]{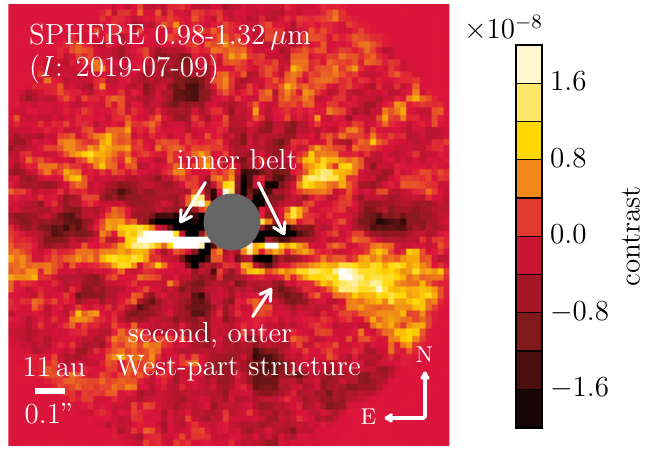}

    \caption{ Images depicting (from top-left to bottom-middle) the total intensity (“$I$”)  in different sub-coverages of the bands Y and J, reduced using PCA~ADI. The bottom-right image corresponds to the average of the PCA~ADI processed images over the full YJ bands, and that we spatially binned ($3\times3$\,pixels$^2\rightarrow1\times1$\,pixel$^2$) to reveal the western part of the second, outer structure. North is up, east is left for all the images. }
    \label{fig:im_red_ti_IFS}
\end{figure*}

Concerning SPHERE/IFS observations, we show our PCA~ADI processing of the best epoch 2019-07-09 in Fig.~\ref{fig:im_red_ti_IFS}. There are five IFS images with sub-coverages corresponding to binning of six spectral channels in the Y or J bands. The sixth IFS image is a mean of these thirty channels, thus covering YJ. We did not consider the first and last channels of the IFS, nor a few that are more impacted by artifacts.

\section{Morphological parameters of the belt based on SPHERE data \label{app:morpho_SPHERE}}

In this appendix, we include materials regarding our morphological analysis of the inner dust belt carried out in Sect.~\ref{sec:disk_modelling}.

\begin{figure}[h!] \centering
    \includegraphics[width=0.49\linewidth]{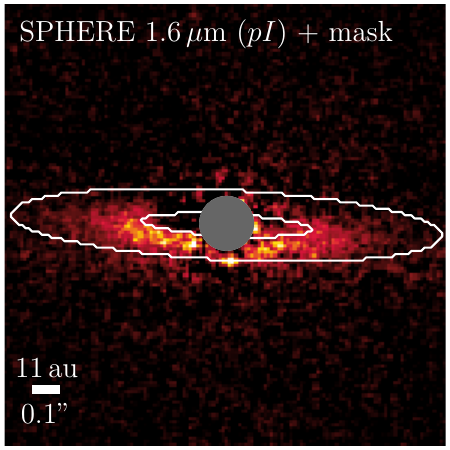} \hfill
    \includegraphics[width=0.49\linewidth]{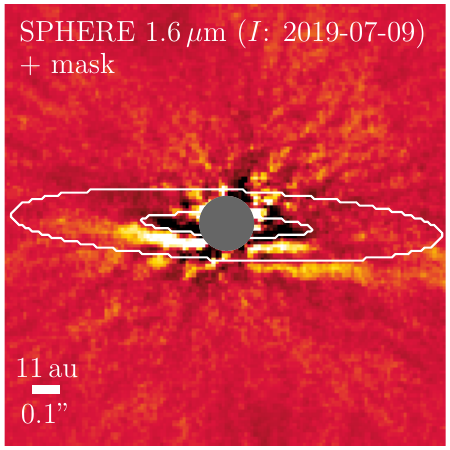} 
    \caption{ Region of the SPHERE polarized (left) and total (right) intensity images considered to model the inner belt in Sect.~\ref{sec:disk_modelling}.}
    \label{fig:im_disk_modelling_mask}
\end{figure}

\begin{figure*}[h]
    \includegraphics[width=\linewidth]
    {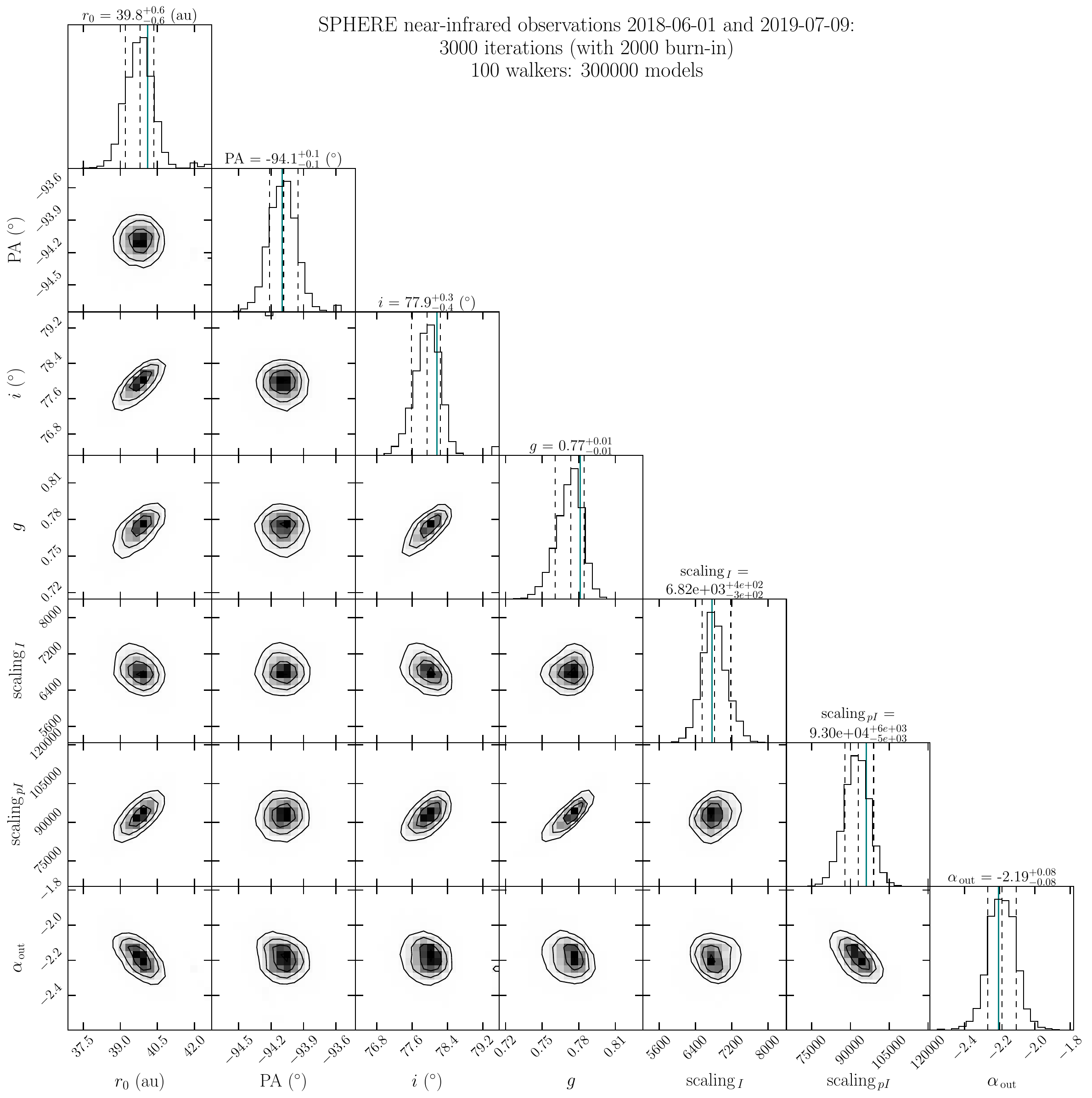}
    \caption{Posteriors of the \texttt{MCMC} exploration of the parameter space for the disk seen in NIR ($1.6~\mic$) SPHERE observations in polarized and total intensity. The vertical dashed lines represents the values of the median and at $\pm1\sigma$, which are defined as the $15.9\%$ and $84.1\%$ percentiles to encompass $68.2\%$ of the solutions for a given parameter. The blue vertical line represent the best-fitting parameters, which are the values minimizing the $\chi_r^2$. }
    \label{fig:cornerplot_morpho_sphere}
\end{figure*}

First, Fig.~\ref{fig:im_disk_modelling_mask} shows the areas considered to model the inner dust belt.
Second, we describe below the equations setting the geometry of the disk, show the posteriors of our simulations jointly modeling the total and polarized intensity observations in Fig.~\ref{fig:cornerplot_morpho_sphere}. Third, in Tables~\ref{tab:mcmc_results_nir_indpt_I} and~\ref{tab:mcmc_results_nir_indpt_pI}, we display the best-fitting parameters obtained when we independently modeled either the total or polarized intensity observations, respectively.

Following \citet{augereau_hr_1999}, the dust density distribution may be expressed in cylindrical coordinates as,
\begin{align}
    \rho(r,\theta,z)
    &~=~\rho_0~\times~R(r,\theta)~\times~ Z(r,z)\,,
    \label{eq:DD_dust_density_distri_1}
\end{align}
where $R(r,\theta)$ and $Z(r,z)$ describe the radial and vertical distribution profile, respectively, and $\rho_0$ is the normalized grain density.

This can be expressed as 
\begin{align}
    \rho(r,\theta,z)
    &~=~\rho_0 \times \pac{ \frac{2}{ \pa{\frac{r}{\tilde{r}(\theta)}}^{-2\alpha_\text{in}} + \pa{\frac{r}{\tilde{r}(\theta)}}^{-2\alpha_\text{out}} }}^{1/2} \times~\exp{\pac{-\pa{\frac{z}{\xi(r)}}^\gamma}}\,,
    \label{eq:DD_dust_density_distri_2}
\end{align}
where $\gamma$ accounts for the shape of the vertical profile, either gaussian ($\gamma=2$), spread out ($\gamma>>2$) or stepped  ($\gamma<<2$), $\xi(r)$ is the scale height of the ring. It is parameterized  with $\beta$ the flaring of the disk $a$ the semi-major axis, and $e$ the eccentricity
as 
\begin{equation}
    \xi(r)\;=\;\xi_0\,\pa{\frac{r}{a\,(1\,-\,e^2)}}^\beta\,.
    \label{eq:DD_scale_height}
\end{equation}
If the disk is circular, $\tilde{r}(\theta)$ is equal to the constant reference radius $r_0$, and otherwise parameterized following \citet{milli_nearinfrared_2017} via 
\begin{equation}
    \tilde{r}(\theta) \;=\; \frac{a\,(1\,-\,e^2)}{1\,+\,e\cos(\theta)}\,,
    \label{eq:DD_ellipse}
\end{equation}
where $\theta$ is the polar angle.

In NIR observations, disk structures are seen via the scattering of the stellar light on the dust particles of the disk. The scattering phase function of a debris disk in total intensity light may be modeled for the sake of convenience with the Henyey-Greenstein (HG) function \citep{henyey_diffuse_1941} as follows
\begin{equation}
    \text{HG}(g,\phi)~=~\frac{1}{4\pi}\times\frac{1\,-\,g^2}{\pa{1\,-\,2\,g\cos{(\phi)}\,+\,g^2}^{3/2}}\,,
    \label{eq:HG_1compo_ti}
\end{equation}
which is parameterized by the HG scattering anisotropy parameter $g$. If there is only backward scattering, $g$ is equal to $-1$, if only forward scattering,  $g$ is equal to $1$, and if the scattering is isotropic, $g$ is equal to $0$. For debris disks, $g$ is typically greater than $0$ because small dust grains are known to be forward scattering.

For polarimetric observations, we multiply the SPF by the Rayleigh scattering function as follows
\begin{equation}
    \text{HG}(g,\phi)~\rightarrow~ \text{HG}(g,\phi) \;\times\; \frac{1\,-\,(\,\cos(\phi)\,)^2}{1\,+\,(\,\cos(\phi)\,)^2}\,.
    \label{eq:HG_1compo_polar}
\end{equation}

\section{Morphological parameters of the belt based on ALMA data \label{app:morpho_ALMA}} 

\begin{figure*}[h] \centering
    \includegraphics[width=0.7\linewidth]{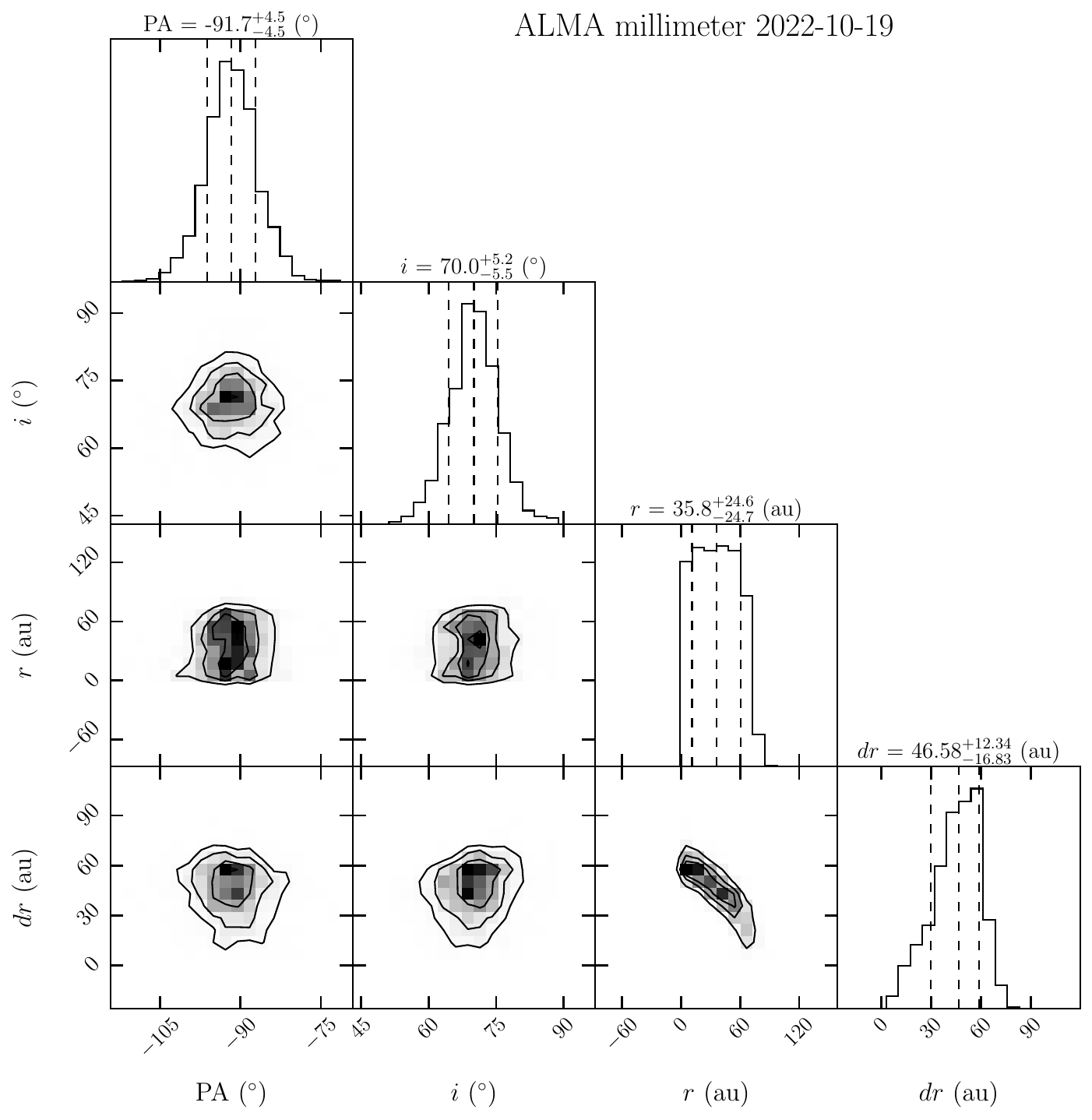}
    \caption{Posteriors of the \texttt{MCMC} exploration of the parameter space for the disk seen in continuum millimeter ($1.3~\mm$) interferometric ALMA observations. The parameters are: position angle (PA), inclination ($i$), radius ($r$), and width ($dr$). }
    \label{fig:cornerplot_morpho_alma}
\end{figure*}

Figure~\ref{fig:cornerplot_morpho_alma} shows some of the posteriors of the simulations constraining the morphology of the debris disk based on the ALMA data (Sect.~\ref{sec:alma_morpho}).

\begin{table}[h] 
    \setlength{\tabcolsep}{4pt}
    \centering \small
    \caption{Morphology of the belt based on the \texttt{MCMC} exploration fitting the SPHERE NIR total intensity (2019-07-09) data only. } 
    \begin{tabular}{cCCC} 
    \hline \hline  \noalign{\smallskip}
        Parameter & \text{Range} & \text{Best fit}~\chi^2_\text{r,min} & \text{Median} \pm 1\sigma \\
    \noalign{\smallskip} \hline \hline  \noalign{\smallskip}
        $r_0$ (au) & [10, 130] & 42.8 & 43.2^{+1.0}_{-0.9} \\ \noalign{\smallskip}
        $PA$ ($\dego$) & [-120, -60] & -94.5 & -94.5^{+0.1}_{-0.2} \\ \noalign{\smallskip}
        $i$ ($\dego$) & [0, 90] & 79.4 & 79.4\pm0.3 \\ \noalign{\smallskip}
        $g$ & [0.05, 0.999] & 0.81 & 0.81\pm0.01 \\ \noalign{\smallskip}
        scaling ($I$) & [100, 10^8] & 6.2 \times 10^3 & (6.0^{+0.4}_{-0.3}) \times 10^3 \\ \noalign{\smallskip}
        $\alpha_\text{out}$ & [-20, -1.1] & -2.27 & -2.34^{+0.10}_{-0.13} \\
     \noalign{\smallskip} \hline \hline  \noalign{\smallskip}
    \end{tabular}
    \label{tab:mcmc_results_nir_indpt_I}
\end{table}

\begin{table}[h] 
    \setlength{\tabcolsep}{4pt}
    \centering \small
    \caption{Morphology of the belt based on the \texttt{MCMC} exploration fitting the SPHERE NIR polarized intensity (2018-06-01) data only.  } 
    \begin{tabular}{cCCC} 
    \hline \hline  \noalign{\smallskip}
        Parameter & \text{Range} & \text{Best fit}~\chi^2_\text{r,min} & \text{Median} \pm 1\sigma \\
    \noalign{\smallskip} \hline \hline  \noalign{\smallskip}
        $r_0$ (au) & [10, 130] & 38.3 & 38.5^{+1.1}_{-0.7} \\ \noalign{\smallskip}
        $PA$ ($\dego$) & [-120, -60] & -93.5 & -93.6^{+0.2}_{-0.3} \\ \noalign{\smallskip}
        $i$ ($\dego$) & [0, 90] & 77.0 & 77.1^{+0.6}_{-0.5} \\ \noalign{\smallskip}
        $g$ & [0.05, 0.999] & 0.73 & 0.74^{+0.03}_{-0.02} \\ \noalign{\smallskip}
        scaling ($pI$) & [100, 10^8] & 7.8 \times 10^4 & (7.9^{+0.4}_{-0.6}) \times 10^4 \\ \noalign{\smallskip}
        $\alpha_\text{out}$ & [-20, -1.1] & -2.01 & -2.00\pm0.12 \\
     \noalign{\smallskip} \hline \hline  \noalign{\smallskip}
    \end{tabular}
    \label{tab:mcmc_results_nir_indpt_pI}
\end{table}

\newpage

\section{The background contaminant\label{app:PMD}}

\begin{figure} \centering 
    \includegraphics[width=0.9\linewidth]{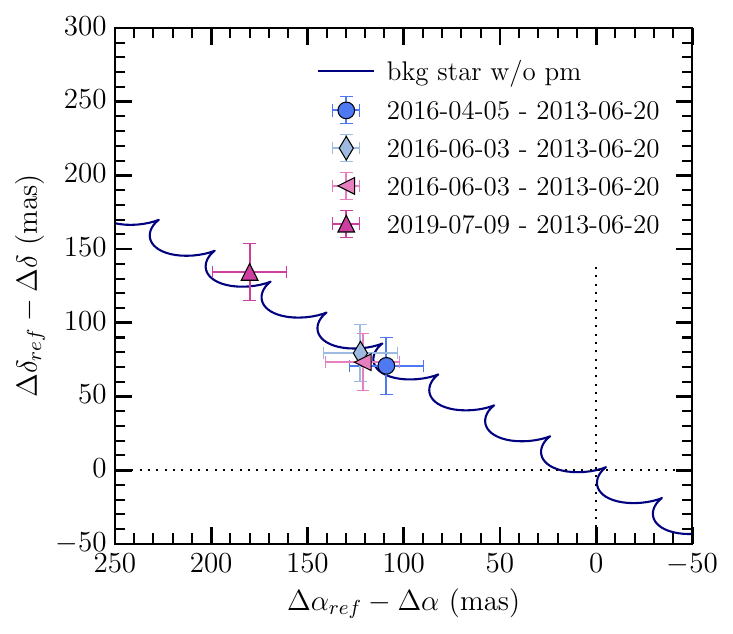}
    \caption{Difference of the relative astrometry of the bright point-source lying on the spiral-like feature (see Fig.~\ref{fig:im_HST_flux}) between the HST/STIS epoch (2013-06-20, used as reference) and the SPHERE/IRDIS epochs (2016 to 2019, see Table~\ref{tab:obslog}). The blue solid curve shows the expected track of a background object, based on the parallactic and proper motions of the primary star HD\,120326. The point source is clearly identified as a background object.
    }
    \label{fig:PMD}
\end{figure}

We show here that the bright point-like source on the extended spiral-like feature (see Fig.~\ref{fig:im_HST_flux}) is a background  star and investigate whether it could explain the asymmetry of the extended feature seen with HST/STIS. This bright point-like source was detected in the five epochs HST/STIS and SPHERE/IRDIS from 2013 to 2019. By comparing the relative motion of this bright point source for different epochs, we show in Fig.~\ref{fig:PMD} that this  source follows the track of a background object. This track is determined by  using the parallactic and proper motions of the primary star. HD\,120326 has a proper motion of  ($-29.0$~mas/yr in right-ascension, and $-21.0$~mas/yr  in declination \citep{gaiacollaboration_gaia_2021}.

Regarding the photometry of the background object, we derived it based on SPHERE/IRDIS data in J2, J3,  H2, H3, and BBH bands (see Table~\ref{tab:appD_bkg_photometry}). We extracted its consrast using a circular aperture of radius $2.5$~FWHM on the SPHERE/IRDIS non-ADI\footnote{ The non-ADI postprocessing consists in de-rotating with respect to the parallactic angles the coronagraphic images of the temporal sequence, and averaging them. This increases the $S/N$ of the data and avoids self-subtraction effects of the ADI technique \citep{milli_impact_2012}. } postprocessed images by the High-Contrast Data Center \citep[HC-DC;][]{delorme_sphere_2017} produced using the SpeCal pipeline \citep{galicher_astrometric_2018}.
We subtracted the sky contribution in the contrast measurements. We computed the sky contribution in circular apertures of $2.5$ FWHM,  centered at the same separation than the background object, and with a close position angle. At $1\sigma$, the background object does not show a significant color in J2--J3 ($0.28\pm0.35$~mag) and  H2--H3 ($0.12\pm0.16$~mag), but it does between J3--H2 ($1.12\pm0.34$~mag). This could hint to 
a M-type object (from M0 to M9) with a J3--H2 color reddened by the presence of dust along the line of sight, or to a L-type object (from L0 to L9), as described in Appendix C in \citet{bonnefoy_gj_2018} and references therein. The absolute magnitude of a M-type object is about between  $6$ and $11$~mag in H2, while for a L-type object it is between $11$ and $15$~mag in H2 \citep{bonnefoy_gj_2018}.  By deriving the distance of the background object to match its absolute magnitude in H2 with that of a M or L-type object, we would expect the object to have a distance between $42$ and $420~\pc$ if M type, or between $6.7$ and $42~\pc$ if L type. Since the object follows the track of a background object, without showing additional proper motion (Fig.~\ref{fig:im_HST_flux}), we would rather expect it to be a background M-type object than a foreground L/M-type object. 

\begin{table}[h] 
    \setlength{\tabcolsep}{4pt}
    \centering \small
    \caption{ Photometry of the background object based on SPHERE/IRDIS observations.  } 
    \begin{tabular}{cCCCCC} 
    \hline \hline  \noalign{\smallskip}
        Band & \lambda & d_\lambda & \text{Contrast} & \text{Rel. app. mag.} & \text{App. mag.}\\
         & (\mic) & (\mic) & (\times 10^{-3})& \text{(mag)} & \text{(mag)}\\
    \noalign{\smallskip} \hline \hline  \noalign{\smallskip}
        J2 & 1.190 & 0.042 & 0.74\pm0.04 & 7.82 \pm 0.06 & 15.53 \pm 0.07 \\
        J3 & 1.273 & 0.046 & 0.96\pm0.06 & 7.54 \pm 0.34 & 15.25 \pm 0.34 \\
        H2($\dagger$)& 1.593 & 0.055 & 2.37\pm0.09 & 6.56 \pm 0.04 & 14.13 \pm 0.05 \\
        H3($\dagger$) & 1.667 & 0.056 & 2.64\pm0.07 & 6.45 \pm 0.15 & 14.01 \pm 0.15 \\
        H2($\ddagger$)& 1.593 & 0.055 & 2.46\pm0.10 & 6.52 \pm 0.05 & 14.09 \pm 0.05 \\
        H3($\ddagger$) & 1.667 & 0.056 & 2.83\pm0.20 & 6.37 \pm 0.23 & 13.94 \pm 0.23 \\
        BBH & 1.625 & 0.290 & 2.25\pm0.10 & 6.62 \pm 0.04 & 14.19 \pm 0.05 \\
    \noalign{\smallskip} \hline \hline  \noalign{\smallskip}
    \end{tabular}
    \tablefoot{The two last columns correspond to the relative apparent magnitude and the apparent magnitude. The magnitude of the host star used is  $7.708\pm0.026$~mag in J2 and J3, and  $7.568\pm0.049$~mag in H2, H3 and BBH filters \citep{cutri_vizier_2003}. Two SPHERE/IRDIS observations exist in H2 and H3, see Table~\ref{tab:obslog}. The results based on the epoch 2016-04-05 are indicated with ($\dagger$), those based on the epoch 2016-06-03 with ($\ddagger$).}
    \label{tab:appD_bkg_photometry}
\end{table}

\begin{figure} \centering 
    \includegraphics[width=\linewidth]{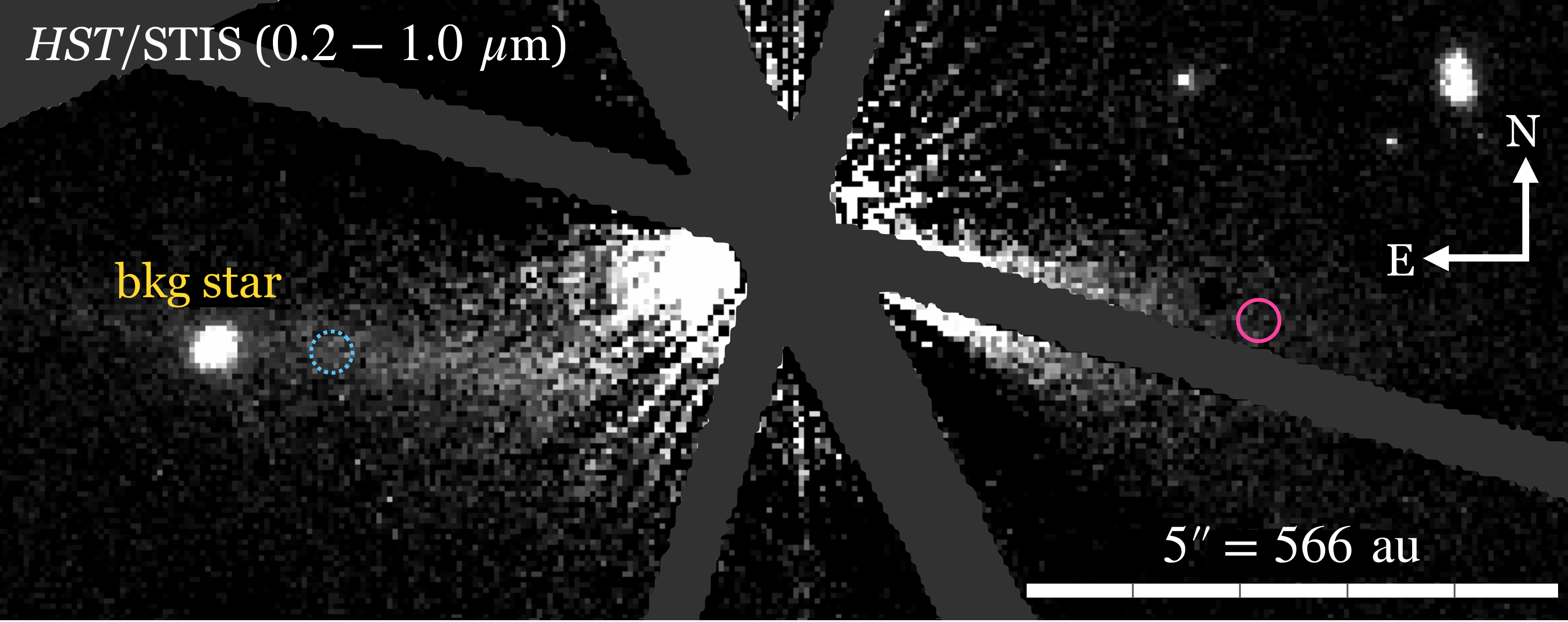}
    \caption{ Location of the two regions of interest of dust grains in the extended dust emission seen in the HST/STIS reduced image. The dotted blue circle corresponds to the region located at a projected distance close to the background star, while the solid pink circle corresponds to the region located a projected distance  far to the background star. Unlike Fig.~\ref{fig:im_HST_flux}, no spatial binning is applied on this image.
    }
    \label{fig:hst_appD}
\end{figure}

\begin{figure} \centering 
    \includegraphics[width=0.7\linewidth]{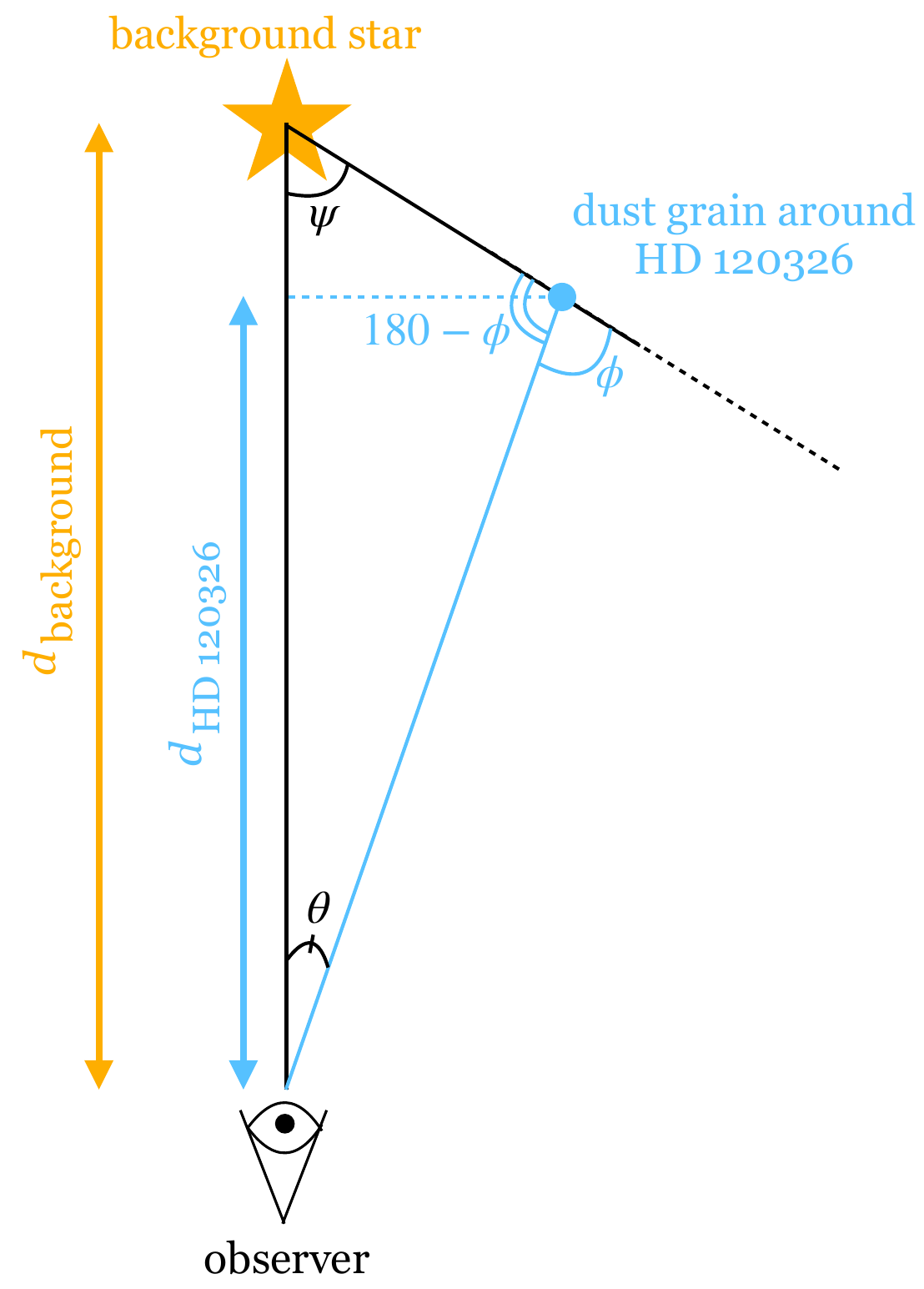}
    \caption{ Sketch representing the parameters (angles and distances) used to express the scattering angle $\phi$ of the background stellar light scattered by a dust grain in the system HD\,120326. 
    }
    \label{fig:hst_sketch_appD}
\end{figure}

\medskip

Below, we investigate whether the alignment of the background object with the southeastern part of the dust seen with HST/STIS could explain why the disk is brighter in the southeastern part than in the southwestern part, which would results in a visual (and not dynamical) asymmetry. We stress that such investigation is limited because the southwestern part of the disk lacks of spatial coverage in the HST/STIS image.

We consider two regions of interest where there are dust grains, one located at a close projected separation of the background star and another one located further away (see Fig.~\ref{fig:hst_appD}).
By deriving the mean average flux within each of this circular aperture (of radius five pixels, i.e., $254$~mas), we obtained a flux of $0.94\pm0.03~\microjy$ and $0.52\pm0.03~\microjy$, respectively.

The first goal is to express the scattering angle ($\phi$) as a function of the distance between the system HD\,120326 and the observer ($d_\text{HD\,120326}$), the distance between the background object and the observer ($d_\text{background}$), the angular separation between the background object and the dust grain of interest ($\theta$, which also represents the angle between the line of sight observer-background object and the scattered wave on the dust grain of interest), see  Fig.~\ref{fig:hst_sketch_appD}.
We assume that the distance between the dust grains around HD\,120326 seen with HST/STIS (Fig.~\ref{fig:hst_appD}) and the observer is  approximately equal to the distance between the host star of HD\,120326 and the observer,  denoted as $d_\text{HD\,120326}$.

By considering the triangle whose corners are the background star, the dust grain of interest and the observer, and the fact that the sum of angles in a triangle is equal to $180\dego$, one can express 
\begin{align}
    \theta \,+\, \psi \,+\, (180 \,-\, \phi) \; &= \; 180  \,, \\ 
    \psi \; &= \; \phi \,-\, \theta\,, \label{eq:background_psi_180}
\end{align}
where  $\psi$ is the angle between the line of sight observer-background star and the incident wave from the background star on the dust grain of interest.
As $\psi$ is very small, $\tan(\psi) \sim \psi$, hence one can express
\begin{equation}
    \psi \; \sim \; \frac{d_\text{HD\,120326} \times \theta}{d_\text{background} \;-\; d_\text{HD\,120326} }\,, \label{eq:background_psi_toa}
\end{equation}
and so by injecting Eq.~\ref{eq:background_psi_180} in Eq.~\ref{eq:background_psi_toa}, one gets
\begin{equation}
    \phi \; \sim \; \theta \,+\, \frac{d_\text{HD\,120326} \times  \theta}{d_\text{background} \;-\; d_\text{HD\,120326}}\,.
\end{equation}
This results in the following expression for the  scattering angle $\phi$,
\begin{equation} \label{eq:phi_dbkg}
    \phi \; \sim \; \frac{ \theta}{1 \;-\; d_\text{HD\,120326} \,/\, d_\text{background}}\,.
\end{equation}

By assuming that the background object is located at $118.3~\pc$, namely, $5~\pc$ further away than the system HD\,120326 ($d_\text{HD\,120326} = 113.3~\pc$), we derived scattering angles $\phi = 0.0073\dego$ and $\phi = 0.065\dego$, for dust grains at the close ($\theta = 22~\pixs = 1.1$" $= 0.00031\dego$\footnote{ The STIS plate scale is equal to $0.05078$~pixels/arcsec$^2$}) and far ($\theta = 195~\pixs = 9.9$" $= 0.0028\dego$) projected separations in Fig.~\ref{fig:hst_appD}, respectively.
The distance of the background object is very likely underestimated, but this is done on purpose to maximize the difference of the scattering angles, and so to maximize the difference of the scattering phase function values, to be able to explain such a strong flux asymmetry.

By assuming Mie's theory, astrosilicate grains distributed from $0.1$ to $100~\mic$ with a power law in $-3.5$, we computed the scattering phase function using the open-access tool \texttt{optool}\footnote{ available here \url{https://github.com/cdominik/optool/}} \citep{dominik_optool_2021} at the central wavelength of the STIS filter, $0.59~\mic$. Due to sampling constraints, we obtained with the code \texttt{optool}  values of the scattering phase function of  $1.97 \times 10^{6}$ and $1.54 \times 10^{6}$ for scattering angles of $0\dego$ and $0.0625\dego$, respectively. This corresponds to a ratio of $1.28$, so a positive increase of $28\%$. By considering the scattering angles of interest, $0.0073\dego$ and $0.065\dego$, the positive increase correspond to  is therefore  $28\% \times (0.065 - 0.0073) / (0.0625 - 0) = 26\%$. Consequently, we expect that the dust grains with a scattering angle of $0.065\dego$ scatter $26\%$ less starlight than those with a scattering angle of $0.0073\dego$. This results in an expected flux of $ 0.94 / 1.26 = 0.75 ~\microjy$ at  $0.065\dego$. This is higher than the flux ($0.52\pm0.03~\microjy$) measured in the STIS image at such a location.

One can formalize the expression of the flux expected at the far location $F_\text{far,expected}$  as a function of $\phi_\text{far}-\phi_\text{close}$, and so as a function of $d_\text{background}$ using the Eq.~\eqref{eq:phi_dbkg}, as follows:
\begin{align} \label{eq:ratio_theory}
    &F_\text{far,predic} \notag \\ 
    \;&=\; \frac{F_\text{close,predic}}{\pa{\displaystyle \frac{ \texttt{SPF}_\texttt{optool}\texttt{(}\phi_1\texttt{)}}{ \texttt{SPF}_\texttt{optool}\texttt{(}\phi_2\texttt{)}}-1} \pa{\displaystyle \frac{\phi_\text{far}-\phi_\text{close}}{\phi_2-\phi_1}}\;+\;1}  \smallskip \\
    &=\; \frac{F_\text{close,predic}}{\pa{\displaystyle \frac{ \texttt{SPF}_\texttt{optool}\texttt{(}\phi_1\texttt{)}}{ \texttt{SPF}_\texttt{optool}\texttt{(}\phi_2\texttt{)}}-1} 
    \times
    \displaystyle \frac{\pa{\theta_\text{far}-\theta_\text{close}}}{\pa{\displaystyle 1-\frac{d_\text{HD\,120326}}{d_\text{background}}} \pa{\phi_2-\phi_1}}\;+\;1}
\end{align}
By isolating $d_\text{background}$, we get
\begin{align} \label{eq:dbkg}
    &d_\text{background} \notag \\ 
    \;&=\; \frac{d_\text{HD\,120326} \times \pa{\frac{\displaystyle F_\text{close,obs}}{\displaystyle F_\text{far,obs}}-1}}{ 
    \pa{\frac{\displaystyle F_\text{close,obs}}{\displaystyle F_\text{far,obs}}-1}
    -
    \pa{\frac{\displaystyle \texttt{SPF}_\texttt{optool}\texttt{(}\phi_1\texttt{)} }{\displaystyle \texttt{SPF}_\texttt{optool} \texttt{(}\phi_2\texttt{)} } -1}  
    \pa{\frac{\displaystyle \theta_\text{far}-\theta_\text{close}}{\displaystyle  \displaystyle \phi_2-\phi_1}} 
    }\,. 
\end{align}
Using the values given previously, we found that the distance of the background object is expected to be $114.9~\pc$ in order to reproduce the observed flux asymmetry between the close and far projected locations shown in Fig.~\ref{fig:hst_appD}.

We also tested different assumptions on the distribution of dust grains ($0.1$--$1~\mic$, $0.1$--$10~\mic$, $0.1$--$50~\mic$, $1$--$10~\mic$, $1$--$50~\mic$,  $1$--$100~\mic$). Regarding the distributions of dust grains $0.1$--$50~\mic$, $1$--$50~\mic$, and, $1$--$100~\mic$, the distance of the background object is predicted to be smaller, between $113.5$ and $114.7~\pc$. In the case of the other distributions of dust grains ($0.1$--$1~\mic$, $0.1$--$10~\mic$, and, $1$--$10~\mic$), the scattering phase function is actually slightly higher at $\phi_2$ than $\phi_1$, so the flux predicted at $\phi_\text{far}$ would be slightly higher than the flux at $\phi_\text{close}$. 

To conclude, based on our investigation presented in this appendix, we do not expect that the alignment of the background object with the southeastern part of the dust could explain why the flux asymmetry observed with HST/STIS is that strong between the southeastern part and the southwestern part of the debris disk around HD\,120326.

\end{appendix}

\end{document}